\documentclass[11pt, letterpaper]{article}

\usepackage[left=1in, right=1in, top=1in, bottom=1in]{geometry}
\usepackage[T1]{fontenc}
\usepackage{bm}
\usepackage{type1cm}
\usepackage{lettrine}
\usepackage{amsmath,amssymb,amsthm}
\usepackage{moreverb}
\usepackage{mathtools}
\usepackage{amsmath}
\usepackage{amssymb}
\usepackage{algorithmic}
\usepackage{graphics}
\usepackage{graphicx}
\usepackage{subfigure}
\usepackage{caption}
\usepackage{extarrows}
\usepackage{color}
\usepackage{framed}
\usepackage{wrapfig}
\usepackage{bm}
\usepackage{mathrsfs}
\usepackage{mathabx}
\usepackage{multirow}
\usepackage{longtable}
\usepackage[unicode=true,bookmarks=true]{hyperref}
\usepackage{paralist}
\usepackage{indentfirst}
\usepackage{relsize}
\usepackage{extarrows}

\usepackage{upgreek}
\usepackage{bm}
\usepackage{mwe}
\usepackage[dvipsnames,table]{xcolor}
\usepackage{booktabs}
\usepackage{authblk}
\usepackage{lettrine}
\usepackage{type1cm}
\usepackage{threeparttable}
\usepackage[sort&compress,numbers]{natbib}
\usepackage[figurename=Figure]{caption}
\usepackage[normalem]{ulem}

\graphicspath{ {./Figures/} }
\usepackage[font=footnotesize,labelfont=bf]{caption}

\providecommand{\keywords}[1]{\textbf{\textit{Keywords: }} #1}

\hypersetup{
bookmarksopen=true,
bookmarksnumbered=true,
pdftoolbar=true,
pdfmenubar=true,
pdffitwindow=false,
pdfstartview={FitH},
pdftitle={Modeling Fluid Dynamics with Physics Pre-trained Operator Learning},
pdfauthor={Rui Zhang, Qi Meng, Han Wan, Mengtao Yan, Zeyu Li, Wang Han, Yang Liu, Zhi-Ming Ma, Hao Sun},
pdfsubject={Computational fluid dynamics and neural operators},
pdfkeywords={OmniFluids, neural operator, physics-informed learning, computational fluid dynamics},
pdfnewwindow=true,
colorlinks=true,
linkcolor=blue,
citecolor=blue,
filecolor=blue,
urlcolor=blue
}

\makeatletter
\AtBeginDocument{\check@mathfonts}
\makeatother

\usepackage{tikz-cd,placeins,siunitx,fancyhdr,algorithm}
\usepackage[figuresright]{rotating}

\definecolor{mygray}{gray}{0.8}
\ifdefined\pdfsuppressptexinfo\pdfsuppressptexinfo=15\fi
\begin{document}

\title{\textbf{{Modeling Fluid Dynamics with Physics Pre-trained Operator Learning}}}

\author[1,$^\dag$]{Rui Zhang}
\author[2,$^\dag$]{Qi Meng}
\author[1]{Han Wan}
\author[1]{{Mengtao Yan}}
\author[3]{{Zeyu Li}}
\author[3]{{Wang Han}}
\author[4,5]{Yang Liu}
\author[2]{\\Zhi-Ming Ma}
\author[1,$^*$]{Hao Sun}

\affil[1]{\small Gaoling School of Artificial Intelligence, Renmin University of China, Beijing, China}
\affil[2]{\small Academy of Mathematics and Systems Science, Chinese Academy of Sciences, Beijing, China}
\affil[3]{\small {School of Astronautics, Beihang University, Beijing, China}}
\affil[4]{\small School of Engineering Science, University of Chinese Academy of Sciences, Beijing, China}
\affil[5]{\footnotesize State Key Laboratory of Nonlinear Mechanics, Institute of Mechanics, Chinese Academy of Sciences, Beijing, China \vspace{18pt}}

\affil[$\dag$]{Equal contribution}
\affil[*]{Corresponding author}

\date{}

\maketitle

\normalsize

\vspace{-18pt} 
\begin{abstract}
\small
Computational fluid dynamics (CFD) drives progress in numerous scientific and engineering fields, {yet accurate simulations remain computationally prohibitive}. While machine learning approaches offer computational acceleration, they typically specialize in single physical systems or require extensive training data, hindering their applicability in highly nonlinear and 3D flow scenarios. To overcome these limitations, we propose OmniFluids, a pure physics pre-trained model that captures fundamental fluid dynamics laws and adapts efficiently to diverse downstream tasks with minimal data. We develop a training framework combining physics-only pre-training, coarse-grid operator distillation, and few-shot fine-tuning. This enables OmniFluids to retain broad physics knowledge while delivering fast and accurate predictions. Architecturally, OmniFluids integrates a mixture of operators, a multi-frame decoder, and factorized Fourier layers, seamlessly incorporating physics-based supervision while allowing efficient and scalable modeling of diverse tasks. Extensive tests on various 2D and 3D benchmarks show that OmniFluids outperforms state-of-the-art AI-driven methods in terms of flow field prediction and turbulence statistics. {It delivers $9.9$--$21.2\times$ inference speedups over traditional solvers on 2D and 3D Navier--Stokes benchmarks while maintaining comparable accuracy}, and achieves strong performance on downstream tasks such as inverse problems, reconstruction, and control. This work demonstrates the potential of training a unified CFD solver exclusively from physics knowledge, offering an efficient approach for generalizable modeling across complex fluid systems.
\end{abstract}

\keywords{Neural operator, pre-trained model, physics-informed learning, computational fluid dynamics}

\vspace{12pt} 
\section*{Introduction}

Fluid dynamics plays a fundamental role in various scientific and engineering fields. Understanding and predicting fluid behavior is essential for designing aircraft~\cite{masini2020analysis}, forecasting weather~\cite{bauer2015quiet}, and simulating biological flows~\cite{yu2019active}. Over the past century, traditional computational fluid dynamics (CFD) methods have been developed as primary approaches for simulating fluid dynamics~\cite{lomax2002fundamentals}. However, these numerical methods require fine-grained spatiotemporal discretization to achieve high accuracy and stability, making large-scale 3D and strongly nonlinear problems computationally expensive, especially for real-time applications~\cite{karniadakis2021physics}.

Over the past decade, advances in artificial intelligence (AI) have motivated researchers to explore data-driven approaches to solving partial differential equations (PDEs) in fluid mechanics with improved efficiency. One of the most representative developments is the physics-informed neural network (PINN)~\cite{raissi2019physics,raissi2020hidden}. PINNs use automatic differentiation to construct loss functions from PDE formulations, avoiding the complicated procedures of mesh generation and offering flexibility in dealing with complex geometries~\cite{raissi2020hidden, zhongkai2024pinnacle}. Moreover, PINNs demonstrate advantages in solving inverse problems, especially in scenarios with noisy or sparse data~\cite{chen2021physics, zhang2022drvn, xu2023discovery}. However, PINNs require considerable time to retrain when dealing with new CFD tasks (e.g., altering initial conditions or forcing terms). The training process is also sensitive to hyperparameter tuning, and optimizing the loss function can be unstable~\cite{krishnapriyan2021characterizing, grossmann2024can}.

Unlike PINNs, neural operators aim to learn mappings between function spaces, such as mapping initial conditions to their solution fields~\cite{azizzadenesheli2024neural}. For example, DeepONets~\cite{lu2021learning,kontolati2024learning, kopanivcakova2025deeponet} leverage branch and trunk networks to embed initial conditions and spatiotemporal coordinates for solving PDEs. Fourier Neural Operator (FNO)~\cite{li2021fourier,tran2023factorized, wang2024prediction} employs the Fast Fourier Transform (FFT) to perform learning in the spectral domain, achieving effective functional space mapping. Graph neural networks~\cite{brandstetter2022message, zeng2025phympgn,mi2025conservation} or Transformers~\cite{li2023scalable, li2024transformer} extend neural operators to complex geometry problems. Beyond deterministic models, generative models have been explored for modeling and inpainting fluid flows~\cite{du2024conditional,gao2024generative}. However, the majority of existing neural operators demand large amounts of training data to achieve reasonable performance, and acquiring such high-fidelity labeled data can be extremely costly. 

To alleviate data dependency, physics-aware neural operators seek to enable few-shot or zero-shot learning by integrating physical knowledge~\cite{faroughi2024physics}. Depending on how they incorporate physical priors, these methods can be divided into physics-informed and physics-encoded methods. Physics-informed methods embed PDE residuals or classical discretization schemes as soft-constraint regularizers in the loss function (e.g., PI-DeepONet~\cite{wang2021learning, MANDL2025117586}, PINO~\cite{li2024physics, eshaghi2025variational}, MCNP~\cite{zhang2025monte}). Meanwhile, physics-encoded methods incorporate a (learnable) traditional numerical solver (finite difference~\cite{long2019pde,rao2023encoding,wang2024pcnet, liu2024multi,wang2025multipdenet}, finite volume~\cite{kochkov2021machine, sun2023neural, horie2024graph, yan2025learnable}, or finite element~\cite{franco2023mesh}) as part of the network architecture and employ a neural network to refine the coarse-grid solution. These physics-aware approaches still face critical limitations, including marginal speed gains and unstable training dynamics~\cite{mcgreivy2024weak}. Additionally, most neural operator models remain limited to 2D and single-physics problems, lacking generality and adaptability in real-world scenarios.

Over the past two years, the rapid development of AI foundation models has led to significant progress across  disciplines~\cite{moor2023foundation,bodnar2025foundation}, and motivated the development of unified simulators. For instance, DPOT~\cite{hao2024dpot}, MPP~\cite{NEURIPS2024_d7cb9db5}, and OmniArch~\cite{chen2025building} leverage large-scale multi-physics datasets to pre-train foundation neural operators, enabling adaptation to diverse physical systems. S$^3$GM~\cite{li2024learning} introduces a pre-trained generative model that learns the distribution of spatiotemporal dynamics, and performs conditional sampling to reconstruct full fields. However, several challenges remain when deploying them in real-world scenarios. First, these approaches rely heavily on large labeled datasets, making them difficult to scale to 3D turbulent scenarios where data acquisition comes at a high cost. Second, handling diverse CFD tasks by a single model requires substantially large parameter capacity, while increasing model size inevitably degrades inference speed, thus resulting in a trade-off between efficiency and accuracy.

To this end, we propose OmniFluids, a pre-trained operator learning framework that is purely supervised by physics, capable of solving various CFD problems and adapting to diverse tasks with broad generalization capabilities based on minimal data. On the training side, it combines physics-only pre-training, coarse-grid operator distillation, and few-shot fine-tuning, enabling fast inference, accurate prediction, and strong generalization with limited or zero data supervision. Architecturally, OmniFluids incorporates key components such as (1) a mixture of operators (MoO) to balance model capacity and efficiency, (2) a multi-frame decoder for efficient computation of physics losses, and (3) factorized Fourier layers to support efficient and scalable 3D modeling.

Across diverse 2D and 3D benchmarks with varying Reynolds numbers, initial conditions, and external forces, OmniFluids consistently outperforms existing AI-driven CFD methods in predictive accuracy and consistency with turbulence statistics. Even in the zero-shot regime, OmniFluids can accurately predict flow fields and their corresponding kinetic energy spectrum after physics-only pre-training. On 2D and 3D Navier--Stokes benchmarks, OmniFluids delivers $9.9$--$21.2\times$ inference speedups over traditional solvers while maintaining comparable accuracy. Moreover, it demonstrates potential in various downstream tasks such as inverse problems, reconstruction, and control, delivering accurate forecasts and broad applicability.

\section*{Results}
\subsection*{The overall framework of OmniFluids}

We introduce a challenging scenario in CFD: developing pre-training models purely from physics that can efficiently solve a wide range of CFD problems and adapt to diverse tasks with minimal data, particularly in the 3D setting.

Fig.~\ref{fig:1}\textbf{a} illustrates the schematic of OmniFluids. The operator architecture of OmniFluids is constructed on a mixture of operators (MoO) framework for solving various CFD problems. A routing network takes the PDE parameters as input and dynamically assigns each task to the appropriate specialized operators. During training, an abundant library of specialized operators enhances model capacity, while during inference, we consolidate this ensemble into a single personalized operator to ensure runtime efficiency. In the final layer, we use a multi-frame decoder to forecast multiple future steps in parallel, ensuring temporal continuity among subsequent physical fields. Furthermore, we incorporate factorized Fourier layers~\cite{tran2023factorized} for dimension-wise decomposition, enabling efficient extension to 3D scenarios.

Fig.~\ref{fig:1}\textbf{b} illustrates OmniFluids' three-stage training workflow: physics-only pre-training, operator distillation, and few-shot fine-tuning. In the pre-training phase, the teacher network learns solely from diverse governing equations by predicting multiple future time steps in parallel. This multi-frame loss breaks free from the sequential time-stepping bottleneck of traditional solvers and other physics-aware methods. Next, during operator distillation, we distill knowledge from the high-resolution teacher to a student model by matching its outputs at coarser spatial and temporal resolutions. Finally, in the few-shot fine-tuning stage, the student model is briefly adapted using a small set of high-fidelity samples. Unlike joint physics-data training, this pre-train-then-fine-tuning strategy more robustly reconciles coarse-grid priors with fine-resolution datasets. This training paradigm establishes a unified framework for learning surrogate models under unsupervised or few-shot settings. 

To evaluate OmniFluids, we conduct experiments on PDEs representing classical problems in fluid dynamics. We consider the 2D Kuramoto-Sivashinsky equations (KSEs), the 2D incompressible Navier-Stokes equations (NSEs) in velocity-vorticity form, and the 3D incompressible NSEs in velocity-pressure form. More details are given in the \textcolor{blue}{Methods} section.

\subsection*{Physics-only pre-training for CFD tasks}

In the physics-only pre-training stage, OmniFluids can be trained in an unsupervised regime using only physics priors. We evaluate the performance of the pre-trained model on the aforementioned three benchmark datasets and compare it against the representative physics-informed methods, e.g., PINO~\cite{li2024physics}.

\textbf{2D KSE system.} In the physics-only pre-training stage, we discretize the spatial domain into a $64 \times 64$ grid. For temporal evolution, we construct the loss function over the interval $t \in [0, 0.2]$ and use a multi-frame decoder to divide the time dimension into 100 intervals. During testing, we extend the predictions to $t = 5$ to evaluate the model's ability to perform long-term extrapolation. The error evolution curves indicate that OmniFluids outperforms PINO in error accumulation, maintaining a much lower relative $\ell_2$ error over time (Fig.~\ref{fig:2}\textbf{a}). The corresponding predicted snapshots show that OmniFluids preserves the fine-scale structures of the solution, closely matching the reference. 

\textbf{2D vorticity-velocity NSE system.}  In this case, we train OmniFluids using a spatial discretization of $256 \times 256$ grid points. For the temporal domain, the loss function is constructed over the interval $t \in [0, 0.2]$, divided into $50$ bins. The predictions are extrapolated to $t = 10$ to evaluate the long-term forecasting ability of each model. This case demands that the model simultaneously handle varying Reynolds numbers, external forcings, and initial conditions.

The error evolution curves reveal that the performance gap between PINO and OmniFluids becomes increasingly pronounced as time progresses (Fig.~\ref{fig:2}\textbf{b}). OmniFluids successfully captures the fine-scale structures of the vorticity field and preserves their spatial integrity throughout the temporal evolution. In contrast, PINO struggles with long-term extrapolation. Furthermore, we check how well the models can keep the flow's spectral and statistical features at the last time step (Extended Data Fig.~1\textbf{a}). The kinetic energy spectrum $E(k)$ characterizes the energy distribution across spatial frequencies. OmniFluids closely follows the reference across all wavenumbers, accurately capturing both large-scale motions and small-scale turbulent features. The probability density function (PDF) of the normalized vorticity $\omega$ further illustrates the statistical fidelity of each model. OmniFluids matches the reference PDF well, reproducing the sharp peaks and heavy tails associated with intermittent turbulent structures. 

\textbf{3D velocity-pressure NSE system.} For the 3D NSE case, we extend OmniFluids to a spatial resolution of $128^3$ grid points. The training time domain spans $t \in [0, 0.5]$, discretized into 25 intervals, and predictions are extrapolated to $t=10$. Compared to the 2D settings, 3D fluid modeling is inherently more challenging due to vortex stretching, anisotropic multi-scale dynamics, and a forward energy cascade~\cite{lienen2024zero}. We choose the spatiotemporal resolutions such that the pre-training DNS trajectories lie in a coarse-DNS regime. For instance, at $\text{Re}=2000$, the spatial resolution satisfies $\Delta x/\eta \approx 3.6$, while the time step $\Delta t = 0.02$ yields $\Delta t/\tau_\eta \approx 0.05$, where $\eta$ and $\tau_\eta$ denote the Kolmogorov length and time scales associated with the smallest dissipative eddies. Therefore, OmniFluids is trained to satisfy the governing PDE on a coarse grid while focusing its capacity on learning effective dynamics for the large-scale and inertial-range structures.

As shown in Fig.~\ref{fig:2}\textbf{c} and Extended Data Fig.~2, PINO fails to maintain physically consistent predictions beyond the short-term horizon, while OmniFluids accurately preserves the fine-scale features of the flow. Quantitatively, OmniFluids maintains a stable relative $\ell_2$ error below $0.30$ over the entire extrapolation domain. In contrast, PINO's error grows rapidly, exceeding $1.0$ after $t > 3$, indicating that its predictions deviate substantially from the reference dynamics. Spectral analysis further reveals the superior accuracy of OmniFluids (Extended Data Fig.~1\textbf{b}). OmniFluids matches the reference solution across all resolved wavenumbers, with only minor deviations at the highest frequencies. In contrast, PINO deviates significantly across all spectra, from low to high frequencies, indicating an inability to capture energy distribution across scales. 

\subsection*{Adaptation to coarser resolutions via distillation and fine-tuning}

After physics-only pre-training, we perform operator distillation to transfer knowledge from a high-resolution teacher model to a student model with coarser spatiotemporal resolutions. In this setting, the distilled student not only operates on a coarser spatial grid, but also advances the solution with a much larger effective temporal stride compared to the fine time step used by the reference solver. For a specific downstream task, the distilled student model is fine-tuned using a small amount of data from the target distribution. We evaluate the adaptability of OmniFluids on 2D KSE, 2D NSE, and 3D NSE, including in-distribution (ID) and out-of-distribution (OOD) adaptation scenarios. For 2D cases, we compare OmniFluids with various methods, including neural operators like PINO~\cite{li2024physics}, DeepONet~\cite{lu2021learning}, FNO~\cite{li2021fourier}, FFNO~\cite{tran2023factorized}, FactFormer~\cite{li2023scalable}, and CNO~\cite{raonic2023convolutional}; a large-scale data-driven pre-trained foundation model (DPOT~\cite{hao2024dpot}); and direct numerical simulation (DNS) based on spectral methods (detailed in \textcolor{blue}{Supplementary Note B}). For the 3D case, we use PINO~\cite{li2024physics}, FNO~\cite{li2021fourier}, DPOT~\cite{hao2024dpot}, {FactFormer}~\cite{li2023scalable}, and DNS as baselines.

\textbf{2D KSE system.} During operator distillation, we compress the model to a coarser spatiotemporal resolution of $32 \times 32$ with a time step $\Delta t =1.0$. We use selected values of $\alpha = 0.2, \beta = 0.5$ for ID, and $\alpha = 1.0, \beta = 1.0$ for OOD. The task is comparatively easier in the ID scenario, where the effects of fourth-order dissipation and nonlinear advection are relatively weak. OmniFluids successfully reconstructs spatial structures and preserves fine-scale features, achieving the lowest relative $\ell_2$ error among all AI-driven methods (Extended Data Fig.~3\textbf{a} and \textbf{c}). In contrast, purely data-driven baselines underperform due to their dependency on extensive training data. PINO struggles to enforce accurate physics-based constraints because inconsistencies often arise between the physics-informed loss and the high-fidelity data, leading to suboptimal training. DPOT achieves the best performance after OmniFluids among neural operator baselines, thanks to its pre-training on extensive datasets. However, the absence of physical knowledge limits its ability to capture delicate spatiotemporal dynamics. Under the OOD conditions (Extended Data Fig.~3\textbf{b} and \textbf{d}), baseline models' errors increase significantly, often growing by an order of magnitude. OmniFluids accurately captures the dominant physical modes and maintains coherent spatial patterns, while other data-driven methods exhibit overfitting during training. {At the same spatial resolution, the rollout times are 0.034~s for OmniFluids and 3.495~s for the pseudospectral solver (see Extended Data Fig.~3\textbf{e}). Because the KSE contains a fourth-order derivative, the numerical solver uses a fine time step for stability.}

\textbf{2D vorticity-velocity NSE system.} In this case, we compress the pre-trained model to a coarser resolution of $128 \times 128$ with $\Delta t =1.0$. We evaluate the model on ID and OOD settings, using selected Reynolds numbers of $\text{Re} = 2000$ for ID and $\text{Re} =  4000$ for OOD, respectively.  The predicted vorticity fields $\omega$ are shown in Fig.~\ref{fig:3}\textbf{a} and \textbf{b}. We also reconstruct the velocity fields ($\mathbf{u} = [u, v]$) and visualize them alongside the vorticity. Among these baselines, DeepONet, FNO, and CNO fail to generate meaningful predictions due to limited training data and insufficient model expressivity. FactFormer and FFNO, while equipped with stronger representation capacity, only manage to localize a few primary vortices and exhibit apparent discrepancies from the reference solution. DPOT, which benefits from data-driven pre-training, shows moderate adaptability to new regimes via fine-tuning. However, it fails to capture vortex positions and intensity accurately due to a lack of physical priors. PINO performs better than most AI baselines but suffers from blurring and distortion when modeling fine-scale vortex features. In contrast, OmniFluids consistently delivers accurate reconstructions, capturing intricate vorticity structures even in turbulence with a high Reynolds number. Temporal correlation curves (Fig.~\ref{fig:3}\textbf{e} and \textbf{f}) illustrate OmniFluids' ability to maintain accurate long-term flow predictions. It consistently achieves the highest correlation with the reference solution over time, whereas other models exhibit rapid degradation.

Fig.~\ref{fig:3}\textbf{c} and \textbf{d} present the kinetic energy spectra and normalized vorticity PDFs for the ID and OOD settings. OmniFluids surpasses the coarse DNS baseline at the same resolution. This superior spectral accuracy is attributed to the model's physics-only pretraining on a finer spatial grid, followed by distillation and task-specific fine-tuning. OmniFluids faithfully recovers the reference vorticity distribution, capturing both the sharp peak and the heavy-tailed behavior. In contrast, other AI-driven methods exhibit noticeable deviations, with shifted peaks and distorted tails. At the same spatial resolution, the measured rollout times are 0.136~s for OmniFluids and 3.627~s for the pseudospectral solver (Fig.~\ref{fig:3}\textbf{g}; accuracy-qualified results in Extended Data Fig.~4). Although baselines such as FNO and DeepONet run even faster, they fail to offer reasonable predictions, with correlation coefficients with the reference at the final time step falling below 0.2.

\textbf{3D velocity-pressure NSE system.} We evaluate model performance on the 3D NSE with coupled velocity-pressure dynamics. Compared to the 2D case, modeling 3D turbulence is more challenging due to the increased degrees of freedom and nonlinearity. To retain the spatial fidelity, we distill OmniFluids on the same $128^3$ physical resolution as in physics-only pre-training, while using an encoder to downsample the 3D fields into a $64^3$ latent space where the Fourier layer is applied, which substantially reduces computational cost. Furthermore, the student operates with a much larger step size of $\Delta t = 1.0$, corresponding to $\Delta t/\tau_\eta \approx 2.5$. We consider ID and OOD adaptation settings for evaluation, with Reynolds numbers of $\text{Re} = 2000$ and $\text{Re} = 3000$.

Fig.~\ref{fig:4}\textbf{a} and \textbf{b} show isosurface renderings of the predicted 3D velocity magnitude, extracted from the center region for ID and OOD settings, respectively. The full velocity visualizations are provided in Extended Data Fig.~5. OmniFluids achieves superior accuracy in predicting the 3D velocity field over time, maintaining close agreement with the reference solution even at later stages. Specifically, OmniFluids consistently achieves a correlation coefficient above or around 0.8 with the reference solution from the initial to the final timestep, performing similarly to coarse DNS results (Fig.~\ref{fig:4}\textbf{e} and \textbf{f}). In contrast, PINO, FNO, FactFormer, and DPOT exhibit rapid degradation in predictive accuracy, with temporal correlation coefficients falling below 0.8 within the first one or two steps. Kinetic energy spectra and velocity PDFs in Fig.~\ref{fig:4}\textbf{c} and \textbf{d} reveal the model's ability to recover multiscale dynamics. Only OmniFluids and DNS accurately reproduce the reference spectrum across the resolved range of wavenumbers. {At the same spatial resolution, the measured rollout times are 0.843~s for OmniFluids and 9.382~s for the pseudospectral solver (see Fig.~\ref{fig:4}\textbf{g}).}

{\textbf{Accuracy-qualified computational efficiency.} We compare OmniFluids with resolution-swept pseudospectral solvers on two OOD NSE benchmarks using an accuracy-qualified protocol. For each system, we select the fastest spatial configuration among the tested resolutions at the prescribed time step for which both evaluated relative $\ell_2$ errors are no greater than the corresponding OmniFluids values. The neighboring tested resolutions bracket each qualification threshold: $96^2$ fails whereas $100^2$ qualifies for 2D NSE, and $112^3$ fails whereas $120^3$ qualifies for 3D NSE (see Extended Data Fig.~4\textbf{a}--\textbf{b}). As a result, OmniFluids yields rollout speedups of $21.2\times$ for 2D NSE and $9.9\times$ for 3D NSE (see Extended Data Fig.~4\textbf{c}). Details of the test setup are provided in the \textcolor{blue}{Methods} section.}

\subsection*{{Wall-bounded turbulent channel flow}}

To further evaluate OmniFluids on complex wall-bounded fluid dynamics beyond idealized periodic benchmarks, we introduce a 3D turbulent channel-flow problem. This case contains key challenges of wall-bounded turbulence, including no-slip boundaries, strong near-wall shear, anisotropic velocity fluctuations, coherent streaks, and vortical structures. We compare OmniFluids with PINO, FNO, FactFormer, and DPOT on an unseen trajectory with viscosity $\nu=2.5\times10^{-4}$. All models are initialized from the same turbulent state and rolled out autoregressively to predict the subsequent flow evolution. More details are given in the \textcolor{blue}{Methods} section.

As shown in Fig.~\ref{fig:5}\textbf{a}, most baselines recover the dominant streamwise motion, but they differ markedly in the near-wall flow organization. PINO and FNO tend to oversmooth the near-wall streaks, while FactFormer introduces visible wall-parallel artifacts. DPOT reconstructs richer spatial variations, but still deviates from the reference structures. In contrast, OmniFluids better preserves the high- and low-speed streaks and produces a velocity field that is more consistent with the reference. The positive $Q$-criterion isosurfaces further show that OmniFluids maintains a more reference-like distribution of vortical structures under the same autoregressive evaluation protocol (Fig.~\ref{fig:5}\textbf{b}).

The wall-normal statistics in Fig.~\ref{fig:5}\textbf{c} provide a more stringent evaluation than qualitative snapshots alone. Besides the mean velocity profile $U^+$, we compare Reynolds stresses and fluctuation intensities, including $\overline{u'u'}/u_\tau^2$, $-\overline{u'v'}/u_\tau^2$, $u_{\mathrm{rms}}/u_\tau$, $v_{\mathrm{rms}}/u_\tau$, and $w_{\mathrm{rms}}/u_\tau$. Across these profiles, OmniFluids gives the closest overall agreement with the reference, while the baselines show larger deviations in either the near-wall peak, the Reynolds shear stress, or the wall-normal and spanwise fluctuation intensities. These results demonstrate that OmniFluids can extend beyond periodic fluid benchmarks and capture physically meaningful statistics in a complex wall-bounded flow.

\subsection*{{Downstream tasks of fluid systems with OmniFluids}}

{In this section, we evaluate the performance of OmniFluids across several downstream tasks, including inverse analysis, control, and sparse reconstruction on 2D NSEs, as well as a transfer study on experimental cylinder-wake measurements (Fig.~\ref{fig:6}).}

\textbf{Inverse problems.} We evaluate OmniFluids on inverse problems under three levels of additive observational noise, $\sigma_k \in \{0\%~(\mathrm{E1}), 5\%~(\mathrm{E2}), 10\%~(\mathrm{E3})\}$, where for a given vorticity field $\omega$ each perturbation $\delta \omega^{(k)}$ is sampled i.i.d.\ from a zero-mean Gaussian distribution with standard deviation $\sigma_k \,\overline{|\omega|}$. In these experiments, we assume access to only two frames of flow field data. The spatial resolution of the observed flow fields is \(128 \times 128\), aligning with the student network. Starting with a random guess, we aim to infer two key unknown parameters, including the viscosity coefficient (\(\nu=\text{Re}^{-1}\)) and the external forcing, from the observed data. As shown in Fig.~\ref{fig:6}\textbf{a}, the inferred external forcing fields closely match the reference forcing across all noise levels. Similarly, Fig.~\ref{fig:6}\textbf{b} shows the inferred viscosity coefficients, which remain close to the reference value, with a relative error of only around 3\% even in the 10\% noise scenario.

{\textbf{Flow control.} We also evaluate whether OmniFluids can serve as a surrogate model for flow control. Given an initial condition, we seek a time-dependent external forcing that drives the system towards a desired dissipation behavior, i.e., that reduces the enstrophy as rapidly as possible under a moderate forcing amplitude. The forcing is parameterized as a sequence of control inputs, updated once per time step, and is optimized using the pre-trained OmniFluids model. The resulting control sequence is then applied in OmniFluids and DNS to obtain the predicted and reference flows, respectively. As shown in Fig.~\ref{fig:6}\textbf{c}, the controlled rollouts predicted by OmniFluids closely match the reference DNS trajectories over a long horizon. Fig.~\ref{fig:6}\textbf{d} compares the corresponding enstrophy evolution. The Decay baseline denotes the uncontrolled dynamics, in which enstrophy decreases through viscous dissipation and therefore exhibits a much slower decay. In contrast, the OmniFluids-based controller produces an enstrophy trajectory that exhibits a substantially faster reduction than the Decay baseline.}

{\textbf{Sparse reconstruction.} We evaluate OmniFluids' ability to reconstruct flow fields from sparse observations. In this experiment, at each time step, only 5\% of the grid points are observed, sampled uniformly at random over the spatial domain. Conditioning on these partial measurements, OmniFluids infers a full-resolution flow field that remains consistent with the reference solution (Fig.~\ref{fig:6}\textbf{e}). Despite the severe undersampling, the reconstructed flows accurately recover both the large-scale structures and the fine-scale vortical features, highlighting that the learned model provides a strong physical prior for reconstructing unobserved regions.}

{\textbf{Flow prediction based on experimental data.} {To examine whether a physics-pretrained operator can be adapted to noisy laboratory measurements,} we further evaluate it on a time-resolved experimental cylinder-wake dataset acquired via particle image velocimetry (PIV)~\cite{li2024learning} (see Fig.~\ref{fig:6}\textbf{f} and Extended Data Fig.~6\textbf{a}). The measured two-component velocity fields are converted to vorticity fields on a \(128\times128\) grid. We then fine-tuned the distilled 2D NSE student model using only the beginning portion of recorded frames and performed an autoregressive rollout to extrapolate the future flow evolution. OmniFluids successfully reproduces the characteristic vortex street that closely matches the PIV reference over the evaluation horizon, as do the vorticity distribution statistics (Fig.~\ref{fig:6}\textbf{g} and \textbf{h}, and Extended Data Fig.~6\textbf{b}). Implementation details are provided in \textcolor{blue}{Supplementary Note B.5}.}

\section*{Discussion}

This paper introduces OmniFluids, a physics pre-trained operator learning framework for modeling fluid dynamics. OmniFluids leverages a carefully designed model architecture and a three-stage training strategy to overcome key limitations in current CFD surrogate methods. OmniFluids first undergoes a physics-only pre-training phase, where it learns diverse surrogate models purely from the governing PDEs without any labeled data. We then apply operator distillation, compressing the high-resolution pre-trained model into a lower-resolution surrogate that retains fine-scale physical knowledge while supporting fast coarse-grid inference. Finally, a small amount of task-specific data (as few as 2-10 trajectories) is used for few-shot fine-tuning, allowing the model to reconcile discrepancies between the physics prior and real-world measurements. 

OmniFluids demonstrates strong performance across various challenging CFD benchmarks and consistently outperforms current baselines. Notably, OmniFluids can conduct 3D turbulence simulations efficiently, a long-standing challenge due to the extreme dimensionality and multiscale nature. At Reynolds numbers up to 3000, it maintains accurate long-horizon forecasts while preserving spatiotemporal coherence and energy spectra. Beyond forward simulation, OmniFluids also excels at various downstream tasks such as system identification, control, and reconstruction. Furthermore, {on 2D and 3D NSE benchmarks, OmniFluids delivers $9.9$--$21.2\times$ inference speedups over the pseudospectral methods under the accuracy-qualified error criterion}. 

Despite its strong empirical performance and diversity, OmniFluids still has several limitations that suggest directions for future work. First, the current framework assumes that the governing equations are fully known and explicitly specified during pre-training. This assumption may not hold in gray-box or partially known systems, such as real-world settings with missing or uncertain physical terms. While OmniFluids demonstrates some robustness to such cases through adaptation, its performance could degrade without accurate physical priors. Second, the implementation of OmniFluids is tailored to structured Cartesian grids, which limits its direct applicability to domains involving irregular geometries in other CFD problems. Extending OmniFluids to such settings will necessitate introducing geometric inductive biases through graph neural operators, mesh-aware architectures, or coordinate-free representations. Third, although the current pre-training strategy leverages rich PDE dynamics, it does not explicitly encode additional physical constraints such as conservation laws (e.g., mass, momentum, energy), symmetries, or entropy conditions. These constraints are often crucial for ensuring robustness, stability, and generalization in real-world scenarios. Integrating such domain-specific inductive priors could further enhance the framework's reliability across broader applications.

\section*{Methods}
We now introduce the method of the proposed OmniFluids.

\subsection*{Network architecture}
\begin{figure}[t!]
  \centering
   \includegraphics[width=0.9\linewidth]{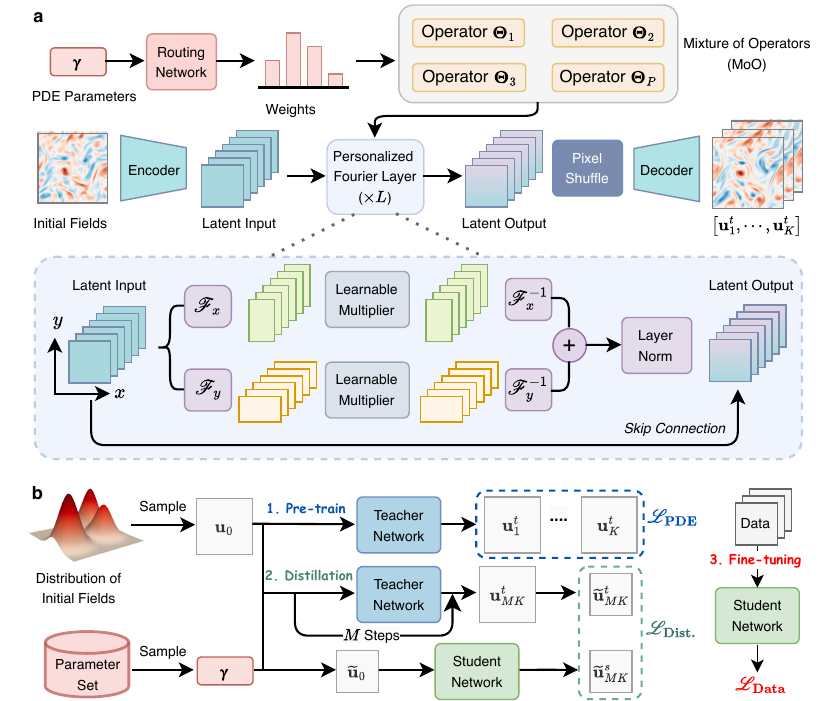}
   \caption{\textbf{Schematic framework of OmniFluids.} \textbf{a},  Teacher/student network architecture. OmniFluids employs a mixture of operators (MoO) backbone to flexibly handle diverse CFD tasks, alongside a multi-frame decoder that predicts $K$ future steps in parallel. Factorized Fourier layers enable efficient and scalable modeling, especially in 3D. \textbf{b}, Training workflow for OmniFluids. (1) \textbf{Physics-only pre-training.} The teacher network is trained using purely physics-based loss functions with no reliance on data. (2) \textbf{Operator distillation.} A compact student model is distilled from the pre-trained teacher at coarser spatial and temporal resolutions, enabling faster inference without sacrificing accuracy. (3) \textbf{Few-shot fine-tuning.} When a small amount of task-specific data becomes available, the student model is fine-tuned in a few-shot regime to bridge any remaining gap between pre-training priors and high-fidelity datasets. Here, ${\mathbf{u}}_{k}^{t}\ (\tilde{\mathbf{u}}_{k}^{t})$ denotes the high-resolution (downsampled) prediction at the $k^{\text{th}}$ frame predicted by the teacher network. $\tilde{\mathbf{u}}_{k}^{s}$ is the low-resolution prediction produced by the student network.}
   \label{fig:1}
\end{figure}

\begin{figure}[t!]
   \centering
    \includegraphics[width=1.0\linewidth]{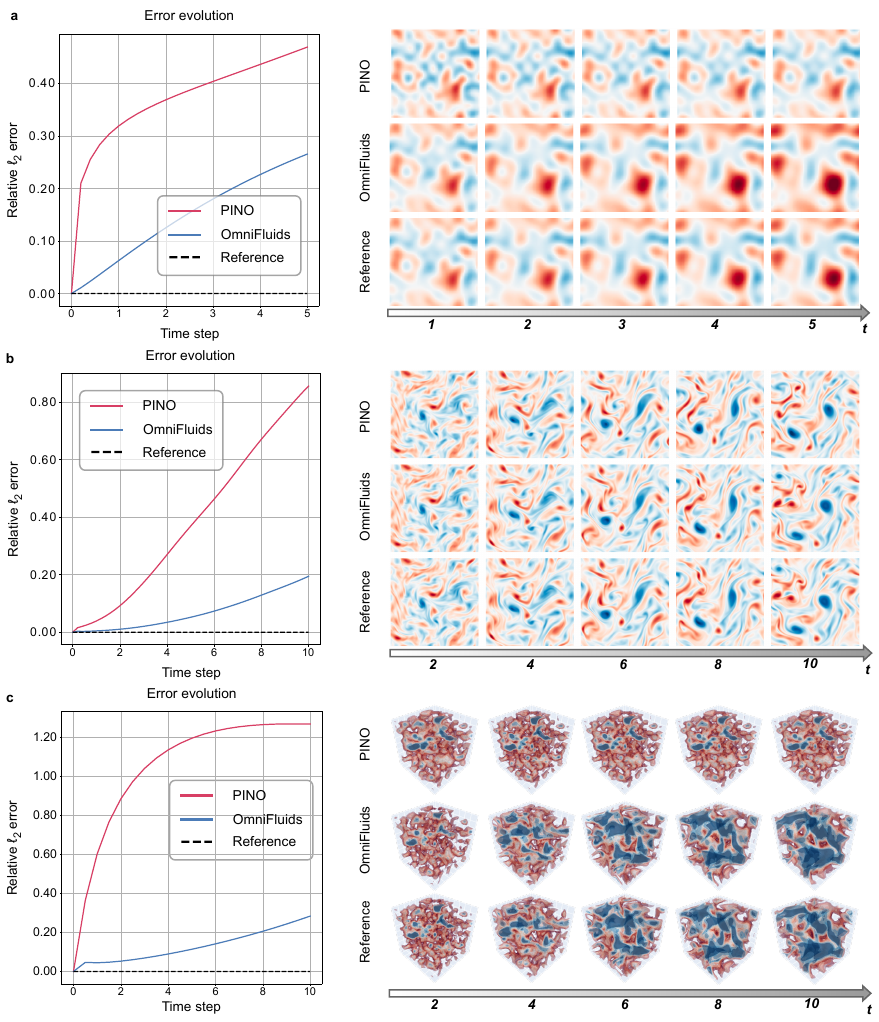}
    \caption{\textbf{Multi-step error and predicted snapshots from PINO and OmniFluids across diverse CFD systems.} \textbf{a}, 2D KSE. \textbf{b}, The vorticity field ($\omega$) of 2D NSE. {\textbf{c}, The velocity magnitude of 3D NSE. Note that only a local flow field extracted from the center region is shown for the 3D NSE in \textbf{c} for better visualization.} The full velocity visualizations for 3D NSEs are provided in Extended Data Fig.~2. The predicted flow evolutions are further depicted in \textcolor{blue}{Supplementary Videos 1--3}.}    \label{fig:2}
 \end{figure}

\begin{figure}[!t]
  \centering
   \includegraphics[width=1\linewidth]{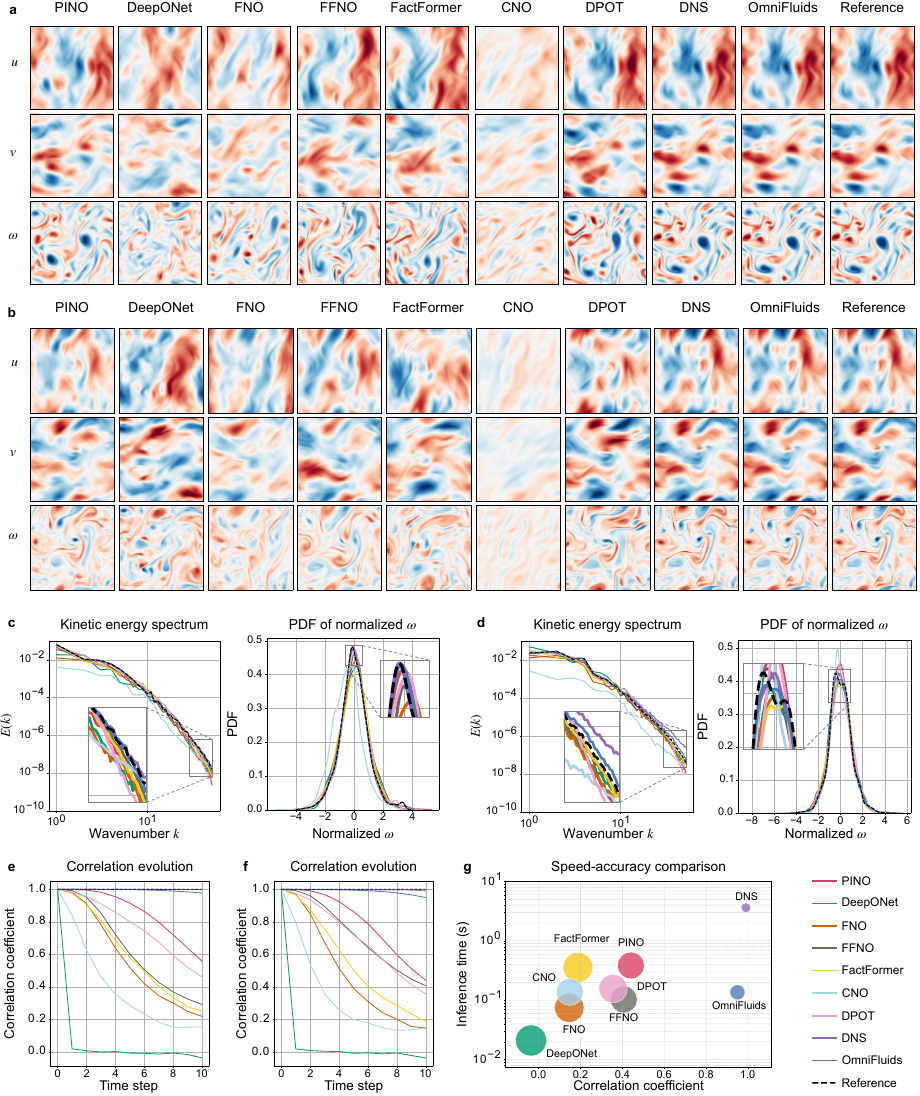}
\caption{\textbf{Predicted snapshots, flow statistics, temporal correlation, and inference speed of OmniFluids and baseline methods on 2D NSE.} \textbf{a} and \textbf{b}, Predicted velocity ($\mathbf{u} = [u,v]$) and vorticity fields ($\omega$) at $t=10$ under ID ($\text{Re}=2000$) and OOD ($\text{Re}=4000$) settings, respectively. \textbf{c} and \textbf{d}, Kinetic energy spectrum and PDFs of normalized vorticity for ID and OOD scenarios. \textbf{e} and \textbf{f}, Temporal evolution of correlation coefficients between predicted and reference vorticity fields for ID and OOD scenarios. \textbf{g}, Speed-accuracy comparison for the OOD case, where marker size represents the relative $\ell_2$ error at the final time.}
\label{fig:3}
\end{figure}

\begin{figure}[t!]
  \centering
   \includegraphics[width=1\linewidth]{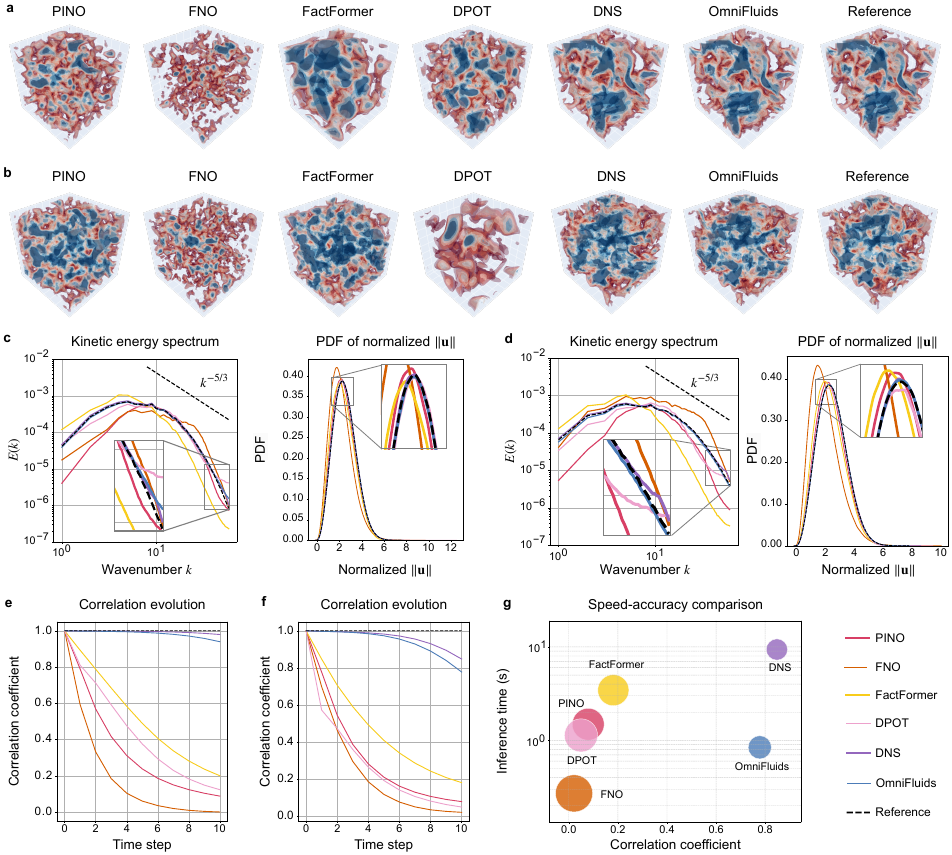}
\caption{\textbf{Predicted snapshots, flow statistics, temporal correlation, and inference speed of OmniFluids and baseline methods on 3D NSE.} \textbf{a} and \textbf{b}, {Isosurface renderings} of the predicted 3D velocity magnitude extracted from the center region at $t=8$ under the ID setting ($\text{Re} = 2000$) and OOD ($\text{Re}=3000$) settings. The full velocity visualizations are provided in Extended Data Fig.~5. \textbf{c} and \textbf{d}, Kinetic energy spectrum and normalized velocity PDFs for ID and OOD scenarios. \textbf{e} and \textbf{f}, Temporal evolution of correlation coefficients between predicted and reference velocity fields for ID and OOD scenarios. \textbf{g}, Speed-accuracy comparison for the OOD case, where marker size represents relative $\ell_2$ error at the final time.}
   \label{fig:4}
\end{figure}

\begin{figure}[t!]
  \centering
  \includegraphics[width=1.0\linewidth]{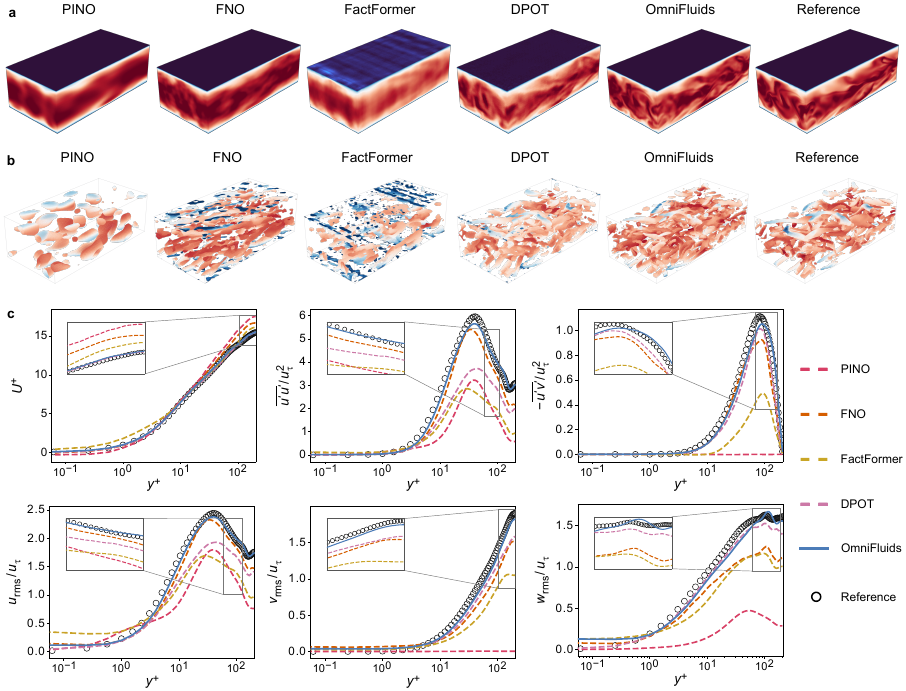}
  \caption{{\textbf{Wall-bounded turbulent channel flow with viscosity $\nu=2.5\times10^{-4}$.}
  PINO, FNO, FactFormer, DPOT, OmniFluids, and the reference simulation are compared on a test turbulent channel trajectory under the same autoregressive evaluation protocol. 
  \textbf{a}, Predicted speed magnitude $\|\mathbf{u}\|_2$ on visible channel faces at the final prediction step. 
  \textbf{b}, Positive $Q$-criterion isosurfaces colored by the streamwise velocity $u$. 
  \textbf{c}, Time-averaged wall-normal profiles of the mean velocity $U^+$, Reynolds normal stress $\overline{u'u'}/u_\tau^2$, Reynolds shear stress $-\overline{u'v'}/u_\tau^2$, and velocity fluctuation intensities $u_{\mathrm{rms}}/u_\tau$, $v_{\mathrm{rms}}/u_\tau$, and $w_{\mathrm{rms}}/u_\tau$. The predicted flow evolutions are depicted in \textcolor{blue}{Supplementary Video 4}.}}
  \label{fig:5}
\end{figure}

\begin{figure}[t!]
  \centering
   \includegraphics[width=0.97\linewidth]{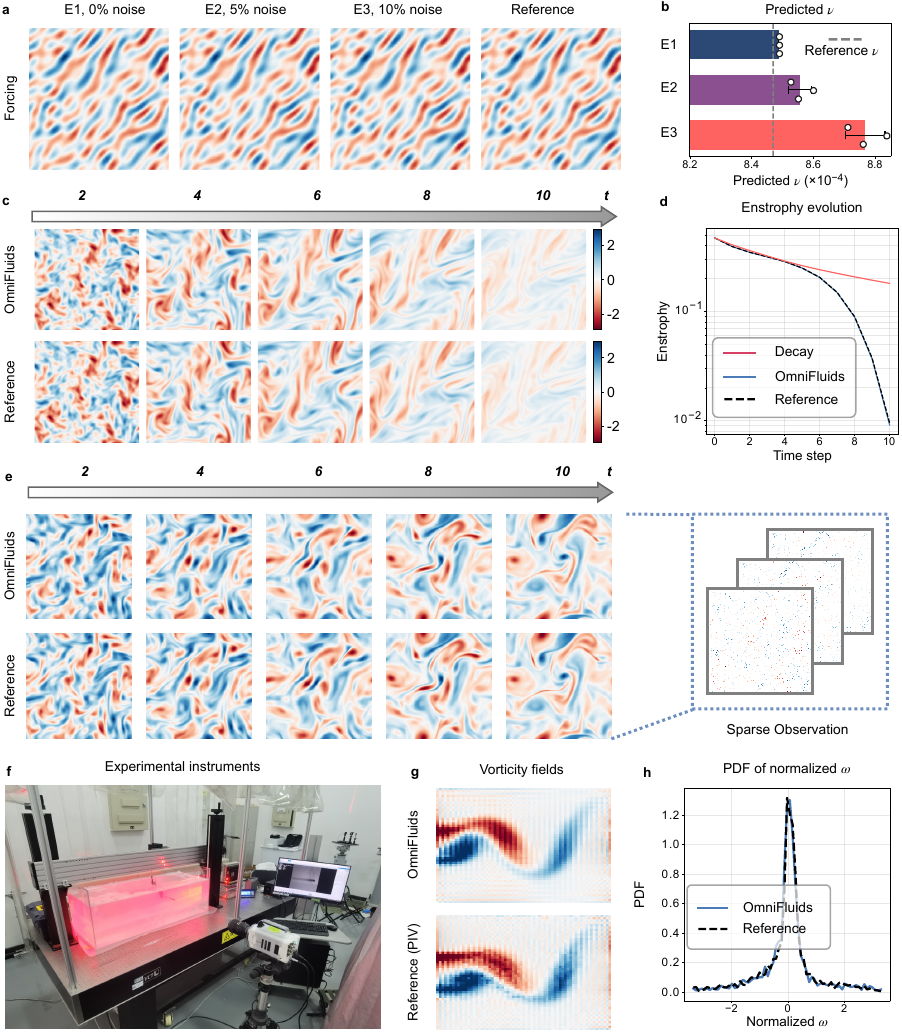}
   \caption{{\textbf{Downstream tasks on 2D NSEs.} \textbf{a} and \textbf{b}, \textbf{Inverse problems.} Predicted external forcing (\textbf{a}) and viscosity coefficients (\textbf{b}) using two frames of flow data under three noise levels. In \textbf{b}, bars show means, error bars show standard deviation, and open circles show individual inversions ($n=3$ technical repeats per noise level on one flow trajectory). Repeats use independently randomized parameter initializations and, for nonzero-noise cases, independent noise realizations to assess run-to-run variability. \textbf{c} and \textbf{d}, \textbf{Flow control.} Controlled flow dynamics from OmniFluids and DNS at multiple time steps (\textbf{c}), and enstrophy evolution for the uncontrolled Decay baseline, the OmniFluids-based controller, and the reference (\textbf{d}). \textbf{e}, \textbf{Sparse reconstruction.} Full-field reconstruction from 5\% randomly sampled observations. \textbf{f}--\textbf{h}, \textbf{Flow prediction based on experimental data}: the laboratory setup \cite{li2024learning} (\textbf{f}), the OmniFluids-predicted and PIV-measured vorticity fields at $t = 8$ (\textbf{g}), and the corresponding distribution of normalized vorticities (\textbf{h}).}}
   \label{fig:6}
\end{figure}

To address the challenges of modeling complex fluid systems, our network architecture is designed to satisfy three key requirements. First, it should enable rapid inference to efficiently integrate physics priors and scale across extensive spatiotemporal domains. Second, it should have sufficient capacity to capture diverse fluid behaviors under varying parameters, initial conditions, and external forcings. Third, the design can efficiently handle 3D CFD tasks, preserving essential turbulence statistics (e.g., the kinetic energy spectrum). Guided by these principles, we design the OmniFluids architecture (Fig.~\ref{fig:1}\textbf{a}), which can deliver accurate, efficient, and generalizable modeling for CFD tasks. Since the teacher and student networks differ mainly in the decoder's resolution, we describe the teacher network in detail.

\textbf{Spatially factorized spectral layer.} We perform primary spatial operations in the Fourier domain to ensure the accurate preservation of high-frequency structures and turbulence statistics. Following the FFNO paradigm, we decompose the transform into two or three separable 1D spectral operations rather than using a 2D or 3D Fourier transformer directly~\cite{tran2023factorized}. For 3D cases, given a latent input $\mathbf{v}(x,y,z)$, we compute its 1D Fourier coefficients along each coordinate axis as follows:
\begin{equation}
    \begin{aligned}
    \hat{\mathbf{v}}_x(k;y,z) &= \mathcal{F}_x\bigl[{\mathbf{v}}(\cdot,y,z)\bigr](k),\\
\hat{\mathbf{v}}_y(\ell;x,z) &= \mathcal{F}_y\bigl[{\mathbf{v}}(x,\cdot,z)\bigr](\ell),\\
\hat{\mathbf{v}}_z(m;x,y) &= \mathcal{F}_z\bigl[{\mathbf{v}}(x,y,\cdot)\bigr](m).
    \end{aligned}
\end{equation}

Each set of coefficients is then modulated by a learnable multiplier ($\mathcal{R}_x(k), \mathcal{R}_y(\ell), \mathcal{R}_z(m)$) to capture spatial correlations in the frequency domain. The inverse transforms along each axis are summed to produce the updated field:
\begin{equation}
\mathbf{v}_{\mathrm{update}}(x,y,z) = \mathcal{F}_x^{-1}\bigl[\mathcal{R}_x\hat{\mathbf{v}}_x\bigr] + \mathcal{F}_y^{-1}\bigl[\mathcal{R}_y\hat{\mathbf{v}}_y\bigr] + \mathcal{F}_z^{-1}\bigl[\mathcal{R}_z\hat{\mathbf{v}}_z\bigr].
\end{equation}

After that, we wrap the separable spectral transform in a residual block with Layer Normalization~\cite{ba2016layer} to stabilize training and improve representational capacity:
\begin{equation}
    \mathbf{v}_{\mathrm{out}} = \mathbf{v} \;+\; \mathrm{LayerNorm}\bigl(\mathbf{v}_{\mathrm{update}}\bigr).
\end{equation}

Compared to the original FNO's full-rank 3D transform~\cite{li2021fourier}, this separable implementation processes each axis independently. By retaining $d$ modes per dimension, we reduce parameter complexity from $\mathcal{O}(d^3)$ to $\mathcal{O}(d)$, which is particularly impactful for 3D fields. 

We employ a patch-based embedding strategy to further accelerate inference and reduce computational costs for large-scale spatial domains. Specifically, the input physical field is downsampled into coarse spatial patches before passing through the factorized spectral layers. This operation reduces the spatial resolution during spectral processing, substantially lowering the computational and memory demand. After spectral modulation, we apply a pixel shuffle-based upsampling layer to reconstruct the original resolution from the latent feature space. This two-stage structure enables OmniFluids to operate effectively in the latent space while preserving critical fine-scale structures.

\textbf{Multi-frame decoder.} To avoid the computation overhead and error accumulation, we employ a 1D convolutional decoder~\cite{brandstetter2022message} that generates multiple future time steps in a single forward pass. By predicting $K$ future states at once, this decoder accelerates inference and enables fully parallel computation of the physics-based loss function across all frames. Moreover, the convolutional kernels enforce temporal continuity between adjacent predictions, significantly enhancing stability during long-term rollouts. 

During physics-only pre-training, the decoder outputs all $K$ frames in parallel, while at the inference or operator distillation stage, we typically require only the final frame for each rollout. Based on this observation, we introduce a split trick that accelerates inference without altering the pre-trained model to eliminate this computational redundancy. Concretely, we factor the decoder into two MLP branches, $\mathrm{MLP}_1$ and $\mathrm{MLP}_2$, followed by a 1D convolution. In the pre-trained stage, the model outputs $K$ frames in parallel as follows:
\begin{equation}
         \bigl[{\mathbf{u}}_{1}^t, {\mathbf{u}}_{2}^t, \dots, {\mathbf{u}}_{K}^t\bigr]
     = \mathrm{Conv1D}\bigl(\mathrm{MLP}_1(\mathbf{v}_{\mathrm{out}}) || \mathrm{MLP}_2(\mathbf{v}_{\mathrm{out}})\bigr),
\end{equation}
where $||$ denotes the concatenation operation along the channel dimension. While during the inference stage, we bypass $\mathrm{MLP}_1$, and directly use $\mathrm{MLP}_2$ to predict only the final frame:
\begin{equation}
         {\mathbf{u}}_{K}^t  = \mathrm{Conv1D}(\mathrm{MLP}_2(\mathbf{v}_{\mathrm{out}})).
\end{equation}

This split trick reduces computation and memory overhead during inference and subsequent operator distillation, yielding remarkable speedups for 3D fluid simulations.

\textbf{Mixture of operators.} A core challenge in designing a network that handles diverse CFD tasks is balancing inference speed against representational capacity. Equipping a single model to address multiple complex regimes demands ample parameter space, which degrades inference efficiency. To this end, we introduce a mixture of operators (MoO) framework. Rather than relying on one monolithic network, we maintain a bank of $P$ expert operator branches, each specialized for distinct physical regimes. At inference time, a lightweight routing network dynamically assigns the task to each operator and integrates its parameters into a unified model, thereby balancing inference speed with representational capacity.

As shown in Fig.~\ref{fig:1}\textbf{a}, the routing network takes the PDE parameters $\mathbf{\gamma}$ as input and encodes them into a set of logits $\{z_p\}_{p=1}^P$. These logits are normalized via a temperature-scaled softmax to produce non-negative weights $\{w_p\}_{p=1}^P$ satisfying $\sum_{p=1}^P w_p = 1$:
\begin{equation}
    w_p = \frac{\exp\bigl(z_p / \tau\bigr)}{\sum_{j=1}^P \exp\bigl(z_j / \tau\bigr)},
\end{equation}
where $\tau$ is a temperature hyperparameter that controls the sharpness of the weight distribution. {A moderate temperature encourages soft preferences over experts rather than hard one-hot assignments, which helps avoid routing collapse.} The final operator used for prediction is a weighted combination of the expert parameters, i.e.,
\begin{equation}
    \bar{\boldsymbol{\Theta}} \;=\; \sum_{p=1}^{P} w_p\,\boldsymbol{\Theta}_p,
\end{equation}
where $\boldsymbol{\Theta}_p$ denotes the $p^{\text{th}}$ operator in the operator pool. In this way, the large operator pool is compressed into a personalized operator with parameters $\bar{\boldsymbol{\Theta}}$ that is responsible for tasks governed by PDE parameters $\mathbf{\gamma}$. This task-adaptive mechanism preserves the expressiveness of a large expert ensemble while ensuring high inference efficiency. By dynamically selecting and integrating specialized subnetworks based on the underlying physical characteristics, the MoO architecture enables accurate and scalable modeling across diverse CFD tasks.

\subsection*{Training framework}
OmniFluids is trained through a three-stage pipeline consisting of physics-only pre-training, operator distillation, and few-shot fine-tuning to address the key challenges of neural operator learning in fluid dynamics (Fig.~\ref{fig:1}\textbf{b}). In the physics-only pre-training stage, a high-capacity teacher network is trained by minimizing PDE residuals to learn the coherent large-scale and inertial-range structures. This encourages the teacher to approximate the underlying PDE operators rather than fit any particular dataset, thereby removing dependence on observational data and embedding strong physics priors. Next, operator distillation transfers this PDE information to an efficient student model that operates at coarser effective resolutions. Unlike standard model distillation, which focuses on compressing parameters while preserving an input-output mapping, operator distillation is formulated through the spatiotemporal discretization view, where the student is trained to match downsampled teacher trajectories, so that it learns effective coarse-grid dynamics that aggregate cross-resolution correction while preserving large-scale structures, inertial-range spectra, and long-horizon correlations. Finally, when a small set of high-fidelity observations becomes available, few-shot fine-tuning adjusts the distilled student using only a handful of trajectories, correcting amplitudes and small-scale details on the pre-trained operator rather than learning the dynamics from scratch. In this way, the final OmniFluids operator combines the inductive biases of PDE-based pre-training and coarse-grained operator distillation with the accuracy of task-specific supervision. {\textcolor{blue}{Supplementary Note C} further provides a mechanistic analysis of the three-stage strategy, characterizing the cross-resolution information captured by physics pre-training and distillation and the residual errors corrected by few-shot adaptation. }

\textbf{Physics-only pre-training.} In the physics-only pre-training stage, we leverage the designed OmniFluids architecture to learn the fluid operator directly from the PDE information without any data. At each training epoch, we sample an initial field $\mathbf{u}_0$ from a prescribed distribution and draw PDE parameters $\mathbf{\gamma}$ from their respective priors, ensuring the model generalizes across diverse physical tasks. Given an initial field $\mathbf{u}_0$ on a regular grid, the network predicts $K$ future states $\{{\mathbf{u}}_k^t\}_{k=1}^K$ in parallel via a multi-frame decoder. For spatial derivatives, we employ the pseudo-spectral method, which exhibits spectral (i.e., exponential) convergence and captures global flow features, making it the state-of-the-art choice on uniform grids~\cite{JML1268,mcgreivy2024weak}. Concretely, for each predicted field ${\mathbf{u}}_k^t$, we compute the right-hand side of the governing PDE ($\mathcal{N}(\hat{\mathbf{u}}^t_k)$) in Fourier space to obtain accurate estimates of advection, diffusion, and forcing terms.

We adopt the Crank-Nicolson scheme to integrate time more accurately than first-order methods. Given a time step $\Delta t$, the physics loss $\mathcal{L}_{\mathrm{PDE}}$ is defined by the mean squared residual of the time-discretized PDE across all frames:
\begin{equation}
    \mathcal{L}_{\mathrm{PDE}} = \frac{1}{K}\sum_{k=1}^{K} \Bigl\|\frac{\hat{\mathbf{u}}_k^t - \hat{\mathbf{u}}_{k-1}^t}{\Delta t} - \frac{1}{2}\bigl(\mathcal{N}(\hat{\mathbf{u}}_k^t)+\mathcal{N}(\hat{\mathbf{u}}_{k-1}^t)\bigr)\Bigr\|_2^2,
\end{equation}
where we abbreviate $\hat{\mathbf{u}}_{0}$ as $\hat{\mathbf{u}}_{0}^t$ for simplicity. Because OmniFluids outputs all $K$ frames in a single forward pass, these computations can be performed fully in parallel, eliminating the need for step-by-step iteration and greatly simplifying training. {For the 2D NSEs, we consider the velocity-vorticity formulation, in which the pressure is eliminated. For the 3D case, although we adopt a velocity-pressure formulation, the network's inputs and outputs consist only of the velocity field. The pressure is not predicted explicitly; instead, its contribution appears only in the physics-based loss, where we enforce the 3D momentum equations via an FFT-based Helmholtz--Hodge projection with a zero-mean pressure convention.}

\textbf{Operator distillation.} In this stage, we aim to compress the high-capacity pre-trained teacher network into an efficient student operator via knowledge distillation~\cite{hinton2015distilling}. The goal is to preserve the teacher's predictive fidelity while significantly reducing computational cost, notably by enabling inference on coarser spatial and temporal grids. To achieve this, we fix the pre-trained teacher and train the student network to match its outputs under downsampled conditions. We sample the same initial condition $\mathbf{u}_0$ and parameters $\mathbf{\gamma}$ as in the pre-training stage, but evaluate the difference between the teacher and student models on a coarsened grid and with a larger time step $M \Delta t$ with $M \gg 1$. The distillation loss aligns the student's predictions $\hat{\mathbf{u}}_{MK}^s$ with the downsampled teacher outputs via a mean squared error:
\begin{equation}
\mathcal{L}_{\mathrm{Dist.}} = \left\|\tilde{\mathbf{u}}_{MK}^s - \mathcal{D}({\mathbf{u}}_{MK}^t)\right\|_2^2,
\end{equation}
where $\mathcal{D}(\cdot)$ denotes the spatial downsampling operator applied to the teacher's output. As a result, operator distillation yields a model adapted to coarse spatial and temporal resolutions, supporting fast deployment while preserving the key information from the high-resolution teacher network.

\textbf{Few-shot fine-tuning.} Once a small set of high-fidelity observations becomes available, we fine-tune the distilled student model to further enhance its accuracy. This stage requires only a handful of ground-truth trajectories (e.g., 2--10), enabling the model to align its predictions more closely with empirical data and correct any residual errors from the physics-driven prior. Fine-tuning is performed end-to-end using a standard mean squared loss on predicted fields ($\mathcal{L}_{\text{Data}}$). We adopt a small learning rate during this stage to ensure stable adaptation and avoid overwriting pre-trained knowledge. This training stage preserves the inductive efficiency and generalization capabilities acquired during physics-only pre-training and operator distillation, while tailoring the model to the specific target task.

\subsection*{{Downstream tasks of fluid systems}}

\textbf{Inverse problems.} Inverse problems play a central role in scientific modeling, enabling the identification of unknown physical parameters from limited or noisy observations. In fluid systems, solving such problems is critical for system identification, model personalization, and downstream control or forecasting tasks. Leveraging the generalization capability of the distilled student model, OmniFluids provides an efficient and principled approach for inverse analysis. The key idea is to treat unknown physical parameters, such as viscosity $\nu$ or external forcing, as differentiable inputs to the model and optimize them by minimizing discrepancies between predicted and observed flow fields. The process begins with an initial guess $\mathbf{\gamma}$ of the unknown parameters. Given an initial condition $\tilde{\mathbf{u}}_0$, the student model performs a rollout to predict the final state $\tilde{\mathbf{u}}_T^s(\mathbf{\gamma})$ after $M$ steps. The inverse loss $\mathcal{L}_{\text{Diff.}}$ measures the difference between the observed final state $\tilde{\mathbf{u}}_T$ and the predicted one:
\begin{equation}
\mathcal{L}_{\text{Diff.}}(\mathbf{\gamma}) = \left\| \tilde{\mathbf{u}}^s_T(\mathbf{\gamma}) - \tilde{\mathbf{u}}_T \right\|_2^2.
\end{equation}

This loss is minimized via gradient descent:
\begin{equation}
\mathbf{\gamma} \leftarrow \mathbf{\gamma} - \eta \nabla_{\mathbf{\gamma}} \mathcal{L}_{\text{Diff.}}(\mathbf{\gamma}),
\end{equation}
where $\eta$ is the learning rate. Thanks to the fully differentiable architecture of OmniFluids, gradients with respect to the physical parameters can be computed via automatic differentiation. Compared to traditional simulation-based inverse solvers, OmniFluids is highly efficient for inverse problems, where all forward and backward steps are computed in a single or a few passes, allowing for rapid optimization over physical parameters. Once the optimal $\mathbf{\gamma}$ is recovered, it can be used for further prediction, simulation, or control, making OmniFluids a powerful tool for data-driven discovery in complex fluid systems.

\textbf{{Flow Control.}} {Beyond parameter identification, OmniFluids can be used to design time-dependent control inputs that guide the system toward desired states. In our setting, we model the external forcing sequence as a set of differentiable control variables $\mathbf{f} = \{\mathbf{f}_t\}_{t=0}^{M-1}$ that enter the student model as inputs at each rollout step. Starting from an initial condition $\tilde{\mathbf{u}}_0$ and a candidate control sequence $\mathbf{f}$, the student model generates a trajectory $\{\tilde{\mathbf{u}}_t^s(\mathbf{f})\}_{t=1}^M$. We then define a control loss that penalizes deviations of a target global quantity, as well as a regularization term to avoid excessively strong forcing:}
\begin{equation}
{
\mathcal{L}_{\text{Ctrl.}}(\mathbf{f}) = \sum_{t=1}^M \left\| \mathcal{Q}\big(\tilde{\mathbf{u}}_t^s(\mathbf{f})\big) - \mathcal{Q}_t^{\star} \right\|_2^2 \;+\; \lambda \sum_{t=0}^{M-1} \left\| \mathbf{f}_t \right\|_2^2,
}
\end{equation}
{where $\mathcal{Q}(\cdot)$ extracts the quantity of interest (e.g., enstrophy), $\mathcal{Q}_t^{\star}$ denotes the prescribed target profile, and $\lambda$ controls the strength of regularization. The control variables $\mathbf{f}$ are optimized by gradient descent through the differentiable student operator, enabling efficient discovery of control strategies that realize the desired flow behavior.}

\textbf{{Sparse reconstruction.}} {OmniFluids can also perform sparse reconstruction from limited measurements. In this case, the unknowns are the latent flow states that are consistent with the observed partial data. We parameterize the latent variables as $\mathbf{z}$, which represents the initial conditions. Given an observation operator $\mathcal{P}$ that subsamples the flow field with sparse measurements $\{\mathbf{y}_t\}_{t=1}^M$, the student operator produces a trajectory $\{\tilde{\mathbf{u}}_t^s(\mathbf{z})\}_{t=1}^M$ and we define a reconstruction loss}
\begin{equation}
{
\mathcal{L}_{\text{Recon.}}(\mathbf{z}) = \sum_{t=1}^M \left\| \mathcal{P}\big(\tilde{\mathbf{u}}_t^s(\mathbf{z})\big) - \mathbf{y}_t \right\|_2^2 \;+\; \beta \left\| \mathbf{z} \right\|_2^2,
}
\end{equation}
{where the first term enforces consistency with the sparse observations and the second term provides a mild regularization. As with inverse problems and control, the latent variables $\mathbf{z}$ are updated via gradient-based optimization through the student model.}

\subsection*{Benchmark PDEs}
To evaluate the performance of OmniFluids, we conduct experiments on a diverse set of PDEs that represent classical problems in fluid dynamics, ranging from chaotic pattern formation to realistic incompressible flows. Specifically, we consider the 2D Kuramoto-Sivashinsky equations (KSEs), the 2D incompressible Navier-Stokes equations (NSEs) in velocity-vorticity form, and the 3D incompressible NSEs in velocity-pressure form. All datasets are generated with high-resolution spectral solvers, and the detailed settings for each system are provided in \textcolor{blue}{Supplementary Note A}.

\textbf{2D KSE.} The 2D KSE is a fourth-order nonlinear PDE widely used as a canonical model for spatiotemporal chaos in dissipative systems, such as the propagation of flame~\cite{sivashinsky1980flame}. It exhibits complex behaviors, including pattern formation, wave turbulence, and chaotic attractors~\cite{michelson1977nonlinear}, making it an ideal benchmark for understanding chaotic systems. The 2D KSE is given by  
\begin{equation}
\begin{aligned}
&\frac{\partial u}{\partial t} + \alpha(\Delta {u} + \Delta^2 {u}) + \frac{\beta}{2}  |\nabla{u}|^2 = 0,
\end{aligned}
\end{equation}  
where \({u}(\mathbf{x}, t) \in \mathbb{R}\) denotes the solution field with $\mathbf{x}\in [0, 8\pi)^2$ and $t\in[0, 5]$, \(\Delta\) and \(\Delta^2\) represent the Laplacian and biharmonic operators, and $\nabla$ is the spatial gradient operator. The parameters \(\alpha\) and \(\beta\) control the strength of the linear and nonlinear terms. In the physics-only pre-training phase of OmniFluids, we sample $u_0\in\mathcal{G}_{\text{2DKS}}$, \(\alpha \in [0.1, 0.5]\) and \(\beta \in [0.1, 0.5]\) to cover a wide range of chaotic regimes, where $\mathcal{G}_{\text{2DKS}}$ is a 2D Gaussian random field.

\textbf{2D incompressible vorticity-velocity NSE.} The 2D incompressible NSEs in the vorticity-velocity formulation serve as a compact yet expressive model for laminar and turbulent flows. The vorticity form eliminates pressure, reducing the system to a single scalar PDE that maintains essential nonlinear convection and viscous diffusion effects~\cite{guj1993vorticity}. The governing equation is
\begin{equation} 
\begin{aligned} 
\frac{\partial \omega}{\partial t} + \mathbf{u} \cdot \nabla \omega &= \frac{1}{\text{Re}} \Delta \omega + f,\\
\omega &= \nabla \times \mathbf{u},\\
\end{aligned}\label{eq:nse} 
\end{equation}
where $\omega(\mathbf{x}, t)\in \mathbb{R}$ is the vorticity scalar field with $\mathbf{x} \in [0, 2\pi)^2$ and $t \in [0, 10]$, $\mathbf{u}=[u, v]\in\mathbb{R}^2$ is the incompressible velocity field satisfying $\nabla \cdot \mathbf{u} = 0$, $f$ and $\text{Re}$ denote the external forcing and Reynolds number, respectively. In our pre-training setup, we vary $\text{Re} \in [500, 2500]$, $\omega_0\in\mathcal{G}_{\text{2DNS}}$ and $f\in \mathcal{F}$ to induce a range of vortex dynamics, where $\mathcal{G}_{\text{2DNS}}$ and $\mathcal{F}$ are a 2D Gaussian random field and a 2D uniform random Fourier field, respectively.

\textbf{3D incompressible velocity-pressure NSE.} The 3D NSEs are a cornerstone of CFD, encapsulating the fundamental laws governing the motion of viscous, incompressible fluids~\cite{raugel1993navier}. Unlike 2D cases, the 3D form retains the full spatial complexity and nonlinear interactions that characterize real-world turbulent flows. Solving these equations accurately is critical for simulating real-world flows in various engineering and geophysical systems. The governing equation is
\begin{equation} \begin{aligned}
\frac{\partial{\mathbf{u}}}{\partial t} + \mathbf{u} \cdot \nabla \mathbf{u} &= -\nabla p + \frac{1}{\text{Re}} \Delta \mathbf{u} + 0.1\mathbf{u}, \\
\nabla \cdot \mathbf{u} &= 0,
\end{aligned} \end{equation}
where $\mathbf{u}(\mathbf{x}, t) = [u, v, w] \in \mathbb{R}^3$ is the velocity field with $\mathbf{x} \in [0, 2\pi)^3$ and $t \in [0, 10]$, $p(\mathbf{x}, t) \in \mathbb{R}$ is the pressure field. $\text{Re}$ is the Reynolds number. The incompressibility constraint $\nabla \cdot \mathbf{u} = 0$ enforces mass conservation. In our pre-training regime, we sample $\mathbf{u}_0\in\mathcal{G}_{\text{3DNS}}$, $\text{Re}\in[500,\,2500]$ to generate a diverse ensemble of 3D unsteady flow regimes, where $\mathcal{G}_{\text{3DNS}}$ is a 3D Gaussian random field.

{\textbf{3D wall-bounded turbulent channel flow.} To complement the periodic 3D NSE benchmark, we further consider a three-dimensional incompressible turbulent channel flow. This problem is a classical benchmark for wall-bounded turbulence, where the no-slip walls generate strong shear, near-wall streaks, anisotropic velocity fluctuations, and coherent vortical structures. The governing equations are
\begin{equation}
\begin{aligned}
\frac{\partial \mathbf{u}}{\partial t}
+ \mathbf{u}\cdot\nabla \mathbf{u}
&= -\nabla p + \nu \Delta \mathbf{u} + G\mathbf{e}_x, \\
\nabla\cdot\mathbf{u} &= 0,
\end{aligned}
\end{equation}
where $\mathbf{u}(\mathbf{x},t)=[u,v,w]\in\mathbb{R}^3$ is the velocity field, $p(\mathbf{x},t)\in\mathbb{R}$ is the pressure field, $\nu$ is the kinematic viscosity, and $G\mathbf{e}_x$ denotes the constant streamwise pressure-gradient forcing. The computational domain is $\mathbf{x}=(x,y,z)\in[0,2\pi)\times[-1,1]\times[0,\pi)$ with $t\in[0,8]$. Periodic boundary conditions are imposed in the streamwise and spanwise directions, and no-slip boundary conditions are imposed at the two channel walls,
\begin{equation}
    \mathbf{u}(x,y=\pm1,z,t)=\mathbf{0}.
\end{equation}
In our experiment, we set $\nu=2.5\times10^{-4}$ and $G=0.05$. Here $\mathrm{Re}=1/\nu$ denotes the nominal inverse-viscosity Reynolds number used in the nondimensional solver. The data are generated using a Fourier--Chebyshev spectral solver, with Fourier discretization in the periodic directions and Chebyshev discretization in the wall-normal direction. The reference trajectories are generated on a $256\times256\times128$ grid and evaluated on a $128\times128\times64$ grid. OmniFluids is first pre-trained using only the governing-equation residuals and boundary-condition constraints, without any labeled DNS trajectory supervision. During this physics-only pre-training stage, the viscosity is sampled from $\nu\in[1/4000,1/2000]$.

For the wall-normal statistics in Fig.~\ref{fig:5}\textbf{c}, all profiles are averaged over the homogeneous streamwise and spanwise directions. The two channel halves are folded using the wall distance $d_w=1-|y|$, and the inner coordinate is defined as $y^+=u_\tau d_w/\nu$, where $u_\tau$ is the friction velocity estimated from the reference wall shear. The inner-normalized statistics are computed as
\begin{equation}
U^+=\frac{\langle u\rangle}{u_\tau},\quad
u_{\mathrm{rms}}^+=\frac{\sqrt{\langle (u-\langle u\rangle)^2\rangle}}{u_\tau},\quad
v_{\mathrm{rms}}^+=\frac{\sqrt{\langle (v-\langle v\rangle)^2\rangle}}{u_\tau},\quad
w_{\mathrm{rms}}^+=\frac{\sqrt{\langle (w-\langle w\rangle)^2\rangle}}{u_\tau},
\end{equation}
where $\langle\cdot\rangle$ denotes averaging over the corresponding homogeneous directions, selected rollout times, and folded channel halves. The same reference $u_\tau$ is used to normalize all model predictions and the reference simulation.}

{We also report the inner-normalized Reynolds normal and shear stresses,
\begin{equation}
\frac{\overline{u'u'}}{u_\tau^2}
=
\frac{\langle (u-\langle u\rangle)^2\rangle}{u_\tau^2},
\qquad
-\frac{\overline{u'v'}}{u_\tau^2}
=
-\frac{\langle (u-\langle u\rangle)(v-\langle v\rangle)\rangle}{u_\tau^2}.
\end{equation}
These statistics are averaged over the homogeneous directions and the evaluated prediction snapshots using the same reference friction velocity $u_\tau$ for all methods.}

\subsection*{Baseline models}

In the physics-only pre-training stage, we use PINO~\cite{li2024physics} as a baseline, which can learn CFD surrogate models without data supervision. When transferring to specific tasks, for the 2D cases, we compare the performance of OmniFluids with various other approaches, including state-of-the-art neural operators such as PINO~\cite{li2024physics}, DeepONet~\cite{lu2021learning}, FNO~\cite{li2021fourier}, FFNO~\cite{tran2023factorized}, FactFormer~\cite{li2023scalable}, and CNO~\cite{raonic2023convolutional}; a large-scale data-driven pre-trained foundation model (DPOT~\cite{hao2024dpot}); and direct numerical simulation (DNS) based on spectral methods. For the 3D case, we use PINO~\cite{li2024physics}, FNO~\cite{li2021fourier}, {FactFormer~\cite{li2023scalable}}, DPOT~\cite{hao2024dpot}, and DNS as baselines for comparison. The introduction and detailed settings of the baseline models are found in \textcolor{blue}{Supplementary Note B}. 

\subsection*{{Accuracy-qualified comparison with numerical solvers}}

{To perform the accuracy-qualified comparison, we evaluate the OOD 2D NSE and 3D NSE cases on common evaluation grids of $128^2$ and $128^3$, respectively. For each pseudospectral configuration, the initial condition is obtained by periodic Fourier restriction of the fine-grid reference, and the resulting predictions are Fourier-resampled to the common evaluation grid before the errors are computed. We test $N\in\{64,80,96,100,104,112,128\}$ for 2D NSE and $N\in\{64,96,112,120,128\}$ for 3D NSE. Here, $N$ denotes the number of grid points per spatial direction on a uniform periodic Cartesian grid. A numerical configuration is considered accuracy-qualified when both its trajectory-averaged rollout-mean and terminal relative $\ell_2$ errors are no greater than the corresponding OmniFluids values. For each benchmark-resolution pair, we prescribe a time step that allows all test trajectories to complete the full forecast horizon without solver failure or non-finite field values.} 

\subsection*{Evaluation metrics}

We use two primary metrics to quantitatively evaluate the prediction accuracy of each model, including the relative $\ell_2$ error and correlation coefficient.

\textbf{Relative \(\ell_2\) error} measures the normalized difference between the predicted solution \({\mathbf{u}}_{\operatorname{pre}}\) and the reference solution \(\mathbf{u}\). It is defined as
\[
\text{Relative } \ell_2 \text{ error} = \frac{\| {\mathbf{u}_{\operatorname{pre}}} - \mathbf{u} \|_2}{\| \mathbf{u} \|_2} = \frac{\sqrt{\sum_i ({\mathbf{u}_{\operatorname{pre}}}_{(i)} - \mathbf{u}_{(i)})^2}}{\sqrt{\sum_i \mathbf{u}_{(i)}^2}},
\]
where the sums run over all spatial points in the solution.

\textbf{Correlation coefficient} evaluates the linear correlation between the predicted and reference solutions, defined by
\[
\text{Corr}({\mathbf{u}_{\operatorname{pre}}}, \mathbf{u}) = \frac{\sum_i ({\mathbf{u}_{\operatorname{pre}}}_{(i)} - \overline{{\mathbf{u}}_{\operatorname{pre}}})(\mathbf{u}_{(i)} - \bar{\mathbf{u}})}{\sqrt{\sum_i ({\mathbf{u}_{\operatorname{pre}}}_{(i)} - \overline{{\mathbf{u}}_{\operatorname{pre}}})^2} \sqrt{\sum_i (\mathbf{u}_{(i)} - \bar{\mathbf{u}})^2}},
\]
where \(\overline{{\mathbf{u}}_{\operatorname{pre}}}\) and \(\bar{\mathbf{u}}\) denote the mean values of \({\mathbf{u}_{\operatorname{pre}}}\) and \(\mathbf{u}\), respectively.

\section*{Data availability} 
{The KSE, 2D and 3D NSE, and channel-flow datasets are available on Zenodo~\cite{zhang2026omnifluidsdata}. The cylinder-wake PIV dataset is publicly available at \url{https://doi.org/10.24433/CO.6670426.v2}. Source data are provided with this paper.}

\section*{Code availability} 
Code for the core KSE, 2D NSE, 3D NSE and channel-flow pipelines is available at \url{https://github.com/intell-sci-comput/OmniFluids}, with version 1.0.0 archived on Zenodo~\cite{zhang2026omnifluidssoftware}. The release includes data generation, physics-only pre-training, distillation, fine-tuning and periodic-student evaluation.

\clearpage

\noindent\textbf{Funding Statement:}
The work is supported by the Beijing Natural Science Foundation (No. F261002), the National Natural Science Foundation of China (No. 62276269, No. 62506367, No. 62576331, and No. 92270118), and the Strategic Priority Research Program of the Chinese Academy of Sciences (No. XDB0620103). R.Z. was supported by the China Postdoctoral Science Foundation under Grant Number 2025M771582 and the Postdoctoral Fellowship Program of CPSF under Grant Number GZB20250408. Y.L. was supported by the Fundamental Research Funds for the Central Universities (E2EG2202X2). \par\vspace{\baselineskip}

\noindent\textbf{Author contributions:} R.Z., Q.M., and H.S. contributed to the ideation and design of the research; R.Z. and H.W. performed the research; {Z.L. and W.H. provided and analyzed the experimental cylinder-wake data; M.Y. assisted with the CFD-related experimental analysis;} H.S. supervised the project; all authors contributed to the research discussions, writing, and editing of the paper.\par\vspace{\baselineskip}

\noindent\textbf{Correspondence to:} Hao Sun (\url{haosun@ruc.edu.cn}).\par\vspace{\baselineskip}

\noindent\textbf{Competing interests:}
The authors declare no competing interests.\par\vspace{\baselineskip}

\noindent\textbf{Supplementary information:}
The Extended Data figures and Supplementary Information follow the main reference list in this preprint. Supplementary Videos 1--4 and Source Data are provided as ancillary files.

\clearpage
\pdfbookmark[0]{Extended Data}{extended-data}
\setcounter{figure}{0}
\captionsetup[figure]{name=Extended Data Figure}
\renewcommand{\theHfigure}{ED.\arabic{figure}}
\clearpage
\begin{figure}[p]
\centering
\includegraphics[width=\linewidth,height=0.77\textheight,keepaspectratio]{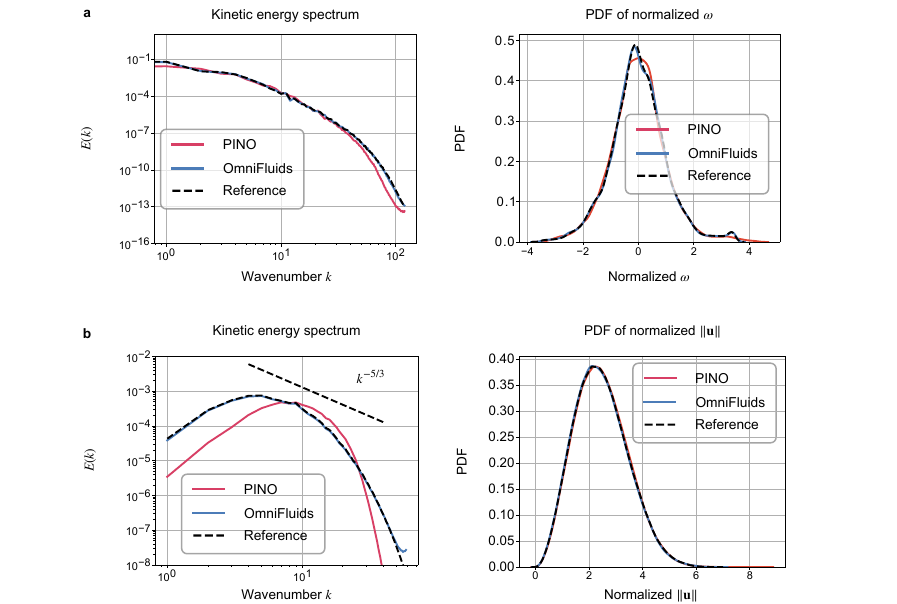}
\caption{\textbf{Flow statistics from PINO and OmniFluids across diverse CFD systems in the physics-only pre-training stage.} \textbf{a} and \textbf{b}, Flow statistics comparison: {kinetic energy spectrum $E(k)$} and probability density function (PDF) of normalized vorticity $\omega$ (for 2D NSE) or velocity magnitude $\|\mathbf{u}\|$ (for 3D NSE) at $t=10$.}
\label{edfig:1}
\end{figure}
\clearpage
\begin{figure}[p]
\centering
\includegraphics[width=\linewidth,height=0.77\textheight,keepaspectratio]{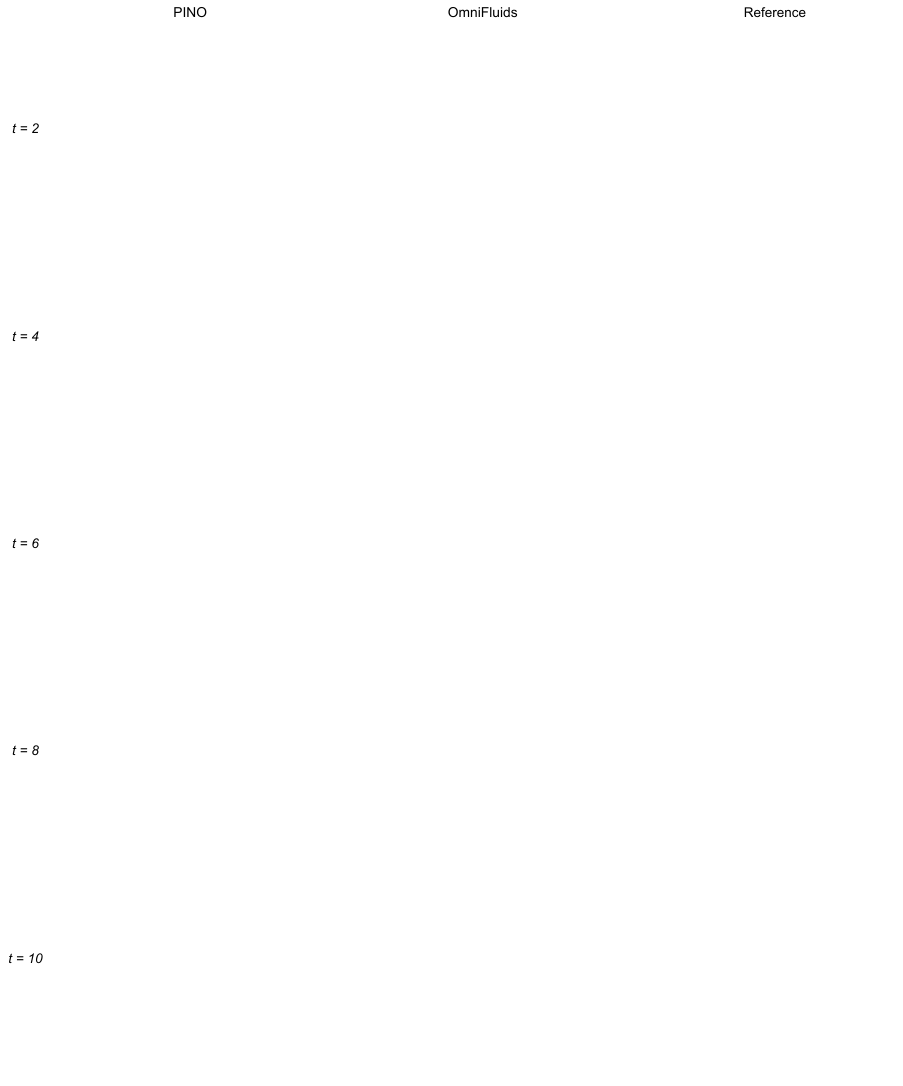}
\caption{\textbf{{Isosurface renderings} of the predicted 3D full velocity magnitude fields in the physics-only pre-training stage.}}
\label{edfig:2}
\end{figure}
\clearpage
\begin{figure}[p]
\centering
\includegraphics[width=\linewidth,height=0.77\textheight,keepaspectratio]{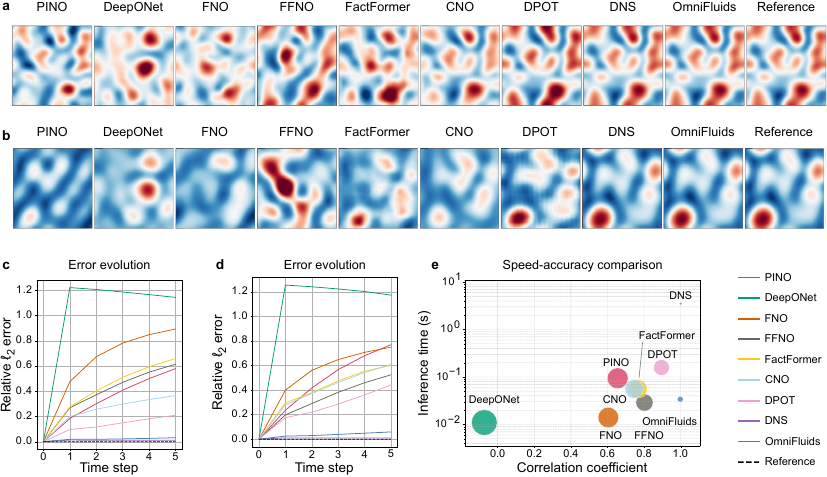}
\caption{\textbf{Predicted snapshots, multi-step errors, and inference speed comparison of baseline methods and OmniFluids on 2D KSE.} \textbf{a} and \textbf{c}, In-distribution snapshots ($t=5$) and relative $\ell_2$ errors with parameters $\alpha=0.2, \beta=0.5$.
\textbf{b} and \textbf{d}, Out-of-distribution snapshots ($t=5$) and relative $\ell_2$ errors with parameters $\alpha=1.0, \beta=1.0$. \textbf{e}, Speed-accuracy comparison for the OOD case, where marker size represents relative $\ell_2$ error at the final time.}
\label{edfig:3}
\end{figure}
\clearpage
\begin{figure}[p]
\centering
\includegraphics[width=\linewidth,height=0.77\textheight,keepaspectratio]{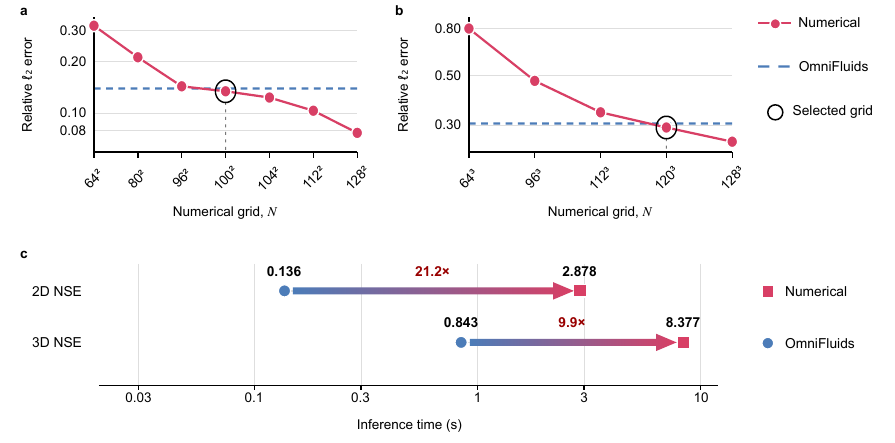}
\caption{{\textbf{Accuracy-qualified comparison with pseudospectral solvers.} \textbf{a} and \textbf{b}, Mean relative $\ell_2$ errors across the tested numerical-grid resolutions for 2D and 3D NSE, respectively. Dashed blue lines show the corresponding OmniFluids errors; open circles mark the first tested grids satisfying both the rollout-mean and terminal-error criteria ($100^2$ and $120^3$, respectively). \textbf{c}, Inference times of OmniFluids and the selected numerical configurations.}}
\label{edfig:4}
\end{figure}
\clearpage
\begin{figure}[p]
\centering
\includegraphics[width=\linewidth,height=0.77\textheight,keepaspectratio]{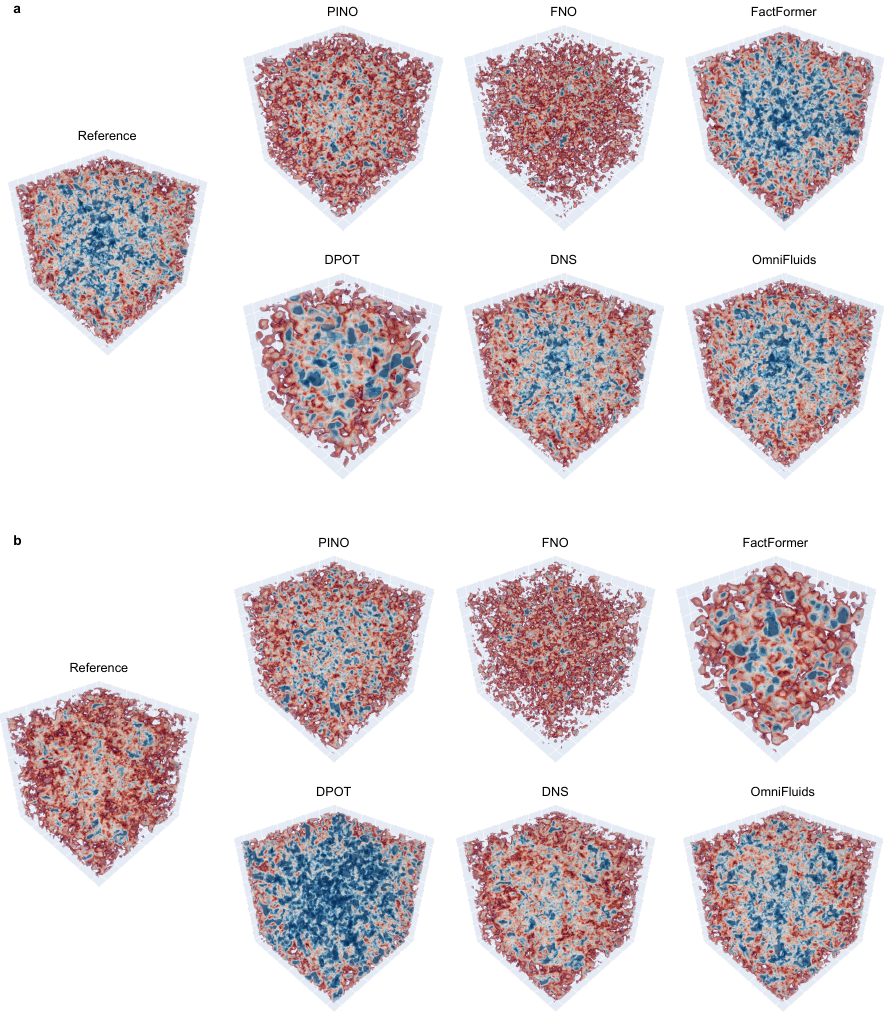}
\caption{\textbf{{Isosurface renderings} of the predicted 3D velocity magnitude at $t=8$.} \textbf{a}, ID setting ($\text{Re} = 2000$). \textbf{b}, OOD setting ($\text{Re} = 3000$).}
\label{edfig:5}
\end{figure}
\clearpage
\begin{figure}[p]
\centering
\includegraphics[width=\linewidth,height=0.77\textheight,keepaspectratio]{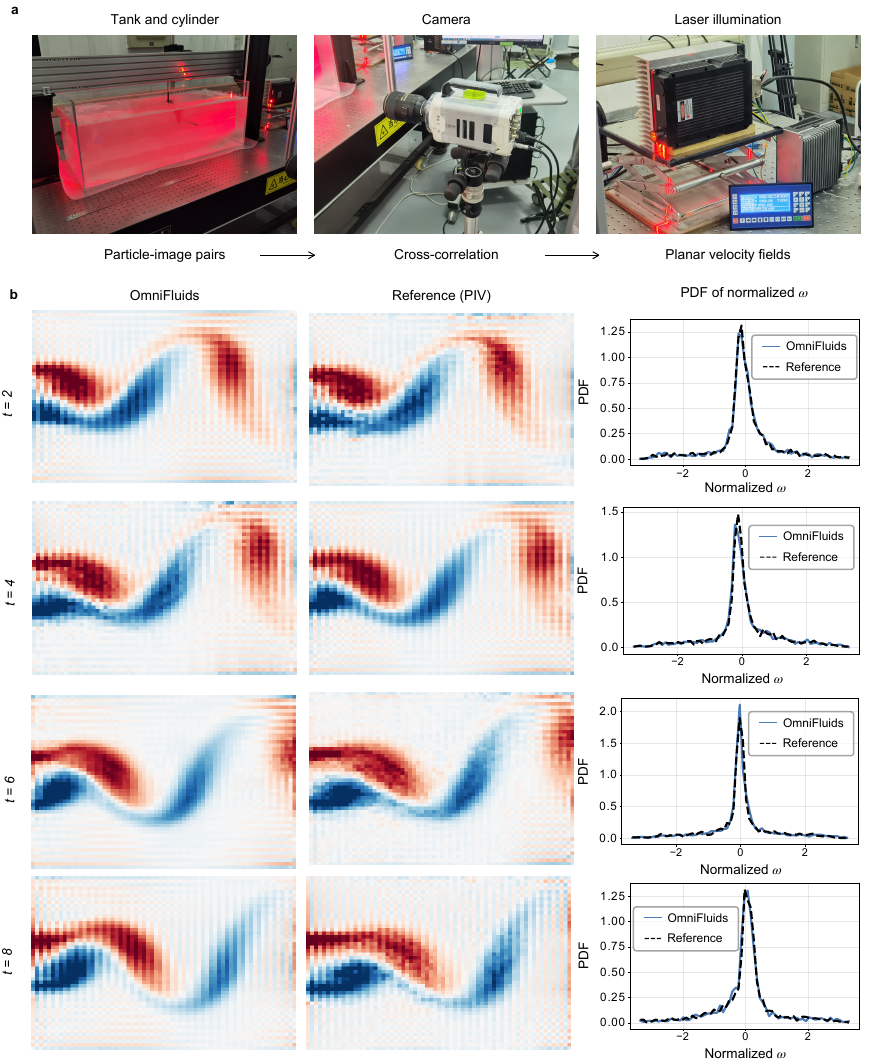}
\caption{\textbf{{Transfer to experimental cylinder-wake PIV measurements.}} \textbf{a}, Photographs of the experimental apparatus: water tank and cylinder (left), camera (middle), and laser illumination unit (right). The workflow below summarizes the processing of particle-image pairs by cross-correlation to obtain planar velocity fields. \textbf{b}, Autoregressive rollout snapshots of the vorticity field predicted by OmniFluids (Left) compared with the PIV reference (Middle) at four time steps. Right: PDFs of normalized vorticity at the corresponding time steps.}
\label{edfig:6}
\end{figure}

\clearpage
\pdfbookmark[0]{Supplementary Information}{supplementary-information}
\section*{Supplementary Information}
\setcounter{section}{0}
\setcounter{figure}{0}
\setcounter{table}{0}
\setcounter{equation}{0}
\renewcommand\refname{Supplementary References}
\renewcommand\thesection{\Alph{section}}
\renewcommand\thesubsection{\thesection.\arabic{subsection}}
\renewcommand\thesubsubsection{\thesubsection.\arabic{subsubsection}}
\renewcommand{\thefigure}{S.\arabic{figure}}
\renewcommand{\thetable}{S.\arabic{table}}
\renewcommand{\theHsection}{SI.\arabic{section}}
\renewcommand{\theHfigure}{SI.\arabic{figure}}
\renewcommand{\theHtable}{SI.\arabic{table}}
\renewcommand{\theHequation}{SI.\arabic{equation}}
\captionsetup[figure]{name=Figure}

{\footnotesize
\tableofcontents
}

\clearpage

\noindent This supplementary document provides a detailed description of the data generation process, details on the implementation of OmniFluids and baseline methods, a discussion of ablation studies, and additional experimental results.

\section{Data generation}

\begin{table}[t!]
\centering
\caption{Computational parameters for datasets generation.}
\resizebox{16cm}{!}{
\begin{tabular}{cccccc}
\toprule
\textbf{Dataset} & \textbf{Phase} & \textbf{Parameters} & $\mathbf{N_x}$ & $\mathbf{N_t}$ & \# \textbf{Train/Val./Test} \\
\midrule
\multirow{3}{*}{\textbf{2D KSE}} 
  & Pre-train & $\alpha \in [0.1, 0.5]$, $\beta \in [0.1, 0.5]$ & $128^2$ &$5\times 10^4$ & 0 / 0  / 10 \\
\cmidrule{2-6}
  & Fine-tuning (ID) & $\alpha = 0.2$, $\beta = 0.5$ & $128^2$ & $5\times 10^4$ & 2 / 2 / 10 \\
\cmidrule{2-6}
  & Fine-tuning (OOD) & $\alpha = 1.0$, $\beta = 1.0$ & $128^2$ & $5\times 10^4$ & 2 / 2 / 10 \\
\midrule
\multirow{3}{*}{\textbf{2D NSE}} 
  & Pre-train & $\text{Re} \in [500, 2500]$, $f \in \mathcal{F}$ & $1024^2$ &$1\times 10^5$ & 0 / 0  / 10 \\
\cmidrule{2-6}
  & Fine-tuning (ID) & $\text{Re} = 2000$, $f \in \mathcal{F}$ & $1024^2$ & $1\times 10^5$ & 10 / 2 / 10 \\
\cmidrule{2-6}
  & Fine-tuning (OOD) & $\text{Re} = 4000$, $f \in \mathcal{F}$ & $1024^2$ & $1\times 10^5$ & 10 / 2 / 10 \\
\midrule
\multirow{3}{*}{\textbf{3D NSE}} 
  & Pre-train & $\text{Re} \in [500, 2500]$ & $512^3$ &$1\times 10^4$ & 0 / 0  / 4 \\
\cmidrule{2-6}
  & Fine-tuning (ID) & $\text{Re} = 2000$ & $512^3$ & $1\times 10^4$ & 4 / 2 / 4 \\
\cmidrule{2-6}
  & Fine-tuning (OOD) & $\text{Re} = 3000$ & $512^3$ & $1\times 10^4$ & 4 / 2 / 4 \\
\bottomrule
\end{tabular}}
\label{SI-tab:data}
\end{table}

To evaluate the performance of OmniFluids, we conduct experiments on a diverse set of PDEs that represent canonical problems in fluid dynamics, ranging from chaotic pattern formation to realistic incompressible flows. Specifically, we consider the 2D Kuramoto-Sivashinsky equations (KSEs), the 2D incompressible Navier-Stokes equations (NSEs) in velocity-vorticity form, and the 3D incompressible NSEs in velocity-pressure form. These systems are physically interrelated and collectively capture a broad spectrum of fluid phenomena--from the emergence of spatiotemporal chaos to 3D turbulence--thus offering a comprehensive evaluation framework for CFD modeling and simulation. Fixed random seeds were used for dataset splitting to ensure reproducibility: seed 0 for training, seed 1 for validation, and seed 2 for testing. All datasets are generated with high-resolution pseudo-spectral solvers, and the detailed settings for each system are provided in Table~\ref{SI-tab:data}.

\textbf{2D Kuramoto-Sivashinsky equations.} The 2D KSE is given by  
\begin{equation}
\begin{aligned}
&\frac{\partial u}{\partial t} + \alpha(\Delta {u} + \Delta^2 {u}) + \frac{\beta}{2}  |\nabla{u}|^2 = 0,\\
&\mathbf{x}\in [0, 8\pi)^2, t\in[0, 5],
\end{aligned}
\end{equation}  
where \({u}(\mathbf{x}, t) \in \mathbb{R}\) denotes the solution field, \(\Delta\) and \(\Delta^2\) represent the Laplacian and biharmonic operators, and $\nabla$ is the spatial gradient operator. The parameters \(\alpha\) and \(\beta\) control the strength of the linear and nonlinear terms. In the physics-only pre-training phase of OmniFluids, we sample $u_0\in\mathcal{G}_{\text{2DKS}}$, \(\alpha \in [0.1, 0.5]\) and \(\beta \in [0.1, 0.5]\) to cover a wide range of chaotic regimes, where $\mathcal{G}_{\text{2DKS}}$ is a 2D Gaussian random field.

\textbf{2D incompressible vorticity-velocity NSE.} The governing equation of 2D incompressible NSEs is
\begin{equation} \begin{aligned} &\frac{\partial \omega}{\partial t} + \mathbf{u} \cdot \nabla \omega = \frac{1}{\text{Re}} \Delta \omega + f,\quad \omega = \nabla \times \mathbf{u},\\ &\mathbf{x} \in [0, 2\pi)^2,\ t \in [0, 10], \end{aligned}\label{SI-eq:nse} \end{equation}
where $\omega(\mathbf{x}, t)\in \mathbb{R}$ is the vorticity scalar field, $\mathbf{u}=[u, v]\in\mathbb{R}^2$ is the incompressible velocity field satisfying $\nabla \cdot \mathbf{u} = 0$, $f$ and $\text{Re}$ denote the external forcing and Reynolds number, respectively. In our pre-training setup, we vary $\text{Re} \in [500, 2500]$, $\omega_0\in\mathcal{G}_{\text{2DNS}}$ and $f\in \mathcal{F}$ to induce a range of vortex dynamics representative of real-world unsteady 2D flows, where $\mathcal{G}_{\text{2DNS}}$ and $\mathcal{F}$ are a 2D Gaussian random field and a 2D uniform random Fourier field, respectively. To generate training data, we adapted the publicly available code in~\cite{SI-li2024physics}, accessible at \url{https://github.com/neuraloperator/physics_informed}.

\textbf{3D incompressible velocity-pressure NSE.} The governing equations of 3D NSE are
\begin{equation} \begin{aligned}
&\frac{\partial{\mathbf{u}}}{\partial t} + \mathbf{u} \cdot \nabla \mathbf{u} = -\nabla p + \frac{1}{\text{Re}} \Delta \mathbf{u} + 0.1\mathbf{u}, \quad \nabla \cdot \mathbf{u} = 0, \\
&\mathbf{x} \in [0, 2\pi)^3,\ t \in [0, 10],
\end{aligned} \end{equation}
where $\mathbf{u}(\mathbf{x}, t) = [u, v, w] \in \mathbb{R}^3$ is the velocity field, $p(\mathbf{x}, t) \in \mathbb{R}$ is the pressure field, $\text{Re}$ is the Reynolds number. The incompressibility constraint $\nabla \cdot \mathbf{u} = 0$ enforces mass conservation. In our pre-training regime, we sample $\mathbf{u}_0\in\mathcal{G}_{\text{3DNS}}$, $\text{Re}\in[500,\,2500]$ to generate a diverse ensemble of 3D unsteady flow regimes, where $\mathcal{G}_{\text{3DNS}}$ is a 3D Gaussian random field. We compute the curl of a 3D Gaussian random field as the initial condition, thereby ensuring the divergence-free condition is satisfied. Data generation is based on the reference implementation provided at \url{https://jipolanco.github.io/PencilFFTs.jl/dev/generated/navier_stokes/#}.

{\textbf{3D wall-bounded turbulent channel flow.} To further evaluate OmniFluids on a canonical wall-bounded turbulent flow, we consider the three-dimensional incompressible channel-flow problem. Unlike the periodic NSE benchmarks, this case includes no-slip walls, strong wall-normal inhomogeneity, near-wall streaks, anisotropic velocity fluctuations, and coherent vortical structures. The governing equations are}
{\begin{equation}\begin{aligned}\frac{\partial \mathbf{u}}{\partial t}+\mathbf{u}\cdot\nabla \mathbf{u}&=-\nabla p+\nu\Delta\mathbf{u}+G\mathbf{e}_x,\\\nabla\cdot\mathbf{u}&=0,\end{aligned}\end{equation}}
{where $\mathbf{u}(\mathbf{x},t)=[u,v,w]\in\mathbb{R}^3$ is the velocity field, $p(\mathbf{x},t)\in\mathbb{R}$ is the pressure field, $\nu$ is the kinematic viscosity, and $G\mathbf{e}_x$ denotes a constant streamwise pressure-gradient forcing. The computational domain is}
{\begin{equation}\mathbf{x}=(x,y,z)\in[0,2\pi)\times[-1,1]\times[0,\pi).\end{equation}}
{Periodic boundary conditions are imposed in the streamwise and spanwise directions, while no-slip boundary conditions are imposed at the two channel walls,}
{\begin{equation}\mathbf{u}(x,y=\pm1,z,t)=\mathbf{0}.\end{equation}}
{In this experiment, we set $\nu=2.5\times10^{-4}$ and $G=0.05$. Here $\mathrm{Re}=1/\nu$ denotes the nominal inverse-viscosity Reynolds number used in the nondimensional solver, giving nominal $\mathrm{Re}=4000$.}
{The reference trajectories are generated on a $256\times256\times128$ grid using Fourier discretization in the periodic streamwise and spanwise directions and Chebyshev discretization in the wall-normal direction. The learned models are trained and evaluated on fields interpolated to a $128\times128\times64$ grid. We use ten trajectories in total: 8 for training, 1 for validation, and 1 for testing.}

\section{Baseline methods and implementation details}

We herein introduce the baseline methods and implementation details.

\subsection{Baseline methods}

When transferring to specific tasks, for the 2D cases, we compare the performance of OmniFluids with various other approaches, including state-of-the-art neural operators such as PINO~\cite{SI-li2024physics}, DeepONet~\cite{SI-lu2021learning}, FNO~\cite{SI-li2021fourier}, FFNO~\cite{SI-tran2023factorized}, FactFormer~\cite{SI-li2023scalable}, and CNO~\cite{SI-raonic2023convolutional}; a large-scale data-driven pre-trained foundation model (DPOT~\cite{SI-hao2024dpot}); and direct numerical simulation (DNS) based on spectral methods. For the 3D case, we use PINO~\cite{SI-li2024physics}, FNO~\cite{SI-li2021fourier}, {FactFormer~\cite{SI-li2023scalable}}, DPOT~\cite{SI-hao2024dpot}, and DNS as baselines for comparison. Unless otherwise specified, all models are trained and evaluated on NVIDIA A800 GPUs using PyTorch~\cite{SI-NEURIPS2019_9015}. We fix the random seed to 0 for training all neural models.

\textbf{Software environment.} The experiments used Python 3.10.14 and PyTorch 2.5.1+cu124 (CUDA 12.4). The verified analysis environment includes NumPy 1.26.4, SciPy 1.15.3 and Matplotlib 3.9.2.

\textbf{PINO}~\cite{SI-li2024physics} is a physics-informed neural operator that incorporates PDE residuals into its loss function, enabling training without labeled data. It extends FNO with physics constraints and serves as a strong baseline for physics-aware operator learning. We use the official implementation at \url{https://github.com/neuraloperator/physics_informed}. 

\textbf{DeepONet}~\cite{SI-lu2021learning} is one of the earliest and most influential neural operator architectures. It uses a two-branch design: the branch net encodes input functions (e.g., initial conditions), while the trunk net encodes spatial coordinates, and their inner product yields the operator output. This structure enables strong function approximation capabilities across different input domains. We use the official implementation at \url{https://github.com/lululxvi/deeponet}. 

\textbf{FNO}~\cite{SI-li2021fourier} is a widely-used operator learning framework based on fast Fourier transforms. It learns mappings in the spectral domain, enabling efficient and scalable approximations of solution operators for a broad class of PDEs. We use the official codebase at \url{https://github.com/neuraloperator/neuraloperator}.

\textbf{FFNO}~\cite{SI-tran2023factorized} improves upon FNO by factorizing the Fourier kernel into separable components, which reduces parameter count and memory consumption while enabling deeper and more expressive networks. We use the implementation from \url{https://github.com/alasdairtran/fourierflow}.

\textbf{FactFormer}~\cite{SI-li2023scalable} is a transformer-based architecture tailored for spatiotemporal operator learning. It introduces factored attention mechanisms to handle large input domains with improved efficiency. FactFormer has demonstrated state-of-the-art performance across several PDE benchmarks. We employ the implementation at \url{https://github.com/BaratiLab/FactFormer}.

\textbf{CNO}~\cite{SI-raonic2023convolutional} offers a convolutional approach to neural operators, combining the translation equivariance and inductive biases of CNNs with resolution invariance. It is particularly well-suited for localized dynamics and sharp solution features. Code is available at \url{https://github.com/camlab-ethz/ConvolutionalNeuralOperator}.

\textbf{DPOT}~\cite{SI-hao2024dpot} is a large-scale data-driven pre-trained operator transformer designed as a foundation model for various physics. It is pre-trained on diverse physical systems and can be fine-tuned with minimal data for downstream tasks. We use the official codebase at \url{https://github.com/HaoZhongkai/DPOT}.

\textbf{DNS} (Direct Numerical Simulation) uses spectral methods consistent with our data generation pipeline. To ensure a fair comparison with neural operator methods, we match the spatial resolution of DNS to that used in the AI-driven approaches. This setting allows us to directly compare spectral characteristics across methods. Additionally, we select the largest feasible time step that yields stable and physically meaningful rollouts.

In Table~\ref{SI-tab:summary}, we summarize resource usage statistics for each model, including the total number of parameters, GPU memory usage, and per-epoch training time.

\begin{table}[!t]
\centering
\caption{\textbf{Resource Usage for Models Across Tasks.} This table details the model size (millions of parameters), GPU memory (MB), per-epoch training time (seconds), and {inference speed (seconds)} for each model during different training stages.}
\label{SI-tab:summary}
\small
\resizebox{16.1cm}{!}{
\begin{tabular}{ll l S[table-format=3.3] S[table-format=5.0] S[table-format=1.3] S[table-format=1.3]}
\toprule
\textbf{Task} & \textbf{Training Stage} & {\textbf{Model}} &
{\textbf{Model Size (M)}} &
{\textbf{GPU Mem (MB)}} &
{\textbf{Time / Epoch (s)}} &
{\textbf{Inference (s)}} \\
\midrule

\multirow{10}{*}{\textbf{2D KSE}} & Pre-train & PINO & 21.252 & 1751 & 0.017 & {0.107} \\
& & OmniFluids & 17.211 & 6848 & 0.101 & {0.245} \\
\cmidrule{2-7}
& Distillation & OmniFluids & 8.811 & 1557 & 0.039 & {0.034} \\
\cmidrule{2-7}
& \multirow{8}{*}{Adaptation} 
    & PINO        & 5.327  & 779  & 0.035  & {0.094} \\
&   & DeepONet    & 3.099  & 755  & 0.025  & {0.011} \\
&   & FNO         & 5.323  & 707  & 0.028  & {0.014} \\
&   & FFNO        & 2.239  & 705  & 0.042  & {0.029} \\
&   & FactFormer  & 6.226  & 1835 & 0.067  & {0.056} \\
&   & CNO         & 2.012  & 613  & 0.072  & {0.056} \\
&   & DPOT        & 122.126 & 3165 & 0.152 & {0.158} \\
&   & OmniFluids  & 2.516  & 845  & 0.057  & {0.034} \\
\midrule

\multirow{11}{*}{\textbf{2D NSE}} & Pre-train & PINO & 151.005 & 7759 & 0.091 & {0.171} \\
& & OmniFluids & 158.134 & 43981 & 1.031 & {1.310} \\
\cmidrule{2-7}
& Distillation & OmniFluids & 79.498 & 9840 & 0.347 & {0.136} \\
\cmidrule{2-7}
& \multirow{8}{*}{Adaptation} 
    & PINO        & 151.017 & 5475  & 0.090  & {0.385} \\
&   & DeepONet    & 13.184  & 10798 & 0.037  & {0.021} \\
&   & FNO         & 151.017 & 4467  & 0.053  & {0.075} \\
&   & FFNO        & 44.951  & 15171 & 0.497  & {0.103} \\
&   & FactFormer  & 11.469  & 65166 & 1.092  & {0.358} \\
&   & CNO         & 2.800   & 3285  & 0.140  & {0.138} \\
&   & DPOT        & 122.126 & 6971  & 0.172  & {0.157} \\
&   & OmniFluids  & 79.498  & 16953 & 0.294  & {0.136} \\

\midrule

\multirow{7}{*}{\textbf{3D NSE}} & Pre-train & PINO & 268.448 &  29553 &0.478& {1.489} \\
& & OmniFluids &  29.690 & 54162 &  1.675 & {6.125} \\
\cmidrule{2-7}
& Distillation & OmniFluids &  33.647 & 20396 &  0.706 & {0.843} \\
\cmidrule{2-7}
& \multirow{4}{*}{Adaptation} & PINO & 268.448 & 39584 & 0.665 & {1.489} \\
& & FNO & 268.440 & 17139 & 0.199 & {0.271} \\
& & {FactFormer} & {1.872} & {73906} & {2.221} & {3.443} \\
& & DPOT & 7.157 & 8413 & 0.112 & {1.132} \\
& & OmniFluids & 4.611 & 29464 & 0.511 & {0.843} \\

\bottomrule
\end{tabular}}
\end{table}

\subsection{Physics-only pre-training stage}
In the physics-only pre-training stage, we include PINO~\cite{SI-li2024physics} as a baseline to learn surrogate models for CFD systems without data supervision. PINO and OmniFluids are trained using the Adam~\cite{SI-kingma2015adam} optimizer for 20,000 epochs with cosine annealing learning rate scheduling. The maximum learning rate is selected from $\{0.001, 0.002, 0.005\}$ based on the lowest validation loss based on physics residual. For the 2D case, we set the batch size to 10 and use the maximum allowable number of spectral modes to ensure the model captures all frequency bands. For the 3D case, we use a batch size of 1. OmniFluids employs 64 modes per axis in 3D settings; however, for PINO, whose parameter complexity scales cubically with the number of modes, we use only 32 modes to maintain tractability. For the number of layers, OmniFluids uses 8 layers for 2D KSE, 12 for 2D NSE, and 6 for 3D NSE. In contrast, PINO uses a fixed 4-layer backbone, consistent with its official implementation, as increasing the depth typically degrades performance in its architecture~\cite{SI-tran2023factorized}. To ensure a fair comparison, we tune the channel dimensions of each model such that the overall parameter count remains approximately equal across models. 

\subsection{Operator distillation}
In the operator distillation stage, the architecture of the student model is kept nearly identical to that of the teacher model, except for a reduced number of spectral modes to match the lower spatial resolution of the input. Additionally, the final multi-frame decoder is replaced with a standard MLP, as we only require predictions of the physical field at the final time step. We train the student models for 20{,}000 epochs with cosine annealing learning rate scheduling. The maximum learning rate is selected as 0.002. The batch sizes are 10 for the 2D KSE and 2D NSE cases, and 2 for the 3D NSE cases.

\subsection{Adaptation to new tasks}

For OmniFluids, we adopt the distilled student model architecture for fine-tuning downstream tasks. We use fixed batch sizes of 2, 10, and 2 for the 2D KSE, 2D NSE, and 3D NSE settings, respectively. Each model is fine-tuned for 1,000 optimizer updates with a fixed small learning rate selected from $\{0.00005, 0.0001, 0.0002\}$, chosen based on validation performance.

We also use a small learning rate for adaptation for DPOT on 2D cases, which has been pre-trained on large-scale physics data. We employ the publicly released \texttt{dpot-m} model provided by the original authors, which is consistent with OmniFluids in terms of the size of parameters. The fine-tuning configuration is aligned with that used for OmniFluids.

For PINO, DeepONet, FNO, FFNO, FactFormer, and CNO, all models are trained from scratch for each target task. These baseline models are trained using the Adam optimizer for 20,000 epochs with cosine annealing learning rate scheduling. We use fixed batch sizes of 2, 10, and 2 for the 2D KSE, 2D NSE, and 3D NSE tasks, respectively. The maximum learning rate is selected from \{0.001, 0.002, 0.005\} based on the lowest validation loss. The key hyperparameters for PINO and FNO follow those used in the physics-only pre-training stage, with the number of Fourier modes adjusted according to the spatial resolution of each task. For DeepONet, the number of layers in the branch and trunk net is selected from \{4, 6, 8\} based on validation performance. For FFNO, we use the maximum number of spectral modes and set the number of layers to 8 for 2D KSE and 12 for 2D NSE, which aligns with OmniFluids. For FactFormer, the model depth is selected from \{4, 6, 8\} and the number of attention heads from \{8, 16\}, while other architectural settings follow the original paper. For CNO, the number of blocks is chosen from \{4, 6, 8\}.

\subsection{Downstream tasks of fluid systems}
For inverse problems, we adopt a fixed learning rate of 0.1 and optimize using the Adam optimizer for 200 epochs. The initial viscosity coefficient {is uniformly sampled from $[1/2500, 1/500]$}, and the external forcing field is initialized by randomly sampling from the forcing space \( \mathcal{F} \). {For flow control, we adopt a fixed learning rate of 0.05 and optimize using the Adam optimizer for 400 epochs. The external forcing field is initialized by randomly sampling from the forcing space \( \mathcal{F} \). For sparse reconstruction, we adopt a fixed learning rate of 0.05 and optimize using the Adam optimizer for 400 epochs. The initial field is initialized by randomly sampling from the Gaussian random field \( \mathcal{G}_{\text{2DNS}} \).} 

\begin{figure}[t]
  \centering
  \includegraphics[width=\columnwidth]{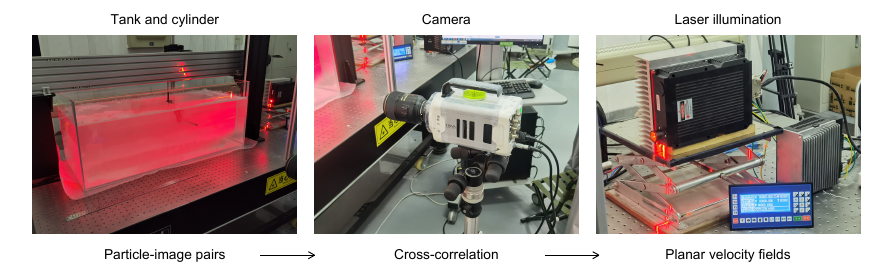}
  \caption{{Experimental apparatus for cylinder-wake PIV measurements. Photographs show the water tank and cylinder (left), camera (middle), and laser illumination unit (right). The workflow below summarizes the processing of particle-image pairs by cross-correlation to obtain planar velocity fields.}}
  \label{SI-fig:appfig0}
\end{figure}

{For laboratory cylinder-wake experiment, a circular cylinder is placed in a water tank and illuminated by a laser sheet; pairs of particle image velocimetry (PIV) images are acquired and processed by a standard cross-correlation algorithm to obtain planar velocity vector fields in the near wake behind the cylinder (Fig.~\ref{SI-fig:appfig0}~\cite{SI-li2024learning}). We use the time-resolved PIV sequence at a moderate Reynolds number ($\mathrm{Re}=109.3$), which provides two-component velocity fields $(u,v)$ on a $50\times 67$ Cartesian grid in the measurement window. The raw PIV data contain a small fraction of missing vectors. We fill these entries using nearest-neighbour interpolation in space and then convert the velocity components into a scalar vorticity field, $\omega = \partial_x v - \partial_y u$, via second-order central differences. The vorticity fields are subsequently resampled to a $128\times 128$ grid using Fourier interpolation, so that they are compatible with the resolution of the 2D NSE student used in the main experiments. We use the first 70 frames of a single PIV sequence for adaptation and the remaining 9 frames for evaluation. For this experiment, we reuse the 2D incompressible NSE student that has been distilled from the physics-only pre-trained teacher, and fine-tune it for 2{,}500 optimization steps with a batch size of 8 using the AdamW optimizer and a OneCycle learning-rate scheduler. Starting from the first test snapshot, we perform an autoregressive rollout over the remaining test horizon, treating consecutive PIV frames as unit time steps for the student model.
}

\section{{Model Analysis}}\label{SI-sec:mechanistic_interpretation}

{In this section, we conduct a detailed mechanistic analysis to clarify the rationale for the proposed design and identify what each training stage learns. Specifically, OmniFluids establishes a resolution-aware pathway from governing equations to efficient, task-specialized fluid operators. The three training stages play distinct roles: physics-only pre-training approximates the flow map on a relatively fine spatiotemporal grid, operator distillation transfers information from this fine-grid operator to an efficient deployment-resolution student, and few-shot fine-tuning calibrates the residual error for a target task. Here, we characterize these roles through cross-resolution and error analyses.}

\subsection{{The role of fine-resolution physics pre-training}}

{In this subsection, we motivate the need for fine-resolution physics pre-training by showing that fine-grid evolution followed by downsampling is not equivalent to direct evolution on the coarse grid. Let $\Phi_h$ and $\Phi_c$ denote flow maps over the same fixed prediction interval on the teacher and coarse grids, respectively, and let $\mathcal{D}$ denote downsampling. Starting from the same fine-grid state, the two routes to the deployment resolution are summarized below.}

\begin{center}
\begin{tikzcd}[column sep=4.2em,row sep=3.0em]
\mathbf{u}_t
  \arrow[r,"\Phi_h"]
  \arrow[d,"\mathcal{D}"']
&
\Phi_h(\mathbf{u}_t)
  \arrow[r,"\mathcal{D}"]
&
\mathbf{u}_{\mathrm{ref}}
\\
\mathcal{D}\mathbf{u}_t
  \arrow[rr,"\Phi_c"']
  \arrow[urr,dashed,"\text{student}"{sloped,above}]
&&
\mathbf{u}_{\mathrm{coarse}}
  \arrow[u,phantom,"\neq"]
\end{tikzcd}
\end{center}

{The upper path ($\mathbf{u}_t \xrightarrow{\Phi_h} \Phi_h(\mathbf{u}_t)
\xrightarrow{\mathcal{D}} \mathbf{u}_{\mathrm{ref}}$) first evolves the state on the teacher grid and then downsamples it. The resulting projected fine-grid evolution is used as the target for operator distillation. By contrast, the lower path ($\mathbf{u}_t \xrightarrow{\mathcal{D}} \mathcal{D}\mathbf{u}_t
\xrightarrow{\Phi_c} \mathbf{u}_{\mathrm{coarse}}$) first downsamples the initial state and then evolves it with a deployment-grid solver. For nonlinear multiscale flows, downsampling and evolution generally do not commute, i.e.,}
\begin{equation}
{
\mathcal{D}\circ\Phi_h
\neq
\Phi_c\circ\mathcal{D},
}
\end{equation}
\begin{equation}
{
\mathbf{c}_\star(\mathbf{u}_t)
\equiv
\mathbf{u}_{\mathrm{ref}}
-\mathbf{u}_{\mathrm{coarse}}
=
\left(
\mathcal{D}\circ\Phi_h
-
\Phi_c\circ\mathcal{D}
\right)(\mathbf{u}_t).
}
\label{SI-eq:finite_time_projected_correction}
\end{equation}

{Here, $\mathbf{c}_\star$ measures the difference between two states represented at the deployment resolution: the projected fine-grid reference state $\mathbf{u}_{\mathrm{ref}}$ and the state $\mathbf{u}_{\mathrm{coarse}}$ obtained by directly evolving the downsampled initial condition on the deployment grid.}

{Downsampling and nonlinear evolution do not commute because fine--coarse scale interactions involve unresolved fine-grid modes that are not explicitly represented along the coarse-grid pathway. The correction $\mathbf{c}_\star$ captures this unresolved-scale feedback together with the differences between the fine- and coarse-grid discretizations over the prediction interval. The nonlinear feedback arises from gradient energy in the KSE and vorticity transport in the vorticity NSE. Operator distillation transfers this correction to the deployment-resolution student, whereas direct coarse-grid residual training does not explicitly supervise it.}

\begin{figure}[!t]
  \centering
\includegraphics[pagebox=cropbox,width=\textwidth]{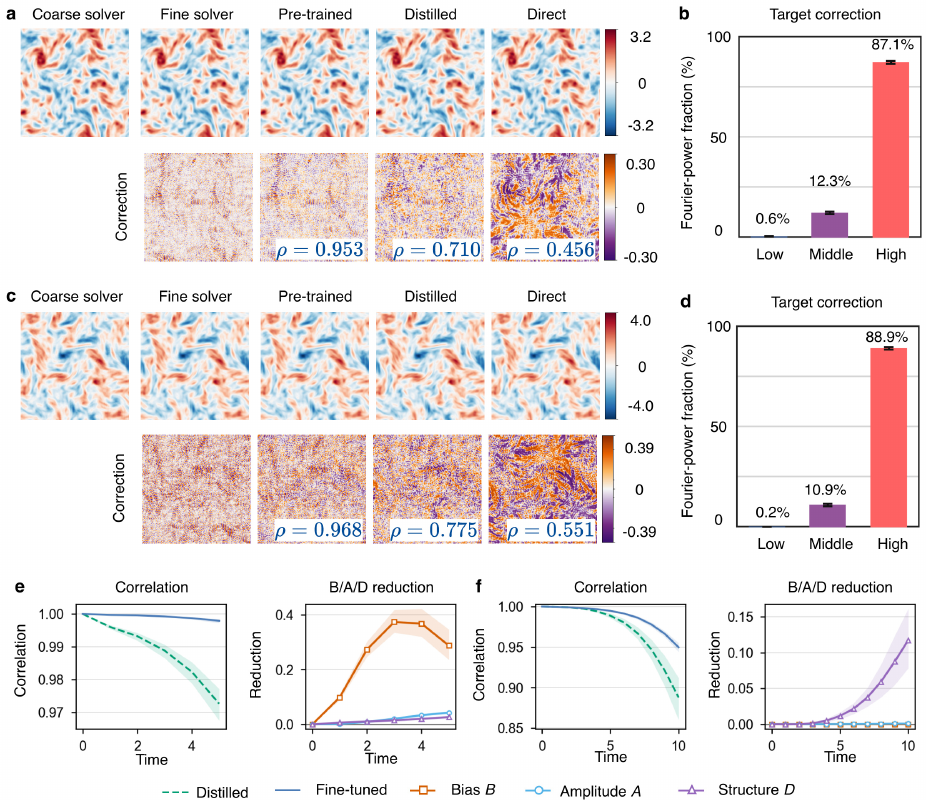}
  \caption{{\textbf{Fine-resolution physics pre-training captures the cross-resolution correction, operator distillation retains a beneficial component of it, and few-shot fine-tuning corrects target-specific errors.}
  \textbf{a} and \textbf{c}, Representative 2D NSE results at $\mathrm{Re}=2000$ and $4000$, respectively. All fields are displayed and compared on a common $128^2$ deployment grid. The Fine solver trajectories originate from DNS generated at $1024^2$ and are projected to $128^2$. The Coarse solver is an independent numerical solve performed directly on the $128^2$ grid. The Pre-trained teacher operates at $256^2$, and its predictions are projected to $128^2$. Distilled operates at $128^2$ and is trained to match these projected teacher outputs. Direct uses the same deployment resolution as Distilled but is trained using deployment-grid physics residuals rather than teacher targets. The first row compares the state fields of Fine solver, Coarse solver, Pre-trained, Distilled and Direct; the second row shows their corrections relative to the Coarse solver. Here, $\rho$ is the spatial Pearson correlation coefficient between the displayed model correction and the Reference correction.
  \textbf{b} and \textbf{d}, Fractions of the Reference correction's Fourier power in the low, middle and high resolved radial bands.
  \textbf{e} and \textbf{f}, Spatial Pearson correlation coefficients and paired reductions in the bias ($B$), amplitude ($A$) and structure ($D$) terms produced by few-shot fine-tuning for OOD KSE ($\alpha=\beta=1$) and 2D NSE ($\mathrm{Re}=4000$), respectively.}}
  \label{SI-fig:projection_correction_mechanism}
\end{figure}

\subsection{{Quantifying the cross-resolution correction transferred by distillation}}

{Physics-only pre-training gives a teacher prediction $\mathbf{u}_{\mathrm{pre}}$ close to the projected fine-grid reference evolution, and distillation trains the student prediction $\mathbf{u}_{\mathrm{dist}}$ to match it:}
\begin{equation}
{
\mathbf{u}_{\mathrm{pre}}\simeq\mathbf{u}_{\mathrm{ref}},
\qquad
\mathbf{u}_{\mathrm{dist}}\simeq\mathbf{u}_{\mathrm{pre}}.
}
\label{SI-eq:correction_transfer_chain}
\end{equation}

{For visualization and quantitative evaluation, we subtract the same coarse-solver output from each model prediction. For a model output $\mathbf{u}_m$, where $m\in\{\mathrm{pre},\mathrm{dist},\mathrm{direct}\}$, the corresponding predicted correction is}
\begin{equation}
{
\mathbf{c}_m
\equiv
\mathbf{u}_m-\mathbf{u}_{\mathrm{coarse}}.
}
\label{SI-eq:model_predicted_correction}
\end{equation}
{We compare the distilled student with Direct, which uses the same deployment resolution but is trained from deployment-grid physics residuals rather than teacher targets. Fig.~\ref{SI-fig:projection_correction_mechanism}\textbf{a--d} compares the predicted states, correction fields and scale distributions.}

{We evaluate correction recovery using the Pearson correlation coefficient $\rho$ and two complementary metrics. The correction-norm ratio is}
\begin{equation}
{
a_C
=
\frac{
\|\mathbf{c}_m\|_2
}{
\|\mathbf{c}_\star\|_2
},
}
\label{SI-eq:correction_amplitude_ratio}
\end{equation}
{which complements it by measuring the overall correction magnitude. Here, $a_C=1$ indicates matched correction norms, whereas $a_C>1$ and $a_C<1$ indicate overestimation and underestimation, respectively.} {To measure the improvement over the uncorrected coarse-grid evolution, we define the coarse-baseline correction skill score}
\begin{equation}
{
S_C
=
1-
\frac{
\|\mathbf{c}_m-\mathbf{c}_\star\|_2^2
}{
\|\mathbf{c}_\star\|_2^2
}
=
1-
\frac{
\|\mathbf{u}_m-\mathbf{u}_{\mathrm{ref}}\|_2^2
}{
\|\mathbf{u}_{\mathrm{coarse}}-\mathbf{u}_{\mathrm{ref}}\|_2^2
}.
}
\label{SI-eq:correction_skill_score}
\end{equation}
{Here, $S_C=1$ denotes exact correction recovery.}

{The results in Table~\ref{SI-tab:correction_recovery} show that the Pre-trained teacher recovers both the structure and magnitude of the target correction, achieving $\rho=0.953$, $a_C=1.038$ and $S_C=0.899$ at $\mathrm{Re}=2000$, and $\rho=0.968$, $a_C=1.041$ and $S_C=0.932$ at $\mathrm{Re}=4000$. The Distilled student retains positive correction skill at both Reynolds numbers, although it overestimates the correction magnitude, particularly at $\mathrm{Re}=2000$. By contrast, Direct substantially overestimates the correction magnitude and yields negative skill scores. These results show that correlation alone is insufficient for evaluating correction recovery and that distillation transfers a correction that improves upon direct coarse-grid evolution.}

\begin{table}[!t]
\centering
\caption{{\textbf{Recovery of the cross-resolution correction on 2D NSE.} The Pearson correlation coefficient $\rho$ measures structural alignment, $a_C$ measures the correction-norm ratio, and $S_C$ measures correction skill relative to the uncorrected coarse-solver state.}}
\label{SI-tab:correction_recovery}
\begin{tabular}{c l c c c}
\toprule
{$\mathrm{Re}$} & {Model} & {$\rho$} & {$a_C$} & {$S_C$} \\
\midrule
{2000} & {Pre-trained} & {0.953} & {1.038} & {0.899} \\
{2000} & {Distilled}   & {0.710} & {1.229} & {0.227} \\
{2000} & {Direct}      & {0.456} & {1.927} & {$-2.238$} \\
\midrule
{4000} & {Pre-trained} & {0.968} & {1.041} & {0.932} \\
{4000} & {Distilled}   & {0.775} & {1.082} & {0.504} \\
{4000} & {Direct}      & {0.551} & {1.557} & {$-0.860$} \\
\bottomrule
\end{tabular}
\end{table}

{The correction is also scale selective. For a radial band $b$, we compute} \begin{equation} { f_b= \frac{\sum_{\boldsymbol{k}\in b}|\widehat{\mathbf{c}}_\star(\boldsymbol{k})|^2} {\sum_{1\leq|\boldsymbol{k}|\leq63}|\widehat{\mathbf{c}}_\star(\boldsymbol{k})|^2}. } \label{SI-eq:correction_fourier_fraction} \end{equation} {As shown in Fig.~\ref{SI-fig:projection_correction_mechanism}\textbf{b} and \textbf{d}, the low ($[1,21]$), middle ($[22,42]$) and high ($[43,63]$) frequency fractions are $0.6\%$, $12.3\%$ and $87.1\%$ at $\mathrm{Re}=2000$, and $0.2\%$, $10.9\%$ and $88.9\%$ at $\mathrm{Re}=4000$. Most of the correction lies in the higher-frequency modes that the deployment grid can still represent. The projected fine-grid reference trajectory therefore differs substantially from the trajectory obtained by direct coarse-grid evolution, particularly in the higher-frequency resolved modes.}

\subsection{{Residual calibration in few-shot adaptation}}

{Distillation already provides a strong zero-shot operator. To identify what the few-shot stage adds, we decompose the remaining prediction error into bias, amplitude, and structure. Let $q=u$ for KSE and $q=\omega$ for 2D NSE, and denote a prediction by $\widehat q$. Here, $\langle\cdot\rangle$ denotes the spatial average over all grid points at a given time. With spatial mean $\mu_q$, standard deviation $\sigma_q$, and spatial Pearson correlation coefficient $\rho$, the squared relative $\ell_2$ error, denoted by $E_{\ell_2}$, admits the exact decomposition}
\begin{equation}
{
E_{\ell_2}
=\frac{\langle(\widehat q-q)^2\rangle}{\langle q^2\rangle}
=
\underbrace{\frac{(\mu_{\widehat q}-\mu_q)^2}{\langle q^2\rangle}}_{B}
+
\underbrace{\frac{(\sigma_{\widehat q}-\sigma_q)^2}{\langle q^2\rangle}}_{A}
+
\underbrace{\frac{2\sigma_{\widehat q}\sigma_q(1-\rho)}{\langle q^2\rangle}}_{D}.
}
\label{SI-eq:bias_amplitude_structure_decomposition}
\end{equation}
{The three terms quantify the spatial-mean bias ($B$), amplitude mismatch ($A$), and structural decorrelation ($D$), respectively. Let $\widehat q_{\mathrm{dist}}(t)$ and $\widehat q_{\mathrm{FT}}(t)$ denote the predictions of the zero-shot distilled student and the few-shot fine-tuned student. For each $Q\in\{B,A,D\}$, we define $Q_{\mathrm{dist}}(t)=Q[\widehat q_{\mathrm{dist}}(t),q(t)]$ and $Q_{\mathrm{FT}}(t)=Q[\widehat q_{\mathrm{FT}}(t),q(t)]$. Fig.~\ref{SI-fig:projection_correction_mechanism}\textbf{e} and \textbf{f} report the paired reduction
$\Delta Q(t)=Q_{\mathrm{dist}}(t)-Q_{\mathrm{FT}}(t)$.
Thus, a positive $\Delta Q(t)$ indicates that few-shot adaptation reduces the corresponding error component relative to the zero-shot distilled operator.}

\noindent{\textbf{KSE: fine-tuning reduces mean drift.} Spatial averaging of the implemented 2D KSE gives}
\begin{equation}
{
\partial_t u
=-\alpha(\Delta u+\Delta^2u)
-\frac{\beta}{2}|\nabla u|^2,
\qquad
\frac{\mathrm d\overline{u}}{\mathrm dt}
=-\frac{\beta}{2}\overline{|\nabla u|^2}.
}
\label{SI-eq:kse_mean_evolution}
\end{equation}
{On the periodic domain, the spatial averages of $\Delta u$ and $\Delta^2u$ vanish. Therefore, the spatial-mean tendency is determined by the nonlinear gradient-energy term. A persistent error in the predicted mean tendency can accumulate over time and produce a systematic mean drift. Because the Pearson correlation coefficient is invariant to a spatially uniform offset, it can remain high even when the predicted mean is inaccurate. Consistent with this mechanism, Fig.~\ref{SI-fig:projection_correction_mechanism}\textbf{e} shows that few-shot fine-tuning mainly reduces the bias component $B$. At $t=5\,\mathrm{s}$, the trajectory-averaged Pearson correlation coefficient increases from $0.973$ to $0.998$. When averaged over all post-initial rollout times, the reduction in $B$ accounts for $88.3\%$ of the total reduction in $E_{\ell_2}$.}

\noindent{\textbf{NSE: fine-tuning reduces structural decorrelation.} The 2D NSE is evolved in vorticity form,}
\begin{equation}
{
\partial_t\omega+\mathbf{u}\cdot\nabla\omega
=\nu\Delta\omega+f,
\qquad
\frac{\mathrm d\overline{\omega}}{\mathrm dt}
=\overline f=0.
}
\label{SI-eq:nse_mean_vorticity}
\end{equation}
{On the periodic domain, incompressibility and zero-mean forcing conserve the spatially averaged vorticity, thereby constraining the bias component $B$. To illustrate how a position error affects structural alignment, consider an idealized prediction that is a slightly displaced version of the centered reference field,
$\widehat{\omega}'(\boldsymbol{x})
=\omega'(\boldsymbol{x}+\boldsymbol\delta)$,
where
$\omega'=\omega-\overline{\omega}$
and $\boldsymbol\delta$ is a small spatial displacement. The Pearson correlation coefficient between the original and displaced fields is}
\begin{equation}
{
\rho(\boldsymbol\delta)
=
\frac{
\left\langle
\omega'(\boldsymbol{x})
\omega'(\boldsymbol{x}+\boldsymbol\delta)
\right\rangle
}{
\left\langle(\omega')^2\right\rangle
},
\qquad
1-\rho(\boldsymbol\delta)
=
\frac{
\left\langle
(\boldsymbol\delta\cdot\nabla\omega')^2
\right\rangle
}{
2\left\langle(\omega')^2\right\rangle
}
+\mathcal{O}(|\boldsymbol\delta|^4).
}
\label{SI-eq:nse_phase_decorrelation}
\end{equation}
{The Taylor expansion shows that a small displacement leaves the spatial mean and variance unchanged but reduces the Pearson correlation coefficient. Such a position or phase mismatch therefore contributes to the structural term $D$, rather than to $B$ or $A$. Consistent with this interpretation, Fig.~\ref{SI-fig:projection_correction_mechanism}\textbf{f} shows that $\Delta D$ grows during the rollout, whereas $\Delta B$ and $\Delta A$ remain near zero. At $t=10\,\mathrm{s}$, the trajectory-averaged Pearson correlation coefficient increases from $0.888$ to $0.950$ after few-shot fine-tuning. When averaged over all post-initial rollout times, the reduction in $D$ accounts for $99.1\%$ of the total reduction in $E_{\ell_2}$.}

{Together, these analyses provide a mechanistic interpretation of the three-stage training strategy in the cases studied here. Fine-resolution physics pre-training provides a teacher approximation to the projected fine-grid flow; operator distillation allows the deployment-resolution student to retain a beneficial but incomplete component of the associated cross-resolution correction; and few-shot fine-tuning reduces the dominant residual error observed for the target task. In the examples considered here, this residual is primarily a spatial-mean drift for KSE and structural decorrelation for 2D NSE.}

\section{Ablation studies}

{In this section, we present a series of ablation studies to analyze the contributions of key components in OmniFluids, investigate the effects of varying the number of fine-tuning trajectories, compare physics-only pre-training with matched supervised PDE pre-training, and examine the accuracy--compute trade-off with data-driven FNO.}

\subsection{{Ablation study on key components of OmniFluids}}
\begin{figure*}[!t]
  \centering
  \includegraphics[width=0.82\textwidth]{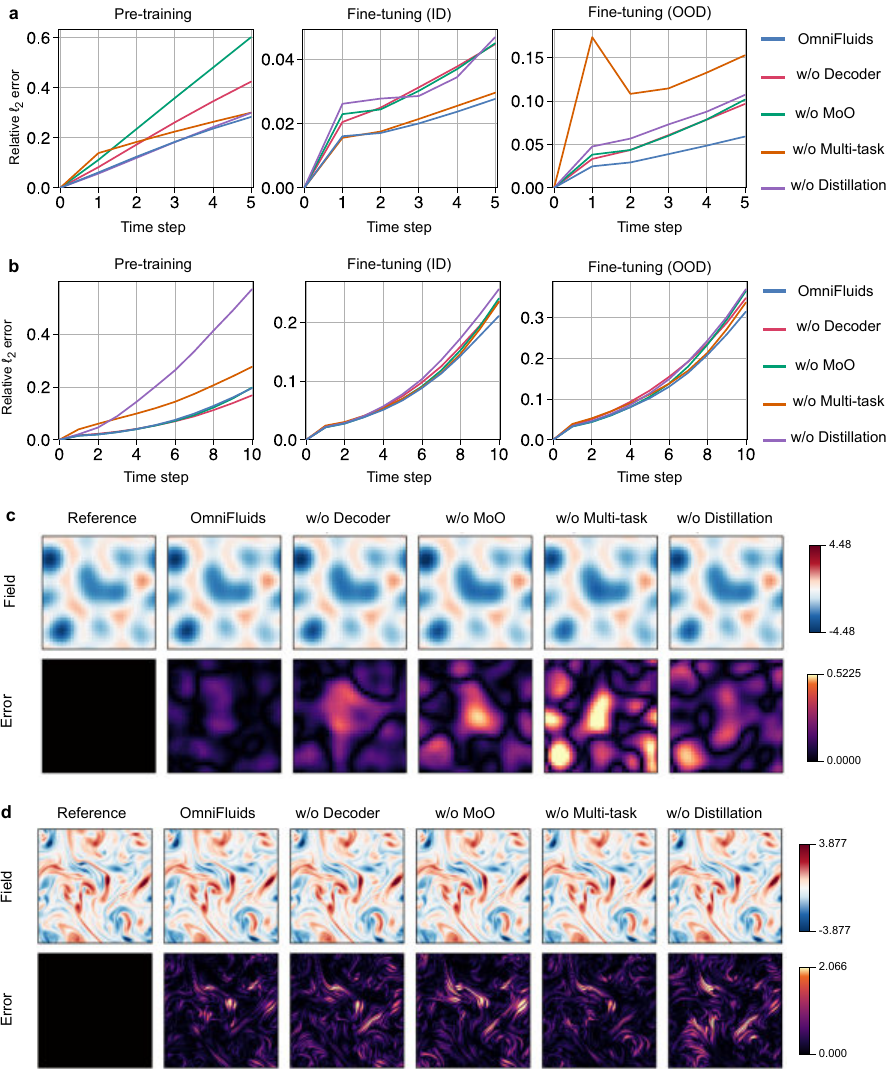}
  \caption{{\textbf{Temporal and qualitative ablations of key OmniFluids components on 2D KSE and 2D NSE.} \textbf{a} and \textbf{b}, Relative $\ell_2$ error over the rollout for 2D KSE (\textbf{a}) and 2D NSE (\textbf{b}) in physics-only pre-training (left), ID fine-tuning (middle), and OOD fine-tuning (right). \textbf{c} and \textbf{d}, Final-step predicted fields (top) and absolute-error maps (bottom) for the OOD KSE (\textbf{c}) and 2D NSE (\textbf{d}) cases.}}
  \label{SI-fig:ablation_components}
\end{figure*}

{We perform component ablations on both 2D KSE and 2D NSE. Specifically, the \textbf{w/o Decoder} variant replaces the multi-frame decoder with a standard MLP; \textbf{w/o MoO} replaces the mixture-of-operators module with standard factorized Fourier layers; \textbf{w/o Multi-task} disables multi-task physics pre-training by using a single physics configuration before transfer; and \textbf{w/o Distillation} removes operator distillation and directly pre-trains the deployment-resolution operator. Fig.~\ref{SI-fig:ablation_components} reports the full rollout-error evolution and final-step field comparisons.}

{The temporal results show that the dominant component depends on both the equation and the training stage. During KSE pre-training, removing MoO or the decoder increases the final-step relative $\ell_2$ error from $0.2831$ to $0.6019$ and $0.4242$, respectively. Under KSE OOD adaptation, disabling multi-task pre-training produces the clearest degradation: the final-step error increases from $0.0590$ for OmniFluids to $0.1525$. For 2D NSE pre-training, operator distillation is the most consequential component, with its removal increasing the final-step error from $0.1955$ to $0.5698$; removing multi-task pre-training also raises it to $0.2761$. Although the no-decoder variant is comparable during the NSE pre-training rollout, the complete model achieves the lowest final-step error after adaptation in both regimes: $0.2108$ under ID and $0.3143$ under OOD, compared with $0.2348$--$0.2563$ and $0.3368$--$0.3692$ for the ablated variants.}

{The field and absolute-error maps (Fig.~\ref{SI-fig:ablation_components}\textbf{c} and \textbf{d}) are consistent with the temporal curves. In the KSE OOD example, removing multi-task pre-training produces the strongest localized error, while the complete model remains closest to the reference. In 2D NSE, the ablated variants exhibit larger filament-wise vorticity discrepancies, whereas the complete model better preserves the reference vortical organization. These results show that the components contribute at different stages rather than through a single uniform mechanism: MoO and the decoder primarily improve representation and rollout quality, multi-task pre-training is especially important for OOD transfer, and distillation is critical for retaining fine-resolution physics at the deployment resolution.}

\subsection{{The effects of varying the amount of fine-tuning data}}

\begin{figure}[!t]
  \centering
  \includegraphics[width=\columnwidth]{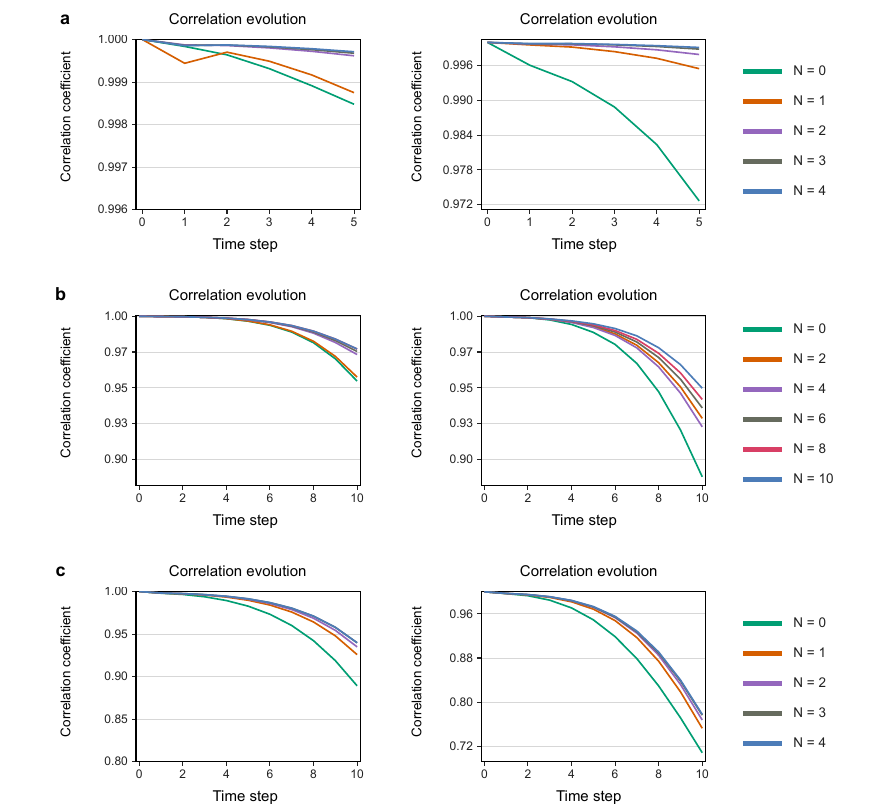}
  \caption{{{\textbf{Correlation evolution versus the number of fine-tuning trajectories.}
  \textbf{a--c}, Correlation coefficient over rollout time for 2D KSE (\textbf{a}), 2D NSE (\textbf{b}), and 3D NSE (\textbf{c}) under ID (left) and OOD (right) conditions.}}}
 \label{SI-fig:data_efficiency}
\end{figure}

{Fig.~\ref{SI-fig:data_efficiency} shows how the long-horizon predictive accuracy of OmniFluids changes with the amount of fine-tuning data. The zero-shot student ($N=0$), which relies only on physics-only pre-training and operator distillation, already captures the early-time dynamics, but its correlation decays more rapidly over the rollout, particularly under OOD conditions. Adding only a few training trajectories substantially slows this degradation, while the differences between successive larger values of $N$ become progressively smaller. For 2D KSE, the correlation remains close to unity after few-shot fine-tuning; for 2D and 3D NSE, increasing $N$ yields a clearer improvement over the longer rollout horizon, with the OOD cases showing greater sensitivity to the amount of adaptation data than the ID cases. These results directly demonstrate that most of the benefit is obtained with a small number of trajectories and that further data provide diminishing improvements.}

\begin{figure*}[!t]    \centering    \includegraphics[width=0.90\textwidth]{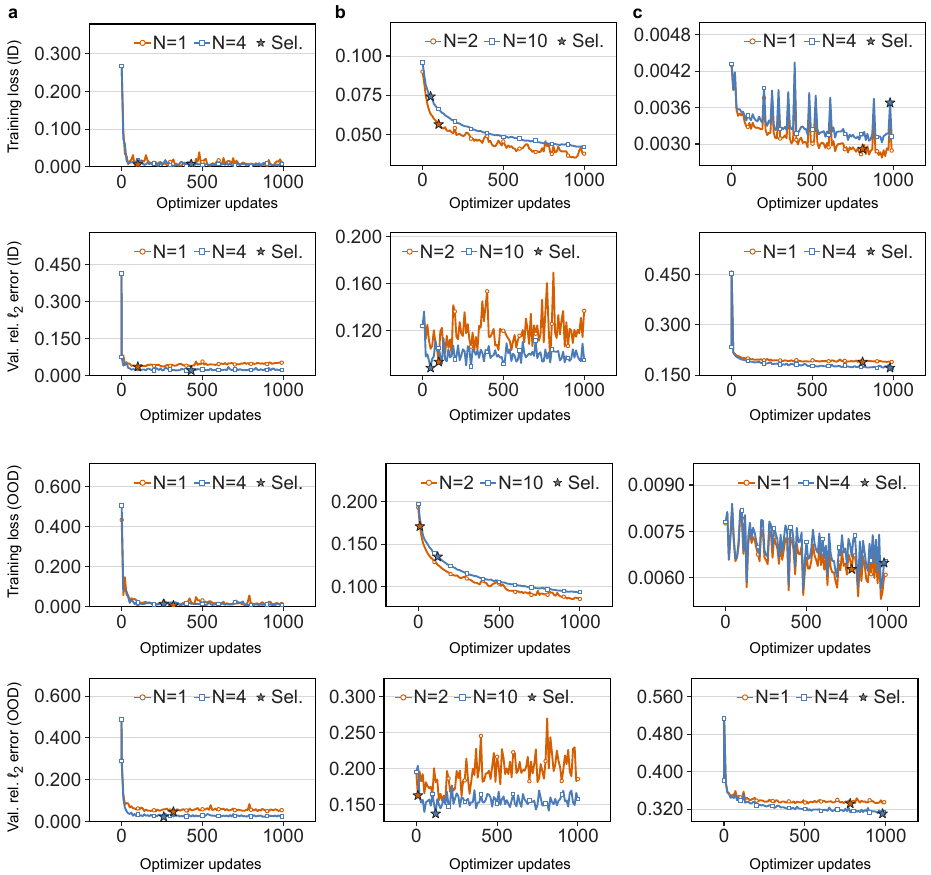}    \caption{{\textbf{Fine-tuning dynamics.} \textbf{a--c}, Results for 2D KSE (\textbf{a}), 2D NSE (\textbf{b}), and 3D NSE (\textbf{c}). From top to bottom, the rows show ID training loss, ID validation relative $\ell_2$ error, OOD training loss, and OOD validation relative $\ell_2$ error.}}    \label{SI-fig:finetuning_iterations}\end{figure*}

{Fig.~\ref{SI-fig:finetuning_iterations} shows the optimization dynamics during the fine-tuning stages. As the number of optimizer updates increases, the validation error first decreases rapidly and then stabilizes or increases, showing a transition from insufficient adaptation to optimal generalization and eventually overfitting. For 2D NSE, the lowest validation errors occur at updates 100 and 50 for \(N=2\) and \(N=10\) under ID conditions, and at updates 10 and 120 under OOD conditions. Continuing to 1000 updates does not further improve validation performance. The optimal points generally occur later for 2D KSE and 3D NSE. These results show that the appropriate training budget depends on both the equation and the amount of training data and should therefore be determined using validation performance.}

\begin{figure}[!t]
  \centering
  \includegraphics[width=12cm]{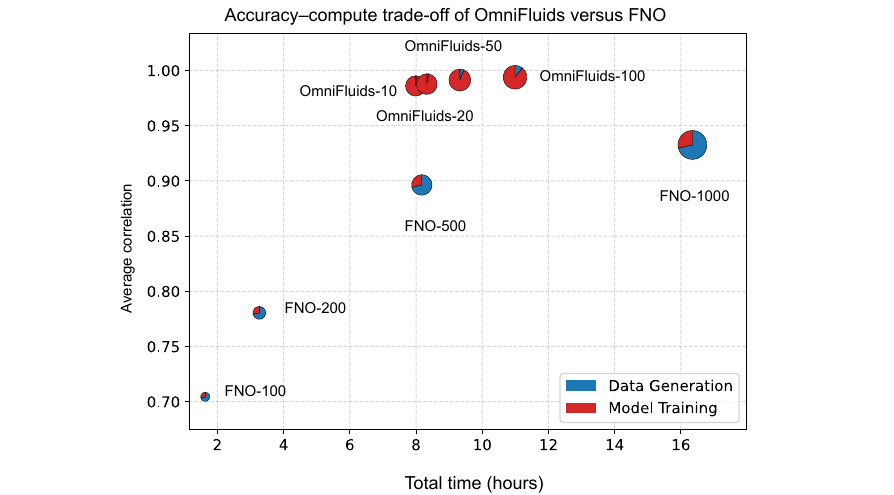}
  \caption{{\textbf{Accuracy--compute trade-off of OmniFluids versus data-driven FNO on 2D NSE.}
  Each disk corresponds to either an FNO trained from scratch with a given number of trajectories (FNO-100/200/500/1000) or an OmniFluids fine-tuned with 10--100 trajectories after physics-only pre-training. The vertical axis reports the average correlation with the reference, while the horizontal axis measures the total time, including both data generation and training. The area of each disk is proportional to the total time, and the blue and red sectors denote the fractions spent on data generation and model training, respectively. The approximate (data generation, model training) times in hours are: FNO-100/200/500/1000: (1.17, 0.47), (2.34, 0.93), (5.85, 2.33), (11.69, 4.65); and OmniFluids-10/20/50/100: (0.12, 7.87), (0.24, 8.09), (0.58, 8.74), (1.17, 9.82), respectively.}}
  \label{SI-fig:appfig1}
\end{figure}

\subsection{{Accuracy--compute trade-off with data-driven FNO}}

{In this part, we study the accuracy--compute trade-off between OmniFluids and a purely data-driven FNO on the 2D NSE case ($\mathrm{Re}=4000$). As shown in Fig.~\ref{SI-fig:appfig1}, each disk corresponds to either an FNO model trained from scratch with a given number of trajectories (FNO-100/200/500/1000) or an OmniFluids model fine-tuned with a small number of trajectories after physics-only pre-training (OmniFluids-10/20/50/100). The vertical axis reports the average correlation with the reference, while the horizontal axis measures the total time, including data generation and model training; the disk area is proportional to this total time. 

Across the entire range of compute budgets, OmniFluids achieves a markedly more favorable trade-off. With only 10--20 fine-tuning trajectories, OmniFluids already matches or surpasses the accuracy of FNO models trained on several hundred trajectories, while incurring substantially less total computational time. Increasing the number of fine-tuning trajectories further improves accuracy with only moderate additional cost, whereas the data-driven FNO requires much larger time budgets to approach comparable correlation levels. Moreover, it is important to note that each FNO is trained for a single Reynolds number and requires retraining from scratch when the target Reynolds number changes. By contrast, the physics-pretrained OmniFluids operator can be rapidly adapted to multiple flow regimes using only a few high-fidelity trajectories, further demonstrating its effective data and compute advantages over classical data-driven neural operators.}

{To further contextualize the comparison under a standardized benchmark protocol, we additionally compare OmniFluids with the 2D and 3D FNO results reported in the original FNO paper~\cite{SI-li2021fourier} on the same Navier--Stokes settings. The FNO relative $\ell_2$ errors are kept as reported in the original benchmark, while all data-generation and training times are re-measured on our hardware under a unified timing protocol. For OmniFluids, the reported training time includes the physics-only pre-training stage and the subsequent fine-tuning/adaptation stage for the target regime. Since this benchmark is relatively simple and already defined at a coarse resolution, we omit the operator-distillation stage and conduct both physics-only pre-training and final prediction on a
$64\times64$ spatial grid with a single-step temporal output. As summarized in Table~\ref{SI-tab:fno_original_benchmark}, OmniFluids uses one to two orders of magnitude fewer labeled trajectories and substantially less total time than data-driven FNO baselines in the lower-viscosity regimes, while achieving comparable or better accuracy.}

\begin{table*}[!t]
\centering
\small
\setlength{\tabcolsep}{7pt}
\renewcommand{\arraystretch}{1.12}
\caption{
{\textbf{Comparison with the original FNO~\cite{SI-li2021fourier} Navier--Stokes benchmark.}
The FNO relative $\ell_2$ errors are taken from the original FNO benchmark, while the
data-generation and training times are re-measured on our hardware under the same timing
protocol. For OmniFluids, the reported training time includes physics-only pre-training and subsequent fine-tuning for the target regime. ``\# Labels'' denotes the number of labeled training trajectories.
}}
\label{SI-tab:fno_original_benchmark}
{\begin{tabular}{lccccc}
\toprule
Method & \# Labels & Rel. $\ell_2$ & Data gen. (h) & Training (h) & Total (h) \\
\midrule
\multicolumn{6}{l}{\textit{$\nu=10^{-3}$, $T=50$}} \\
OmniFluids-FT (10)   & 10    & 0.0172 & 0.089  & 0.735 & \textbf{0.824} \\
OmniFluids-FT (100)  & 100   & 0.0183 & 0.257  & 0.885 & 1.142 \\
FNO-3D-1000         & 1000  & \textbf{0.0086} & 2.216  & 1.162 & 3.377 \\
FNO-2D-1000         & 1000  & 0.0128 & 2.216  & 2.392 & 4.607 \\
\midrule
\multicolumn{6}{l}{\textit{$\nu=10^{-4}$, $T=30$}} \\
OmniFluids-FT (10)    & 10     & 0.1016 & 0.053  & 0.735 & \textbf{0.788} \\
OmniFluids-FT (100)   & 100    & \textbf{0.0791} & 0.154  & 0.885 & 1.039 \\
FNO-3D-1000          & 1000   & 0.1918 & 1.331  & 0.653 & 1.984 \\
FNO-3D-10000         & 10000  & 0.0820 & 13.314 & 2.610 & 15.924 \\
FNO-2D-1000          & 1000   & 0.1559 & 1.331  & 1.369 & 2.701 \\
FNO-2D-10000         & 10000  & 0.0834 & 13.314 & 5.478 & 18.792 \\
\midrule
\multicolumn{6}{l}{\textit{$\nu=10^{-5}$, $T=20$}} \\
OmniFluids-FT (10)   & 10    & \textbf{0.0806} & 0.036 & 0.735 & \textbf{0.771} \\
OmniFluids-FT (100)  & 100   & 0.0808 & 0.103 & 0.885 & 0.988 \\
FNO-3D-1000         & 1000  & 0.1893 & 0.861 & 0.431 & 1.292 \\
FNO-2D-1000         & 1000  & 0.1556 & 0.861 & 0.699 & 1.561 \\
\bottomrule
\end{tabular}}
\end{table*}

\FloatBarrier
\subsection{{Physics-only versus data-based PDE pre-training}}
\label{SI-sec:physics_vs_data_pretraining}

\begin{figure*}[!t]
  \centering
  \includegraphics[width=0.95\textwidth]{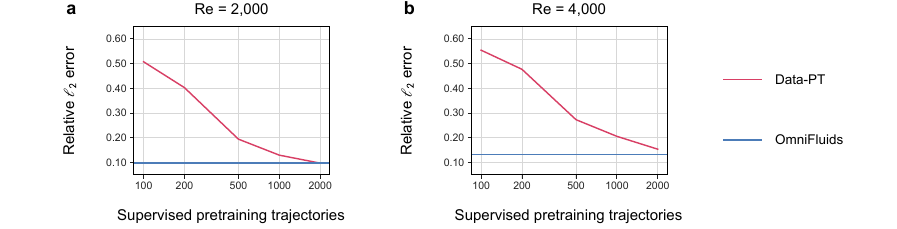}
  \caption{{\textbf{Physics-only versus supervised (Data-PT) pre-training on 2D NSE.} \textbf{a},\textbf{b}, Relative $\ell_2$ error after fine-tuning at $\mathrm{Re}=2000$ (\textbf{a}) and $\mathrm{Re}=4000$ (\textbf{b}) as a function of the number of labeled trajectories $N$ used for Data-PT pre-training. OmniFluids (horizontal line) uses no labeled pre-training data. Both models are fine-tuned on the same ten target trajectories for 1,000 updates, and errors are evaluated on ten held-out test trajectories at validation-selected checkpoints.}}
  \label{SI-fig:physics_vs_data_pretraining}
\end{figure*}

\begin{figure}[!t]
  \centering
  \includegraphics[width=\textwidth]{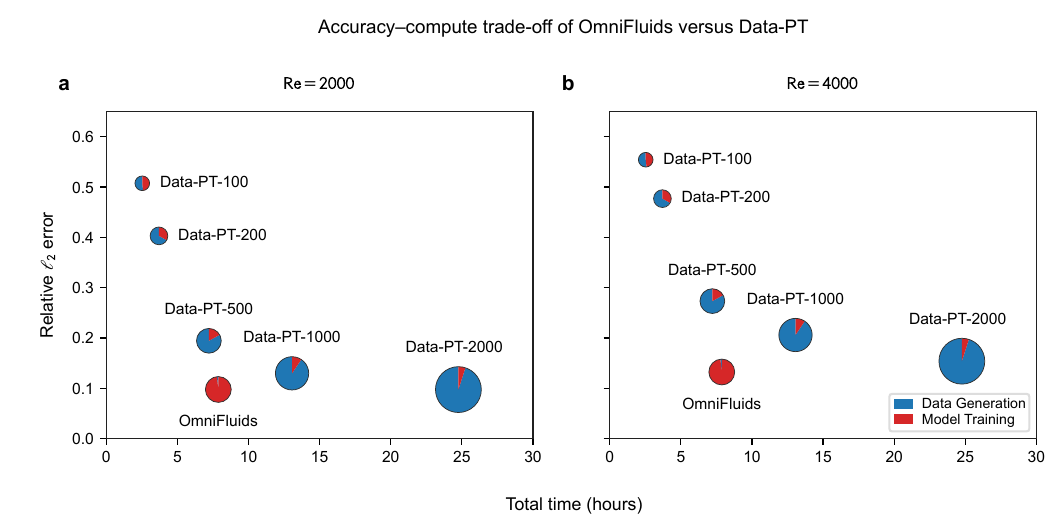}
  \caption{\textbf{Accuracy--cost trade-off of physics-only and supervised pre-training on 2D NSE.} \textbf{a},\textbf{b}, Relative $\ell_2$ error versus total computational time for initialization and adaptation to one target at $\mathrm{Re}=2000$ (\textbf{a}) and $\mathrm{Re}=4000$ (\textbf{b}). Errors are averaged over ten forecast times and ten held-out test trajectories, using validation-selected checkpoints from 1,000-update fine-tuning runs. Disk area is proportional to the total time; blue and red sectors denote data generation and model training, respectively. Stage costs (in hours) are given as (pre-training data generation, pre-training, distillation; target-data generation, fine-tuning); the first three are one-time costs and the last two are incurred for each target. OmniFluids: $(0, 5.73, 1.93;\,0.14, 0.08)$. Data-PT with $N=100$, 200, 500, 1,000 and 2,000: $(1.17, 1.14, 0;\,0.14, 0.08)$, $(2.34, 1.14, 0;\,0.14, 0.08)$, $(5.85, 1.14, 0;\,0.14, 0.08)$, $(11.69, 1.14, 0;\,0.14, 0.08)$ and $(23.39, 1.15, 0;\,0.14, 0.08)$, respectively.}
  \label{SI-fig:physics_data_cost_tradeoff}
\end{figure}

To isolate the contribution of the pre-training signal, we compare OmniFluids with Data-PT, a supervised pre-training baseline that uses the same student architecture and $128\times128$ grid on the 2D NSE benchmark. Data-PT is pre-trained on $N\in\{100,200,500,1000,2000\}$ labeled trajectories with $\mathrm{Re}\in[500,2500]$ and varying forcing, matching the parameter range of OmniFluids physics-only pre-training. Pre-training runs for 20,000 Adam updates with a batch size of 10, no weight decay and a OneCycle learning-rate schedule with a peak of $0.002$. Checkpoints are selected on 50 held-out validation trajectories evaluated every 1,000 updates. For $N=2000$, the validation error changes by less than 2\% over the final 2,000 updates (from $0.01740$ to $0.01707$), indicating convergence within this budget. Both pre-trained models are then fine-tuned at $\mathrm{Re}=2000$ and $4000$ under an identical protocol: the same ten training trajectories, 1,000 Adam updates at a learning rate of $5\times10^{-5}$, and checkpoint selection by the rollout-mean relative $\ell_2$ error on two validation trajectories. Errors are reported on ten held-out test trajectories.

Data-PT improves steadily as $N$ increases (Fig.~\ref{SI-fig:physics_vs_data_pretraining}). With $N=2000$, Data-PT and OmniFluids reach similar errors at $\mathrm{Re}=2000$ ($0.0972$ and $0.0974$, respectively), whereas OmniFluids attains a lower error at $\mathrm{Re}=4000$ ($0.1322$ versus $0.1537$). Fig.~\ref{SI-fig:physics_data_cost_tradeoff} relates these errors to the end-to-end computational cost of each approach. Because both pre-trained models can be reused across target regimes, we report one-time initialization costs (labeled-data generation, pre-training and, for OmniFluids, distillation) separately from per-target adaptation costs (target-data generation and fine-tuning). OmniFluids requires 7.66\,h of one-time initialization (5.73\,h of physics-only pre-training and 1.93\,h of distillation) and no labeled pre-training data. By contrast, the one-time cost of Data-PT is dominated by data generation and increases from 2.31\,h at $N=100$ to 24.54\,h at $N=2000$. Adaptation to each target costs 0.22\,h for both methods.

\section{Additional results}

In this section, we conduct additional experiments to further characterize the behavior and generalization capability of OmniFluids. We first examine the expert activation patterns in the mixture-of-operators (MoO) module to verify that different experts specialize in distinct flow regimes rather than collapsing onto a small subset. We then study the robustness of OmniFluids under more turbulent flows with higher Reynolds numbers and under Kolmogorov forcing. {To further evaluate its performance on inverse-type settings, we also add a sparse reconstruction experiment from only $5\%$ observations and compare OmniFluids with a classical truncated Fourier least-squares baseline with ridge regularization.} {For the experimental cylinder-wake PIV measurements, whose field of view excludes the cylinder surface, we further assess localized reconstruction quality within the measured wake by evaluating the near-wake region and high-vorticity regions associated with separated shear layers and vortex cores.} Finally, we report extended qualitative and quantitative comparisons on both 2D and 3D benchmarks.

\subsection{{Visualization of expert activation in the MoO module}}

\begin{figure*}[!t]
  \centering
  \includegraphics[width=\textwidth]{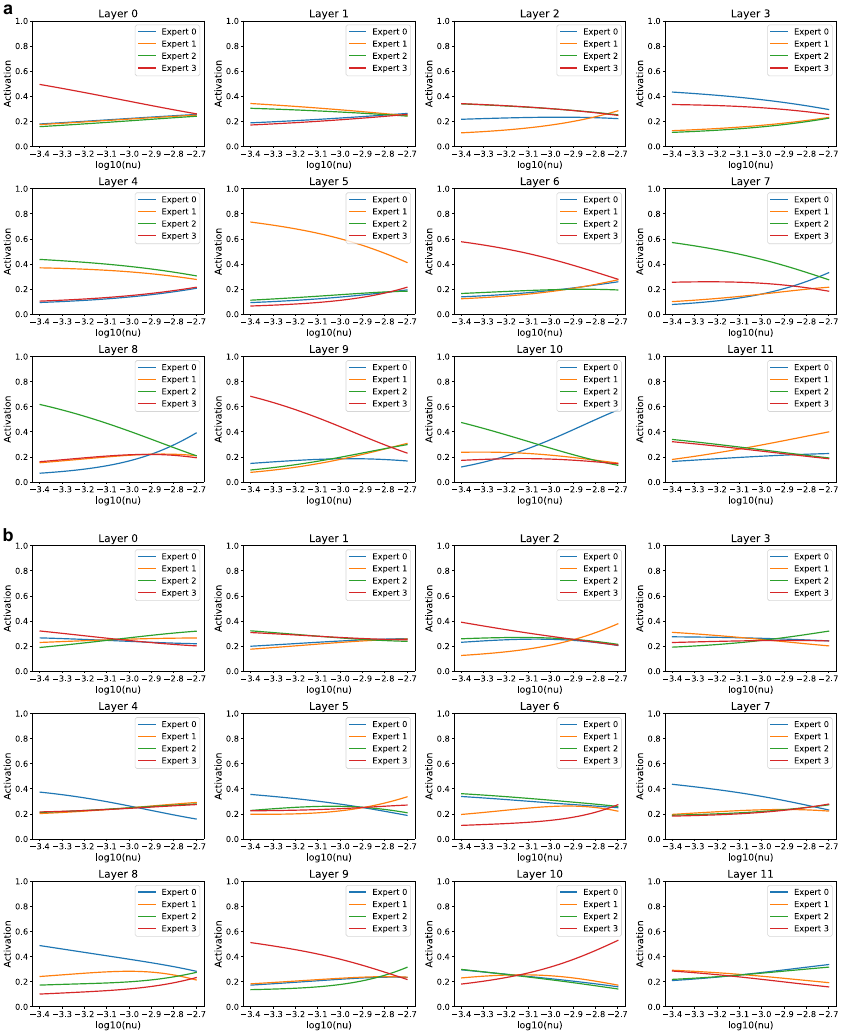}
  \caption{{\textbf{Expert activation patterns in the mixture-of-operators (MoO) module.} For each layer, we plot the routing weights assigned to four experts as a function of $\log_{10}(\nu)$ in the 2D NSE task, where $\nu = 1/\mathrm{Re}$. \textbf{a}, Teacher operator. \textbf{b}, Distilled student operator.}}
  \label{SI-fig:appfig2}
\end{figure*}

{To investigate whether the mixture-of-operators (MoO) router collapses onto a small subset of experts, we explicitly analyze expert usage in the 2D NSE setting. Recall that the router employs a temperature-scaled softmax over expert logits to ensure expert utilization. Specifically, we visualize the routing weights assigned to each expert as a function of the viscosity parameter in the 2D NSE task. Fig.~\ref{SI-fig:appfig2} shows these weights for four experts across all 12 layers of the teacher (Fig.~\ref{SI-fig:appfig2}\textbf{a}) and distilled student (Fig.~\ref{SI-fig:appfig2}\textbf{b}). In both the teacher and student networks, we observe that across all layers, each expert receives a non-degenerate routing weight over various viscosities, indicating that no expert remains inactive. At the same time, the routing patterns reveal clear heterogeneity among experts: some are preferentially selected in low-viscosity regimes, whereas others are predominantly used in high-viscosity regimes, consistent with our design intuition that different experts specialize in different flow regimes.
}

\subsection{{Kolmogorov-forced 2D NSE}}

\begin{figure*}[!t]
  \centering
  \includegraphics[width=\textwidth]{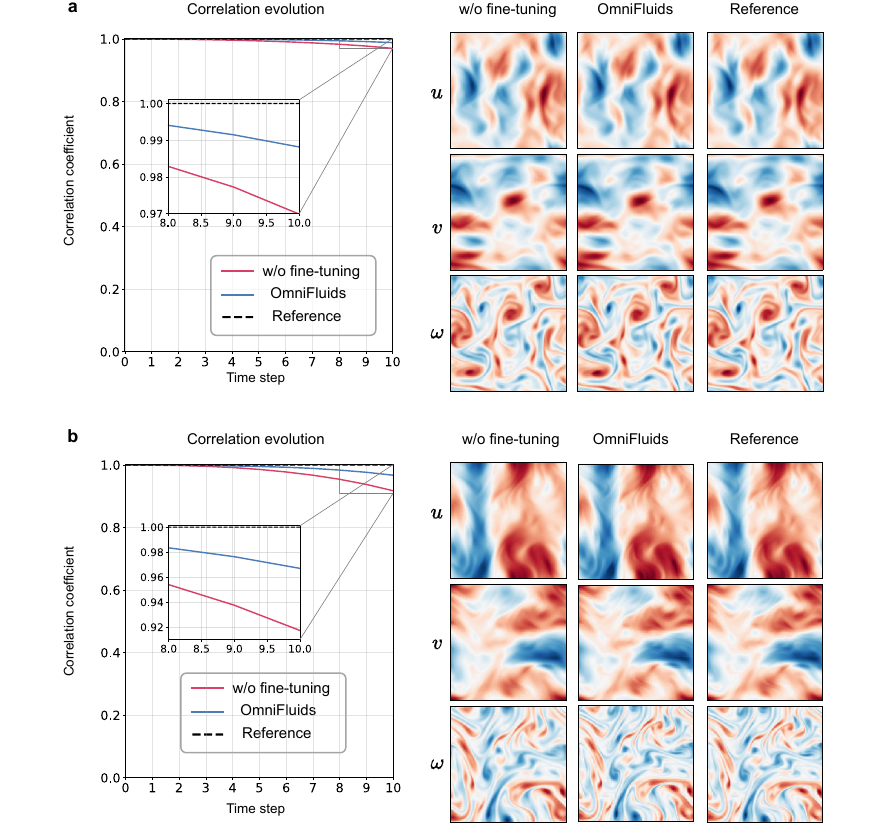}
  \caption{{\textbf{Kolmogorov-forced 2D NSE.} \textbf{a,b}, Correlation evolution (left) and snapshots of velocity $\mathbf{u} = [u, v]$ and vorticity $\omega$ (right) for Kolmogorov forcing with $\mathrm{Re}=2000$ (\textbf{a}) and $4000$ (\textbf{b}). In each case, we compare the distilled student without fine-tuning (`w/o fine-tuning'), the fine-tuned OmniFluids, and the reference.}}
  \label{SI-fig:appfig4}
\end{figure*}

{Beyond the randomly forced 2D NSE considered in the main experiments, we also study flows driven by the classical Kolmogorov forcing~\cite{SI-platt1991investigation,SI-wang2025simulating}. In this setting, the flow is driven by a steady body force applied in the $x$-direction of the form ${f}(x,y) = 0.1 \sin(8y)$. It is worth mentioning that this Kolmogorov forcing never occurs during the physics-only pre-training stage, and thus constitutes an out-of-distribution forcing pattern for OmniFluids.

Fig.~\ref{SI-fig:appfig4} compares the distilled student without any supervised adaptation (`w/o fine-tuning') against the fine-tuned OmniFluids and the reference DNS under two cases with $\mathrm{Re}=2000$ and $4000$, respectively. In both cases, the student model already delivers reasonable zero-shot predictions, but its correlation with DNS gradually decays over long rollouts. After fine-tuning on 10 trajectories, OmniFluids maintains correlations that remain above $0.96$ over the entire horizon and recovers sharper vortical structures that closely match the reference.}

\begin{figure*}[!t]
  \centering
  \includegraphics[width=\textwidth]{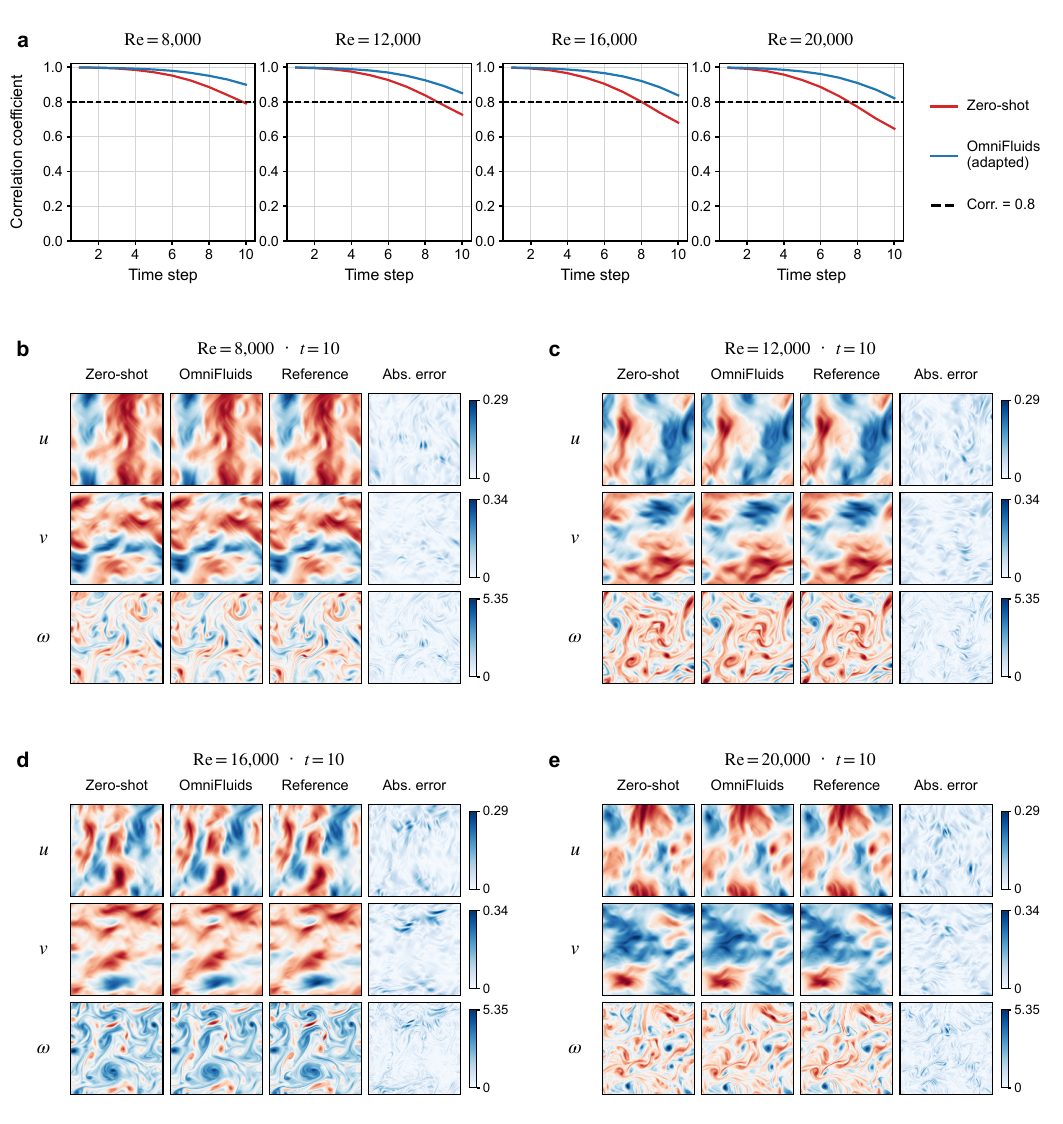}
  \caption{{\textbf{Generalization of OmniFluids to higher-Reynolds-number 2D NSE.} \textbf{a}, Correlation coefficient between the predicted and reference fields over the rollout for the distilled student without fine-tuning (zero-shot; red) and after adaptation with ten target-regime trajectories (blue) at $\mathrm{Re}\in\{8000,12000,16000,20000\}$, averaged over ten held-out test trajectories. \textbf{b}--\textbf{e}, Snapshots of the velocity $\mathbf{u}=[u,v]$ and vorticity $\omega$ at $t=10$ for $\mathrm{Re}=8000$ (\textbf{b}), 12000 (\textbf{c}), 16000 (\textbf{d}) and 20000 (\textbf{e}). Columns show the zero-shot prediction, the adapted prediction (OmniFluids) and the reference, together with the absolute error of the adapted model, with a shared color scale for each variable.}}
  \label{SI-fig:appfig3}
\end{figure*}

\begin{figure*}[!t]
  \centering
  \includegraphics[width=\textwidth]{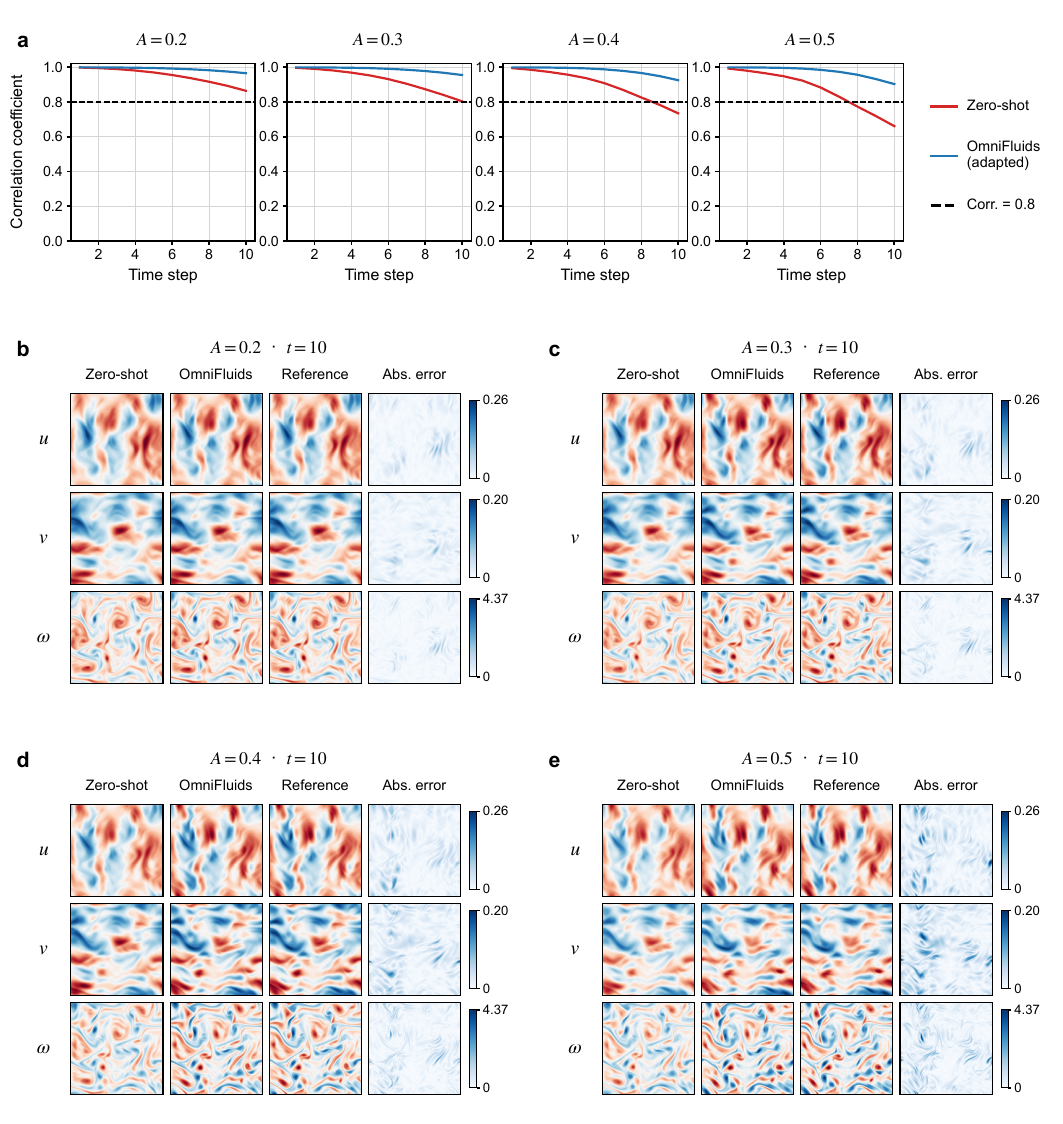}
  \caption{{\textbf{Generalization of OmniFluids to stronger Kolmogorov forcing.} \textbf{a}, Correlation coefficient between the predicted and reference fields over the rollout for the distilled student without fine-tuning (zero-shot; red) and after adaptation with ten target-regime trajectories (blue) at $\mathrm{Re}=4000$ and $k_f=8$ with forcing amplitudes $A\in\{0.2,0.3,0.4,0.5\}$, averaged over ten held-out test trajectories. \textbf{b}--\textbf{e}, Snapshots of the velocity $\mathbf{u}=[u,v]$ and vorticity $\omega$ at $t=10$ for $A=0.2$ (\textbf{b}), 0.3 (\textbf{c}), 0.4 (\textbf{d}) and 0.5 (\textbf{e}). Columns show the zero-shot prediction, the adapted prediction (OmniFluids) and the reference, together with the absolute error of the adapted model, with a shared color scale for each variable.}}
  \label{SI-fig:appfigood}
\end{figure*}

\subsection{{Generalization to out-of-distribution regimes}}

The main-text 2D NSE experiments extend up to $\mathrm{Re}=4000$. To probe behavior beyond this range, we consider two complementary out-of-distribution (OOD) shifts: higher Reynolds numbers and stronger Kolmogorov forcing. For each shift, we distinguish zero-shot prediction by the distilled student, which uses no target-regime data, from adaptation, in which the same student is fine-tuned on ten target-regime trajectories using the adaptation protocol of the main 2D NSE experiments. Correlations and errors are averaged over ten held-out test trajectories.

\textbf{Higher Reynolds numbers.} We evaluate both settings at $\mathrm{Re}\in\{8000,12000,16000,20000\}$. At every Reynolds number, the adapted model maintains a higher correlation with the reference than the zero-shot prediction throughout the rollout (Fig.~\ref{SI-fig:appfig3}\textbf{a}). At $\mathrm{Re}=8000$, the adapted model retains a final-step correlation close to $0.9$, whereas the zero-shot correlation decays to approximately $0.8$. The snapshots in Fig.~\ref{SI-fig:appfig3}\textbf{b}--\textbf{e} show that zero-shot predictions capture the large-scale flow organization but exhibit distorted vortex cores and either over-smoothed or spuriously noisy small-scale features, particularly at higher Reynolds numbers. Adaptation improves agreement with the reference in both the large-scale coherent structures and the fine-scale vortical filaments. The absolute-error fields show that residual errors remain after adaptation and grow with the Reynolds number; at $\mathrm{Re}=20000$, local amplitudes and the smallest eddies deviate from the reference, although the positions and interactions of the dominant vortices are retained.

\textbf{Stronger forcing.} We parameterize the Kolmogorov forcing as $f(x,y)=A\sin(k_f y)$, fix $\mathrm{Re}=4000$ and $k_f=8$, and increase the amplitude from $A=0.1$, as considered above, to $A\in\{0.2,0.3,0.4,0.5\}$, yielding a controlled sweep of increasingly severe shifts. The final-step zero-shot correlation decreases monotonically from $0.86$ at $A=0.2$ to $0.66$ at $A=0.5$ (Fig.~\ref{SI-fig:appfigood}\textbf{a}), and zero-shot predictions develop increasingly visible phase displacements and distorted vortical structures (Fig.~\ref{SI-fig:appfigood}\textbf{b}--\textbf{e}). Adaptation improves agreement with the reference and better preserves both the large-scale flow organization and the fine-scale vortical filaments, although its errors also increase with the forcing amplitude.

\textbf{Summary and scope.} Table~\ref{SI-tab:ood_relative_error_summary} summarizes the rollout-mean relative $\ell_2$ errors. For both shifts, zero-shot errors increase with the magnitude of the shift, from 0.30 to 0.44 across Reynolds numbers and from 0.27 to 0.44 across forcing amplitudes. Adaptation with ten target-regime trajectories reduces these errors to 0.21--0.28 and 0.12--0.18, respectively, although the adapted errors also increase with the shift. These tests do not identify a clear failure boundary for the adapted model, and our conclusions are therefore limited to the tested Reynolds numbers, forcing amplitudes and prediction horizon ($T=10$).

\begin{table}[!t]
\centering
\begin{minipage}{0.83\linewidth}
\caption{\textbf{Relative $\ell_2$ errors under OOD shifts.} Rollout-mean relative $\ell_2$ errors of vorticity, averaged over the ten non-initial forecast times and ten held-out test trajectories. Zero-shot results use the distilled student without target-regime data; adapted results use ten target-regime training trajectories.}
\label{SI-tab:ood_relative_error_summary}
\centering
\small
\setlength{\tabcolsep}{5pt}
\renewcommand{\arraystretch}{1.08}
\begin{tabular}{rrr@{\hspace{16pt}}rrr}
\toprule
\multicolumn{3}{c}{Higher Reynolds numbers} & \multicolumn{3}{c}{Stronger forcing}\\
\cmidrule(lr){1-3}\cmidrule(lr){4-6}
$\mathrm{Re}$ & Zero-shot & Adapted & $A$ & Zero-shot & Adapted\\
\midrule
8000 & 0.2967 & 0.2069 & 0.2 & 0.2718 & 0.1227 \\
12000 & 0.3588 & 0.2508 & 0.3 & 0.3365 & 0.1362 \\
16000 & 0.4028 & 0.2620 & 0.4 & 0.3926 & 0.1632 \\
20000 & 0.4363 & 0.2772 & 0.5 & 0.4427 & 0.1819 \\
\bottomrule
\end{tabular}
\end{minipage}
\end{table}

\subsection{{Lid-driven cavity flow}}

\begin{figure*}[!t]
  \centering
  \includegraphics[width=\textwidth]{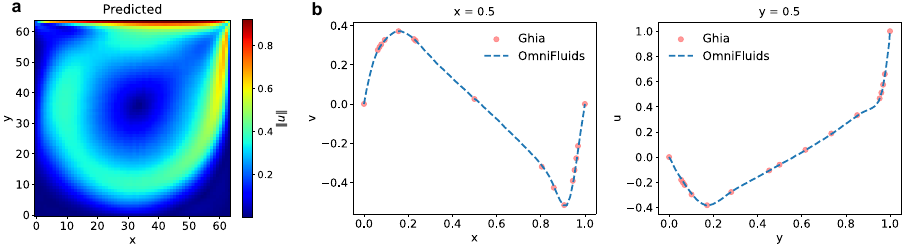}
  \caption{{\textbf{Lid-driven cavity flow at $\mathrm{Re} = 1000$.} \textbf{a}, Predicted velocity-magnitude field $\|\mathbf{u}\|_2$ for OmniFluids. \textbf{b}, Centerline velocity profiles along the vertical line $x = 0.5$ and horizontal line $y = 0.5$, comparing OmniFluids (dashed line) with the reference data of Ghia~\emph{et al.}~\cite{SI-ghia1982high}.}}
  \label{SI-fig:appfig5}
\end{figure*}

{Besides periodic flows, we consider a non-periodic benchmark to test the generalization of OmniFluids. In the lid-driven cavity case, we study the 2D incompressible NSEs on the unit square $\Omega = [0,1]^2$~\cite{SI-ghia1982high,SI-wang2024piratenets}. The flow is driven by a moving top wall with horizontal velocity $u(x,1)=1.0$ and $v(x,1)=0$. The other three walls satisfy no-slip conditions $(u,v)=(0,0)$. For each trajectory, we construct an initial condition by sampling Gaussian random fields and projecting the result onto a divergence-free field that is consistent with the lid boundary conditions. We then advance this field using a high-order MAC solver for a short warm-up period of about $0.5$~s. After warm-up, we integrate up to $T=40$ with time step $\Delta t = 10^{-3}$ and spatial size $256\times256$. In the physics-only pre-training stage, OmniFluids is trained with Reynolds numbers sampled from $\mathrm{Re}\in[500,2500]$, using a spatial discretization of $128\times128$ grid points. The physics-based loss is constructed over the temporal interval $t\in[0,1.0]$, divided into $50$ uniform time bins. We then perform operator distillation onto a student network defined on a coarser $64\times64$ grid with a temporal stride of $\Delta t=5.0$. After distillation, we fine-tune it only on cavity data at $\mathrm{Re}=1000$, using only two trajectories. Fig.~\ref{SI-fig:appfig5}\textbf{a} shows the resulting velocity-magnitude field $\|\mathbf{u}\|_2$ at steady state, which exhibits the characteristic primary recirculation cell and secondary corner vortices. In Fig.~\ref{SI-fig:appfig5}\textbf{b}, we compare the velocity profiles along the vertical centerline ($x=0.5$) and horizontal centerline ($y=0.5$) against the classical reference data~\cite{SI-ghia1982high}. OmniFluids closely matches the benchmark centerline curves despite being adapted from only two trajectories and from a physics-pretrained model, illustrating the ability to handle boundary-driven flows.}

\subsection{{Sparse reconstruction}}

{We further evaluate OmniFluids on sparse flow reconstruction and compare it with a classical spectral reconstruction baseline. Following the sparse-observation setting in the main text, we use a 2D NSE trajectory on a $128\times128$ grid and observe only a small subset of grid points at each time step. The reconstruction is evaluated over $11$ frames from $t=0$ to $t=10$.}

{As a classical baseline, we implement a truncated Fourier least-squares method with ridge regularization, denoted as FLS. For each snapshot, the reconstructed vorticity field is represented by a truncated two-dimensional Fourier expansion
\begin{equation}
    \hat{\omega}(x,y)
    =
    \sum_{k_x=-K}^{K}
    \sum_{k_y=-K}^{K}
    c_{k_x,k_y}
    \exp\left(2\pi i(k_x x + k_y y)\right),
\end{equation}
where $K$ denotes the maximum retained Fourier wavenumber. Given sparse observations
$\mathbf{y}_{\mathrm{obs}}$, the Fourier coefficients are estimated by solving the ridge-regularized least-squares problem
\begin{equation}
    \min_{\mathbf{c}}
    \left\| \mathbf{P} \mathbf{\Phi}_K \mathbf{c}
    - \mathbf{y}_{\mathrm{obs}} \right\|_2^2
    +
    \lambda \|\mathbf{c}\|_2^2 ,
\end{equation}
where $\mathbf{P}$ is the sparse observation matrix, $\mathbf{\Phi}_K$ is the truncated Fourier basis matrix, $\mathbf{c}$ is the vector of Fourier coefficients, and $\lambda$ is the ridge regularization coefficient. The FLS baseline reconstructs each frame independently and therefore only provides a low-frequency smoothness prior, without using any temporal dynamics, PDE constraint, or learned flow-evolution prior.}

{We select the Fourier cutoff $K$ and ridge parameter $\lambda$ by a small hyperparameter search. Specifically, we first sweep $K\in\{10,16,24,32,48,64\}$ with $\lambda=0.1$, and then sweep $\lambda\in\{10^{-4},10^{-3},10^{-2},10^{-1}\}$ at $K=16$. The best searched Fourier baseline is obtained with $K=10$ and $\lambda=0.1$. Larger
values of $K$ lead to worse performance under the highly sparse observation setting, because the inverse problem becomes increasingly ill-conditioned when many high-frequency modes are fitted from only $5\%$ observations.}

\begin{figure*}[!t]
  \centering
  \includegraphics[width=0.82\textwidth]{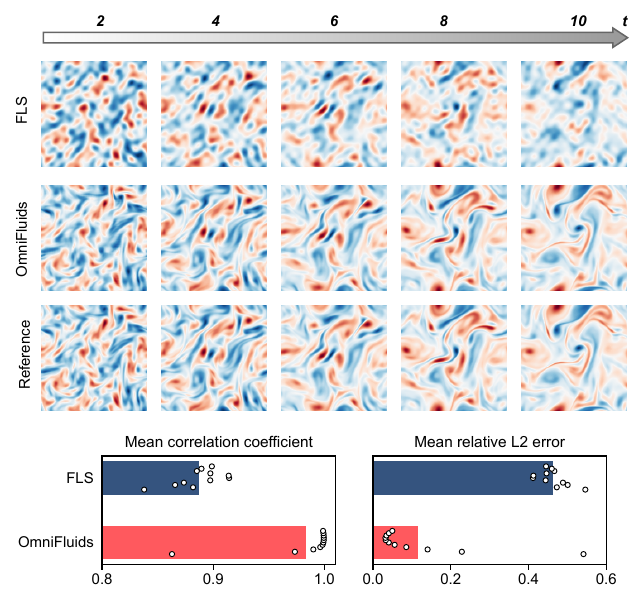}
  \caption{
  {\textbf{Sparse reconstruction from 5\% observations on 2D NSE.}
  Given the same sparse observations, FLS independently fits a truncated Fourier representation to each snapshot, while OmniFluids reconstructs the full flow trajectory using the physics-pretrained operator as a dynamical prior. \textbf{Top:} reconstructed vorticity snapshots at $t=2,4,6,8,10$ for FLS, OmniFluids, and the reference trajectory. \textbf{Bottom:} mean correlation coefficient and mean relative
  $\ell_2$ error over all evaluated frames. Open circles show all $n=11$ temporally dependent frames ($t=0,\ldots,10$) from one trajectory, not independent repeats; bars show arithmetic means.
  }}
  \label{SI-fig:appfig_reconstruction}
\end{figure*}

{Fig.~\ref{SI-fig:appfig_reconstruction} compares the reconstructed trajectories. FLS recovers part of the large-scale structure but produces overly smoothed fields and fails to reconstruct fine vortical patterns. In contrast, OmniFluids uses the physics-pretrained operator as a dynamical prior and recovers both large-scale coherent structures and small-scale vortical features more accurately. These results demonstrate that the physics-pretrained operator provides a strong reconstruction prior beyond classical low-frequency spectral interpolation, especially when only severely sparse observations are available.}

\subsection{{Localized wake-region analysis for experimental cylinder PIV}}

{
We further analyze the reconstruction quality in localized regions of the experimental cylinder-wake PIV measurements. Since the PIV field of view only covers the measured downstream wake behind the cylinder, the cylinder surface and its near-wall boundary layer are outside the camera window. Therefore, a direct evaluation of near-wall boundary-layer errors is not available for this dataset. Instead, we evaluate localized wake-region performance within the measured PIV domain. Specifically, we define a near-wake region as the upstream $30\%$ of the measured PIV window closest to the cylinder. We also define a high-vorticity region at each rollout step as the top $10\%$ of $|\omega_{\mathrm{ref}}|$ from the reference PIV field. This high-vorticity region captures separation-related wake structures, including separated shear layers and vortex cores convected downstream from the cylinder.
}

\begin{figure*}[!t]
  \centering
  \includegraphics[width=0.82\textwidth]{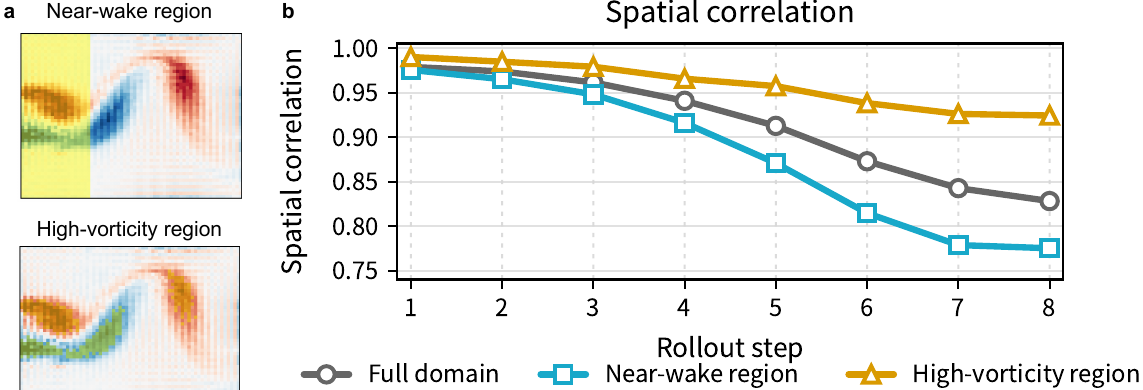}
  \caption{
  {\textbf{Localized wake-region analysis for experimental cylinder PIV measurements.}
  \textbf{a,} Localized evaluation regions in the measured downstream wake. The near-wake region is defined as the upstream $30\%$ of the measured PIV window closest to the cylinder. The high-vorticity region is defined at each rollout step as the top $10\%$ of $|\omega_{\mathrm{ref}}|$ from the reference PIV field, corresponding to separation-related shear layers and vortex cores. \textbf{b,} Spatial correlation between the reconstructed and reference vorticity fields over the 8-step rollout, evaluated over the full measured domain, near-wake region, and high-vorticity region.
  }}
  \label{SI-fig:appfig_cylinder_piv_localized}
\end{figure*}

{
Fig.~\ref{SI-fig:appfig_cylinder_piv_localized} reports the localized wake-region analysis over the 8-step rollout. The model maintains high spatial correlation in all evaluated regions. In particular, the high-vorticity region shows the highest correlation throughout the rollout, indicating that OmniFluids preserves the dominant vortical structures in dynamically challenging separation-related regions. The near-wake region is more challenging due to stronger local gradients and inflow of separated shear layers into the measurement window, but the reconstruction remains structurally consistent with the PIV reference.
}

\subsection{{Additional qualitative and quantitative results}}

\begin{figure}[t!]
  \centering
   \includegraphics[width=1\linewidth]{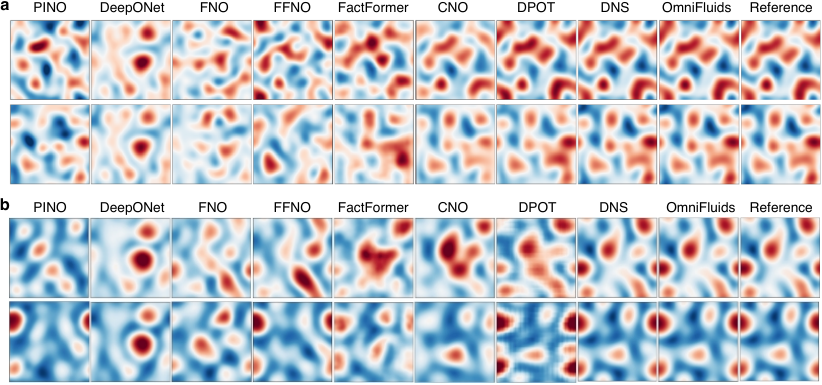}
   \caption{\textbf{Additional predicted snapshots of baseline methods and OmniFluids on 2D KSE.} \textbf{a}. In-distribution snapshots ($t=5$) with parameters $\alpha=0.2, \beta=0.5$.
\textbf{b}. Out-of-distribution snapshots ($t=5$) with parameters $\alpha=1.0, \beta=1.0$.}
   \label{SI-fig:7app}
\end{figure}

\begin{figure}[t!]
  \centering
   \includegraphics[width=1\linewidth]{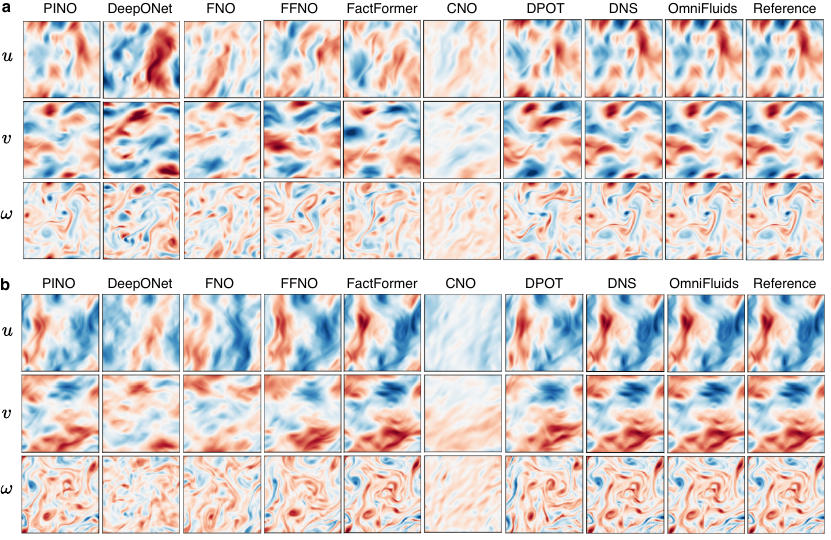}
\caption{\textbf{Additional predicted snapshots of OmniFluids and baseline methods on 2D NSE.} \textbf{a} and \textbf{b}. Predicted velocity ($\mathbf{u} = [u,v]$) and vorticity fields ($\omega$) at $t=10$ under ID ($\text{Re}=2000$) and OOD ($\text{Re}=4000$) settings, respectively.}
\label{SI-fig:8app}
\end{figure}

\begin{figure}[t!]
  \centering
   \includegraphics[width=1\linewidth]{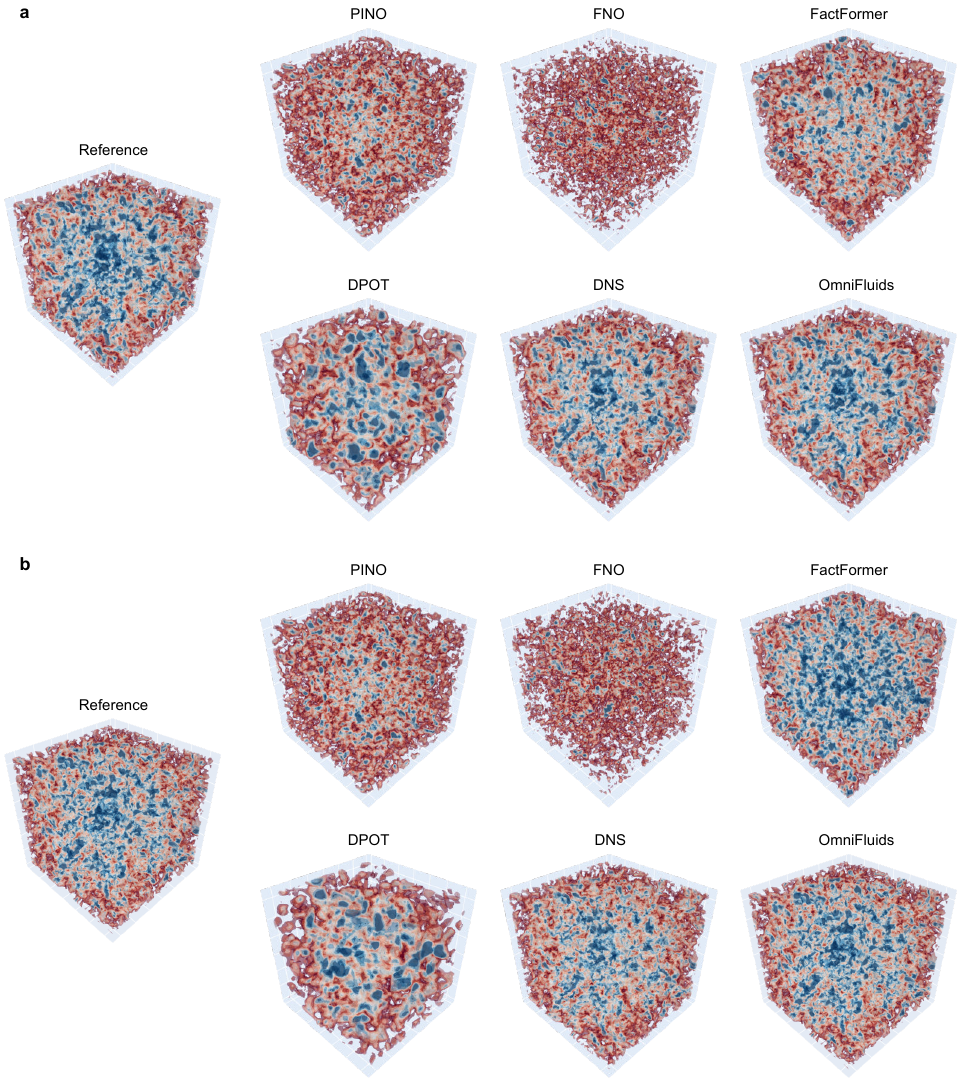}
\caption{\textbf{{Additional isosurface renderings} of the predicted 3D velocity magnitude at $t=8$ under the ID setting ($\mathrm{Re} = 2000$, \textbf{a}) and OOD ($\mathrm{Re}=3000$, \textbf{b}) settings.}}
\label{SI-fig:9app}
\end{figure}

We provide extended results to further demonstrate the robustness, accuracy and generalization capability of OmniFluids across 2D and 3D fluid dynamics benchmarks in Fig.~\ref{SI-fig:7app}, Fig.~\ref{SI-fig:8app} and Fig.~\ref{SI-fig:9app}.

\end{document}